\documentclass[a4paper,fleqn,usenatbib]{mnras}

\usepackage{latexsym}
\usepackage{amssymb}
\usepackage{amsfonts}
\usepackage{amsmath}
\usepackage{bm}
\usepackage{graphicx}
\usepackage{subfigure}
\usepackage{times}
\usepackage{units}
\usepackage{hyperref}
\usepackage{multirow}
\usepackage{comment}
\usepackage{ulem}
\usepackage{hyperref}
\usepackage{graphicx}

\usepackage{pifont} % \cmark \xmark % http://ctan.org/pkg/pifont
\newcommand{\cmark}{\ding{51}}%
\newcommand{\xmark}{\ding{55}}%

\usepackage{enumitem}
\setlist[itemize]{noitemsep, topsep=0pt}

\def\p{\partial}

\def\i{{\rm i}}

\def\kt2{\kappa^\text{T}_2}

\newcommand{\M}{\mathcal M} 
\newcommand{\Msun}{\ensuremath{M_\odot}}

\newcommand{\gccm}{\mathrm{g~cm^{-3}}} 
\newcommand{\Mmax}{\ensuremath{M_\text{max}^\text{TOV}}}

\newcommand{\be}{\begin{equation}}
\newcommand{\ee}{\end{equation}}

\usepackage{color}

\definecolor{cyan}{rgb}{0,0.9,0.9}
\definecolor{orange}{rgb}{0.9,0.5,0}
\definecolor{magenta}{rgb}{1,0,1}
\definecolor{purple}{rgb}{0.8,0.4,0.8}
\definecolor{gray}{rgb}{0.8242,0.8242,0.8242}

\newcommand{\pa}[1]{{\textcolor{red}{AP: #1}}}
\newcommand{\dr}[1]{{\textcolor{purple}{DR: #1} }}
\newcommand{\dredit}[2]{\textcolor{purple}{\sout{#1} #2}}
\newcommand{\vn}[1]{{\textcolor{purple}{VN: #1}}}
\newcommand{\bs}[1]{{\textcolor{orange}{SB: #1}}}
\newcommand{\ea}[1]{{\textcolor{magenta}{AE: #1}}}
\newcommand{\fz}[1]{{\textcolor{green}{FZ: #1}}}

\newcommand{\tbw}[1]{\textcolor{blue}{[Write:] #1}} 
\newcommand{\todo}[1]{\textcolor{red}{$\blacksquare$ TODO: #1}} 
\newcommand{\red}[1]{\textcolor{red}{#1}} 
\newcommand{\newtxt}[1]{\textcolor{green}{#1}} 
\newcommand{\oldtxt}[1]{\red{OLD: #1}}

\def\COMMENTSOFF{} % True
\let\COMMENTSOFF\undefined % false
\ifdefined\COMMENTSOFF
\renewcommand{\pa}[1]{}
\renewcommand{\dr}[1]{}
\renewcommand{\dredit}[2]{}
\renewcommand{\vn}[1]{}
\renewcommand{\bs}[1]{}
\renewcommand{\ea}[1]{}
\renewcommand{\fz}[1]{}
\renewcommand{\tbw}[1]{}
\renewcommand{\todo}[1]{}
\renewcommand{\red}[1]{}
\renewcommand{\newtxt}[1]{}
\renewcommand{\oldtxt}[1]{}
\fi

\renewcommand{\cite}[1]{\citep{#1}}
\title[Accretion-induced prompt BH formation in BNS mergers]{Accretion-induced prompt black hole formation in asymmetric neutron star mergers, dynamical ejecta and kilonova signals}

\author[S. Bernuzzi et al.]{%
Sebastiano Bernuzzi$^{1}$,
Matteo Breschi$^{1}$,
Boris Daszuta$^{1}$,
Andrea Endrizzi$^{1}$,
\newauthor
Domenico Logoteta$^{2,3}$,
Vsevolod Nedora$^{1}$,
Albino Perego$^{4,5}$,
David Radice$^{6,7}$,
\newauthor
Federico Schianchi$^{1}$,
Francesco Zappa$^{1}$,
Ignazio Bombaci$^{2,3}$,
Nestor Ortiz$^{1,8}$
\\
$^1$Theoretisch-Physikalisches Institut, Friedrich-SchillerUniversit\"{a}t Jena, 07743, Jena, Germany\\
$^2$Dipartimento di Fisica, Universit\`{a} di Pisa, Largo Pontecorvo 3, 56127 Pisa, Italy\\
$^3$Istituto Nazionale di Fisica Nucleare (INFN), Largo Pontecorvo 3,56127 Pisa, Italy\\
$^4$Dipartimento di Fisica, Universit\`{a} di Trento, Via Sommarive 14, 38123 Trento, Italy\\
$^5$INFN-TIFPA, Trento Institute for Fundamental Physics and Applications, via Sommarive 14, I-38123 Trento, Italy\\
${}^6$Department of Physics, The Pennsylvania State University, University Park, PA 16802, USA\\
${}^7$Department of Astronomy \& Astrophysics, The Pennsylvania State University, University Park, PA 16802, USA\\
${}^8$Instituto de Ciencias Nucleares, Universidad Nacional Aut\'onoma de M\'exico,
Circuito Exterior C.U., A.P. 70-543, M\'exico D.F. 04510, M\'exico}

\date{%\today
  Accepted XXX. Received YYY; in original form ZZZ
}
\begin{document}
\label{firstpage}
\maketitle

\date{\today}

\begin{abstract}
  We present new numerical relativity results of neutron star
  mergers with chirp mass $1.188\Msun$ and 
  mass ratios $q=1.67$ and $q=1.8$ using finite-temperature
  equations of state (EOS), approximate neutrino 
  transport and a subgrid model for magnetohydrodynamics-induced turbulent viscosity. The EOS
  are compatible with nuclear and astrophysical constraints and
  include a new microphysical model derived from  
  ab-initio calculations based on the Brueckner-Hartree-Fock approach.
  We report for the first time evidence for accretion-induced
  prompt collapse in high-mass-ratio mergers, in which 
  the tidal disruption of the companion 
  and its accretion onto the primary star determine prompt black hole
  formation. As a result of the tidal disruption, an accretion disc of neutron-rich and cold
  matter forms with baryon masses ${\sim}0.15\Msun$, and it is significantly
  heavier than the remnant discs in equal-masses prompt collapse mergers. 
  Massive dynamical ejecta of order ${\sim}0.01\Msun$ also originate
  from the tidal disruption. They are neutron rich and expand from the
  orbital plane with a crescent-like geometry. Consequently, bright, 
  red and temporally extended kilonova emission is predicted from 
  these mergers. 
  Our results show that prompt black hole mergers can power bright
  electromagnetic counterparts for high-mass-ratio binaries, and that
  the binary mass ratio can be in principle constrained from multimessenger
  observations.
\end{abstract}


%%%%%%%%%%%%%%%%%%%%%%%%%%%%%%%%%%%%%%		
\section{Introduction}
\label{sec:intro}

Binary neutron stars (BNS) mergers are key astrophysical laboratories
to explore the fundamental interactions in dynamical and strong
gravity. This was clearly demonstrated by the observation of GW170817
\cite{TheLIGOScientific:2017qsa,Abbott:2018wiz,LIGOScientific:2018mvr}
and its related counterparts \cite{GBM:2017lvd}. After GW170817, a
second event (GW190425) compatible with a BNS source was reported,
indicating a merger rate of 250-2810 Gpc$^3$ per
year \cite{Abbott:2020uma}.  The interpretation of current and
future observations rely on quantitative simulations of
astrophysically relevant binaries in the framework of numerical
relativity (NR). In particular, observational signatures are strongly
dependent on the possible NS masses and the still uncertain equation
of state (EOS). The latter determine the properties of the final
compact object, of the eventual accretion merger remnant and of the
observed gravitational and electromagnetic spectra (see
\citealt{2020arXiv200203863R} and reference therein for a recent review by
some of us.)

The possible NS mass range is ${\sim} 0.9 - 3 \Msun$, where the lower
bound is inferred from the formation scenario (gravitational collapse)
and from current observations
e.g.~\cite{Rawls:2011jw,Ozel:2012ax}. The upper bound is inferred from
a stability argument (Buchdahl limit) and from precise measurements of
${\sim} 2\Msun$ NSs in compact binaries containing a millisecond
pulsar and a white dwarf
~\cite{Demorest:2010bx,Antoniadis:2013pzd,Cromartie:2019kug}.
Coalescing cicularized BNS were long expected to have nearly equal
masses NS with individual masses around $M_A\sim1.35-1.4\Msun$ and
individual spin periods above the millisecond
\cite{Lattimer:2012nd,Kiziltan:2013oja,Swiggum:2015yra}.
For example, the source of GW170817 has a total mass of $M\simeq
2.73-2.77\Msun$ and a mass ratio $q\sim1$ (see below), with the
largest uncertainties coming from the spin prior utilized in the
analysis \cite{Abbott:2018wiz}.  This expectation was however
challenged by GW190425 that is associated to the heaviest BNS source
known to date with $M\simeq 3.2-3.7\Msun$ \cite{Abbott:2020uma}.
Spins distributions in GW170817 and GW190425 are both compatible with
zero~\cite{Abbott:2018wiz,Abbott:2020uma}.

The mass ratio distribution in BNS (here conventionally defined as the
ratio between the most massive primary and the secondary NS,
i.e. $q \equiv M_A/M_B \geq 1$) is very uncertain. BNS population from pulsar
observations indicate mass ratios $1\leq q\lesssim1.4$
\cite{Lattimer:2012nd,Kiziltan:2013oja,Swiggum:2015yra}.
The mass ratio of GW170817 could be as high as $q\sim1.37$
($q\sim1.89$) for low (high) spin priors.  Similarly, the mass ratio
of GW190425 can be as high as $q\sim1.25$ ($q\sim2.5$).  Given the
expected mass values and the recent observations, it is accepted that
BNS mass ratios can reach ``extreme'' mass ratio $q\lesssim2$.  While
these values are not as extreme as those that can be reached in black
hole binaries, significant differences for the remnant and radiation signals are 
expected for BNS with $q\sim1$ and $q\sim2$.

Numerical relativity simulations with microphysical EOS performed so
far focused on comparable-masses cases and mass ratios $q\lesssim1.4$
\cite{Sekiguchi:2011zd,Sekiguchi:2011mc,Neilsen:2014hha,Sekiguchi:2015dma,Palenzuela:2015dqa,Bernuzzi:2015opx,Sekiguchi:2016bjd,Radice:2016dwd,Lehner:2016lxy,Radice:2016rys,Radice:2017lry,Radice:2018xqa,Radice:2018ghv,Radice:2018pdn}. The
highest mass ratios of $q=1.5$ and $q=2$ have been simulated with a
very stiff piecewise polytropic
EOS \cite{Dietrich:2015pxa,Dietrich:2016hky}, that is currently
disfavored by the GW170817 observation.  Mergers of BNS with total
mass $M\sim2.7-2.8\Msun$ and moderate mass ratios up to $q\lesssim1.4$
with a EOS supporting ${\sim}2\Msun$, are likely to produce remnants
that are at least temporarily stable against gravitational collapse to
black hole (BH), as opposed to remnants that collapse immediately to
BH (prompt BH formation). However, the conditions for prompt BH
formation at high-$q$ have not been studied in detail to date.  For a
given total mass, moderate mass ratios can extend the remnant lifetime
with respect to equal mass BNS because of the less violent fusion of
the NS cores and a partial tidal disruption that distribute angular
momentum at larger radii in the
remnant \cite{Bauswein:2013jpa}. However, large mass asymmetries
$q\gtrsim1.6$ can favor BH formation due to the larger mass of
the primary NS.

Tidal disruption in asymmetric BNS can significantly affect the
properties of the dynamical ejecta, favouring a redder kilonova
peaking at late times (see e.g. \citealt{Rosswog:2017sdn,Wollaeger:2017ahm}).
Moderate mass ratios up to $q\sim1.3-1.4$ are found to produce more
massive discs than $q=1$ BNS
\cite{Shibata:2003ga,Shibata:2006nm,Kiuchi:2009jt,Rezzolla:2010fd,Dietrich:2016hky}.
But black hole formation can significantly alter the remnant disc
properties, both in terms of compactness and
composition \cite{Perego:2019adq}. In turn, this can impact the
secular (viscous) ejecta component and the kilonova, with bright
emissions generally favoured by the presence of a long-lived NS
remnant, e.g. \cite{Radice:2018pdn,Nedora:2019jhl}.
Moreover, assuming the Blandford-Znajek mechanism to be the mechanism launching 
the relativistic jet that produces a gamma-ray burst, a more massive
disc is also expected to power a more energetic jet through a more
intense accretion process \cite{Shapiro:2017cny}.
Many of these aspects are currently not well quantified and they
require NR simulations of high mass ratio BNS with microphysics.

In this work, we perform 32 new NR simulations with microphysical EOS,
fixed chirp mass $\M_c=1.188\Msun$ and mass ratios up $q=1.8$ for four
microphysical EOS, including a new microscopic EOS BLh
(Sec.~\ref{sec:eos}).
The simulations show that for sufficiently high value of the mass
ratio (and in a EOS dependent way) the remnant promptly collapses to
BH as consequence of the accretion of the companion on the massive
primary NS (Sec.~\ref{sec:remnant}.)  These prompt collapse dynamics
is not well described by current NR fitting formulas.
By analysing the gravitational waveforms, we further verify current
quasiuniversal NR relations for the merger and postmerger
gravitational waveforms in the high-$q$ limit (Sec.~\ref{sec:gw}.) We
find an overall agreement of the merger relations and characteristic
postmerger GW frequencies. But the accurate modeling of postmerger
waveforms with high-$q$, will require more simulations and improved
methods than those currently employed.
We discuss in detail the differences in the dynamical ejecta between
the $q=1$ and the high-$q$ mergers in terms of overall ejecta masses,
morphology and composition (Sec.\ref{sec:ejecta}.)  High mass ratio
and large chirp mass leading to prompt BH formation maximize the
dynamical tidal ejecta mass, which is expelled with a peculiar
geometry.  The $r$-process nucleosynthesis in these neutron-rich
ejecta result in bright (more luminous than the $q=1$ case), redder,
and temporally extended kilonovae (Sec.~\ref{sec:kn}.)

We employ SI units in most of the paper except for masses, reported in
solar masses ($\Msun$), lengths in km, and densities reported in
$\gccm$.  Nuclear density is indicated as $\rho_0\approx
2.3\times10^{14}$~$\gccm$.  If units are not reported, we then use
geometric units $c=G=1$ in context where those are more appropriate
(e.g. Sec.~\ref{sec:gw} and appendices).

%%%%%%%%%%%%%%%%%%%%%%%%%%%%%%%%%%%%%%		
\section{Equations of state}
\label{sec:eos}

In this work we consider four finite-temperature, composition
dependent EOS: the LS220 EOS \cite{Lattimer:1991nc}, the SFHo
EOS \cite{Steiner:2012rk}, the SLy4-SOR
EOS \cite{daSilvaSchneider:2017jpg}; and the BLh
EOS \cite{Bombaci:2018ksa}.  All these EOS include neutrons ($n$),
protons ($p$), nuclei, electron, positrons, and photons as relevant
thermodynamics degrees of freedom.  Cold, neutrino-less
$\beta$-equilibrated matter described by these microphysical EOS
predicts NS maximum masses and radii within the range allowed by
current astrophysical constraints, including the recent GW constraint
on tidal deformability
\cite{TheLIGOScientific:2017qsa, Abbott:2018wiz, De:2018uhw, Abbott:2018exr} (see below).
All four models have symmetry energies at saturation density within
experimental bounds. However, LS220 has a significantly steeper
density dependence of its symmetry energy than the other models, see
e.g. \cite{Lattimer:2012xj,Danielewicz:2013upa}, and it could possibly
underestimate the symmetry energy below saturation density.

The LS220 EOS is based on a non-relativistic Skyrme interaction with
the modulus of the nuclear bulk incompressibility set to 220
MeV. Non-homogeneous nuclear matter is modelled by a compressible
liquid-drop model including surface effects, and considers an ideal,
classical gas formed by $\alpha$ particles and heavy nuclei. The
latter are treated within the single nucleus approximation (SNA). The
transition between homogeneous and non-homogeneous matter is performed
through a Gibbs construction.

The SFHo EOS combines a relativistic mean field approach for the
homogeneous nuclear matter to an ideal, classical gas treatment of a
statistical ensemble of several thousands of nuclei in Nuclear
Statistical Equilibrium (NSE) for the inhomogeneous nuclear matter.
The transition between the two phases is achieved by an excluded
volume mechanism.

The SLy4 Skyrme parametrization was introduced
in \citet{Douchin:2001sv} for cold nuclear and NS matter. In this work
we employ its extension to finite temperature presented
in \citet{daSilvaSchneider:2017jpg} using an improved version of the
LS220 model that includes non-local isospin asymmetric terms and a
better treatment of nuclear surface properties, and treats the size of
heavy nuclei more consistently. The transition between uniform and
non-uniform phase is achieved by a first order transition,
i.e. choosing the phase with lower free energy.

A main novelty of this work is the use of the BLh EOS, a new finite temperature EOS
derived in the framework of non-relativistic Brueckner-Hartree-Fock (BHF) approach 
(Logoteta et al, in preparation).   
The corresponding cold, $\beta$-equilibrated EOS was first presented in \citet{Bombaci:2018ksa}
and applied to BNS mergers in \citet{Endrizzi:2018uwl}. 
For the homogeneous nuclear phase, this EOS employs a purely microphysical approach 
based on a specific nuclear interaction.  
Consistently with \citet{Bombaci:2018ksa}, the interactions 
between nucleons is described through a potential derived
perturbatively in Chiral-Effective-Field
theory \cite{Machleidt:2011zz}. Specifically, the local potential
reported in \citet{Piarulli:2016vel} and calculated up to next to-next
to-next to-leading order (N3LO) was used as two-body interaction. This 
potential takes into account the possible excitation of a 
$\Delta$-resonance in the intermediate states of the nucleon-nucleon
interaction. The above potential was then supplemented by a
three-nucleon force calculated up to N2LO and including again the
contributions from the $\Delta$-excitation.  
The parameters of the three-nucleon force were determined to
reproduce the properties of symmetric nuclear matter at saturation
density \cite{Logoteta:2016nzc}. For the non-homogeneous nuclear phase
there is no straightforward extension of these microphysical methods
to sub-saturation densities. Thus, the low density part ($n \leq
0.05~{\rm fm^{-3}} $) of the SFHo EOS has been smoothly connected to
the high density BLh EOS. This necessary extension has been tested 
with different finite-temperature, composition-dependented 
tabulated EOS \cite{Hempel:2009mc}. They all use 1) relativistic mean
field approaches for the homogeneous phase, 2) an ideal, classical gas
of a statistical ensemble of several thousands of nuclei in NSE for
the non-homogeneous nuclear phase; 3) an excluded volume mechanism to
model the transition. No appreciable differences
were found in all relevant quantities at subnuclear densities between
different high density treatments.

The LS220 and SLy4-SRO EOS are based on Skyrme effective nuclear
interactions. %\cite{daSilvaSchneider:2017jpg}.
In these models thermal effects are introduced starting from a zero temperature
internal energy functional that contains an explicit nuclear density
dependence. The interaction part of this functional is split into a
quadratic term in the nuclear density (playing the role of a two-body
nucleon-nucleon interaction) plus a term proportional to some power of
the nuclear density. The latter term mimics the effect of many-body
nuclear forces.  The temperature dependence of the effective nuclear
interaction is encoded in the effective mass dependence of the kinetic
energy as well as in the single particle potentials. The latter are
calculated by the variation of the internal energy with respect to the
neutron and proton densities.  Assuming indeed a constant entropy,
smaller effective masses translate into larger kinetic energies and
thus higher matter temperatures.  The LS220 EOS assumes that the
effective nucleon mass is the bare nucleon mass at all densities while
for the SLy4-SRO we have $m_N^*/m_N = 0.695$ at saturation density,
being $m_N^*$ and $m_N$ the effective and the bare nucleon masses
respectively.

In the relativistic Lagrangian underlaying the SHFo EOS, nuclear
interactions are described by $\sigma$-, $\omega$- and $\rho$-meson
exchanges. The resulting Euler-Lagrange equations are then solved in
mean field approximation. In this approach thermal effects are
included by introducing Fermi-Dirac distributions at finite
temperatures for the various nuclear species.  Mesons and nucleon
fields, and consequently all thermodynamical quantities, acquire
automatically a temperature dependence through the self consistent
solution of the mean field equations.

Differently from the other models considered in the present work,
thermal effects enter in a quite different way in the BLh EOS. The
calculation of the Free energy in the BHF approach
\cite{Bombaci:1993uyx} requires first the determination of an effective in-medium nuclear interaction, 
starting from the bare nuclear potential.  This effective interaction
($G$-matrix) is obtained by solving the Bethe-Goldstone integral
equation which describe the nucleon-nucleon scattering in the nuclear
medium and properly takes into account the Pauli principle. Finally,
the nucleon single particle potentials $U_i(k,T)$ ($i=n,\ p$) is
obtained through the integration of the on-shell
$G$-matrix. $U_i(k,T)$ is a sort of mean field felt by a nucleon of
momentum $k$ due to the presence of the surrounding nucleons.  The
determination of $U_i(k,T)$ allows for the calculation of the Free
energy from which all the other thermodynamical quantities can be
derived.  The procedure described above is complicated by the
non-linear and non-local dependence of $U_i(k,T)$ in Bethe-Goldstone
equation.
We finally note that this scheme provides many-body correlations which
are beyond the mean field approximation. Such correlations are not
present in the other EOS models considered in the present paper.

\subsection{EOS Constraints and NS equilibrium models}

\begin{figure*}
  \includegraphics[width=0.49\textwidth]{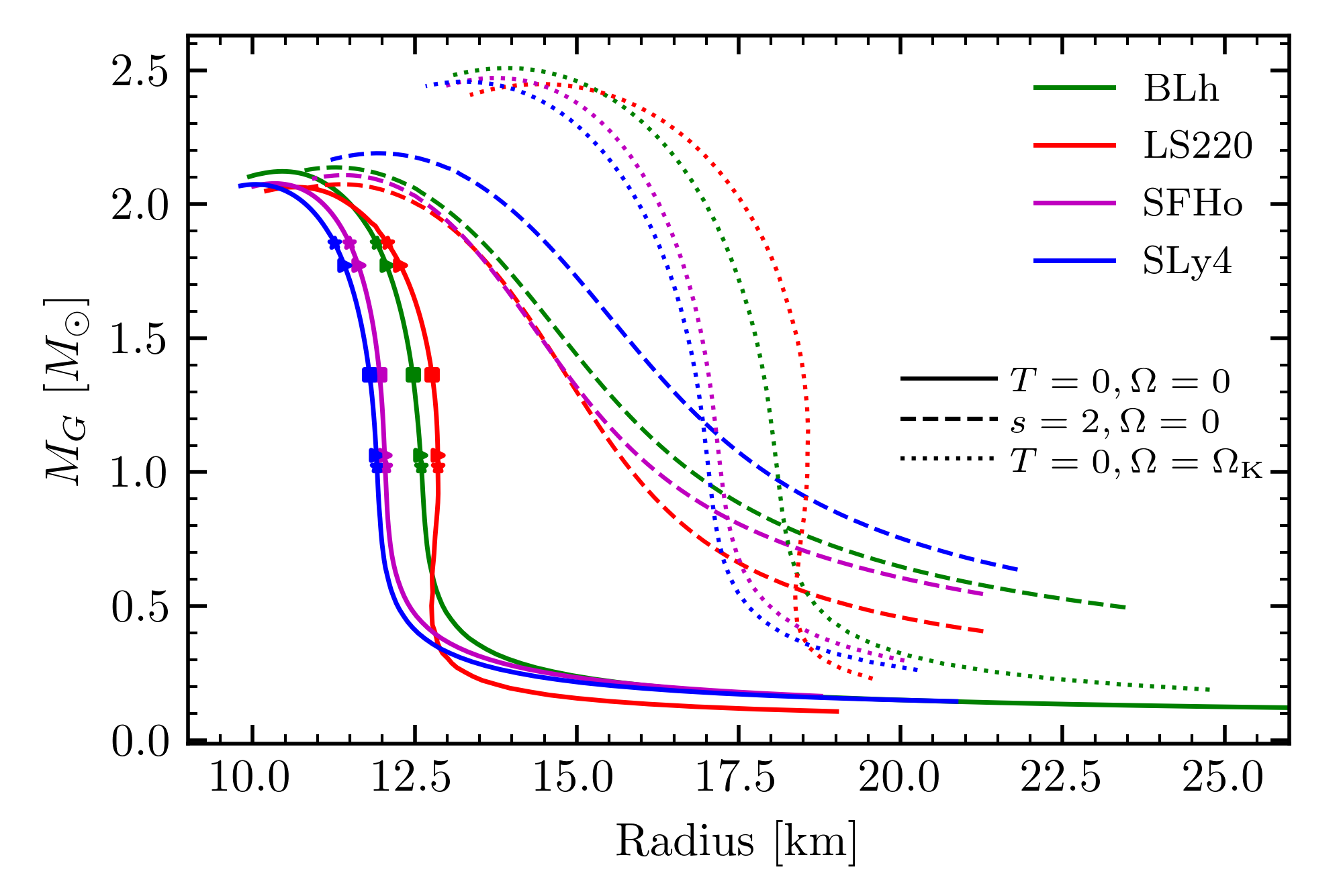}
  \includegraphics[width=0.49\textwidth]{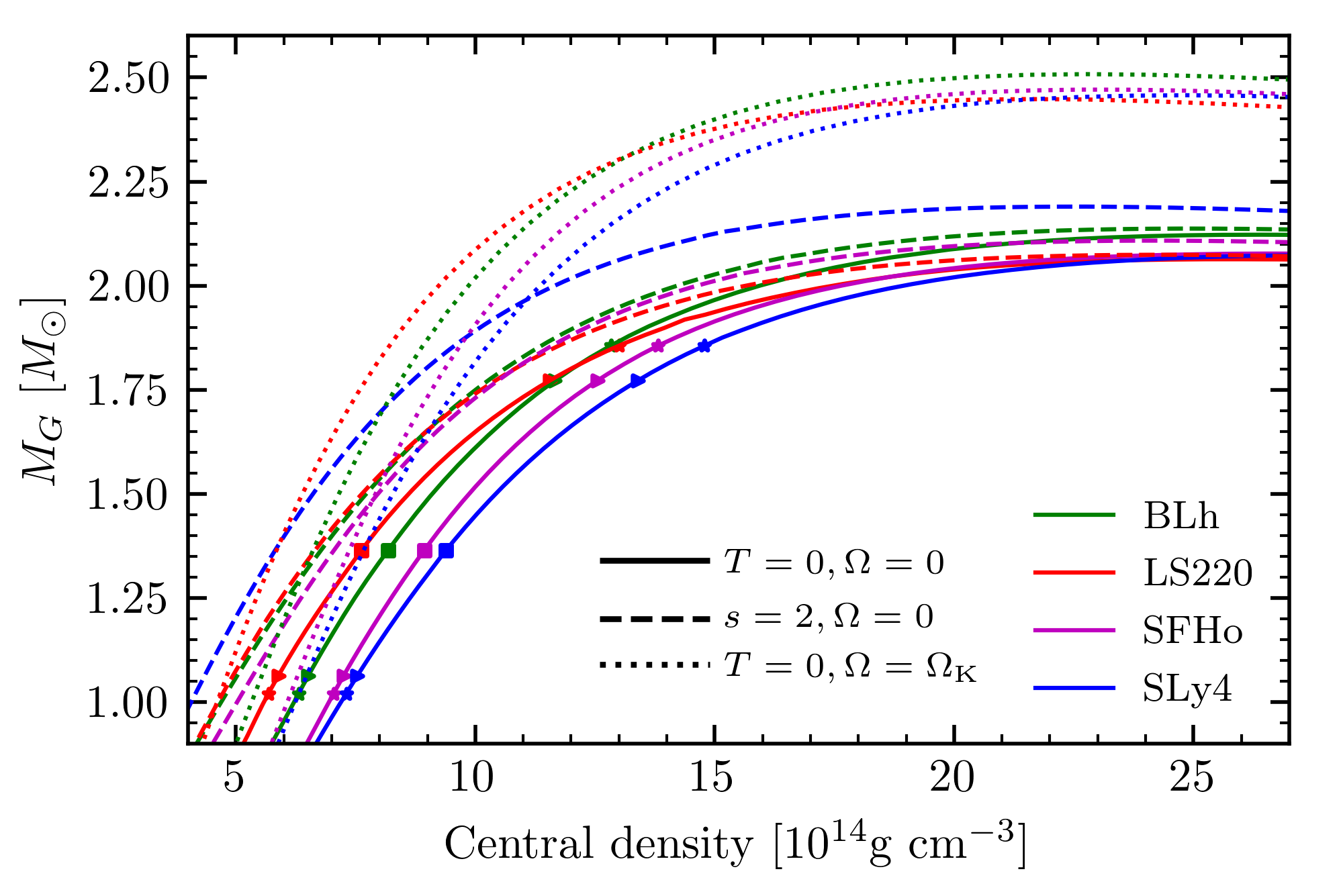}
\centering
\caption{Equilibrium NS sequences obtained for the different EOS used in this work. 
Solid lines correspond to irrotational, $T=0$ configurations;
dashed lines to irrotational, isentropic ($s=2~{k_{\rm B}~{\rm baryon^{-1}}}$) configurations;
dotted lines represent $T=0$, rigidly rotating NS spinning at the mass sheeding limit ($\Omega = \Omega_{K}$). 
Left: gravitational mass versus radius (equatorial for rotating
NS). Right: gravitational mass versus central density. 
Markers along the cold, nonrotating sequences indicate the NSs used in this work: 
squares, triangles and stars refer to $q=1,1.67,1.8$ binaries, respectively.}
\label{fig:NSequilibria}
\end{figure*}

Fundamental differences in the EOS models translate in different NS structures.
For cold, non-rotating NSs, the considered EOS can support maximum 
masses in the range $\Mmax\sim 2.06-2.10\Msun$, while the predicted radii of a 1.4$\Msun$ 
NS lay in the range $R_{1.4}\sim 11.78-12.74$ km. More specifically, LS220, SFHo, SLy4-SRO, and
BLh EOS have $\Mmax$ of 2.04, 2.06, 2.06, and 2.10 $M_\odot$, and $R_{1.4}$ of 12.8, 
12.0, 11.9, and 12.5 km, respectively. The predicted maximum NS masses and the 1.4$\Msun$ NS radii 
are all compatible at one-sigma level with the recent detection of an extremely massive 
millisecond pulsar \cite{Cromartie:2019kug} and with results obtained by the {\it NICER} 
collaboration \cite{Miller:2019cac,Riley:2019yda}, although being systematically on the lower side. 
Note that EOS allowing NS radii $R_{1.4}\gg 13$~km are currently disfavoured by both 
GW BNS and X-ray pulsar observations \cite{Abbott:2018wiz,Miller:2019cac,Riley:2019yda}.

Finite temperature effects introduce additional pressure support. On
the one hand, for the typical central entropies expected for nuclear
matter during a BNS merger ($s \lesssim 2{k_{\rm B}}/{\rm baryon}$),
this additional support is not sufficient to significantly alter the
maximum TOV mass \cite{Kaplan:2013wra} or the central baryon density
due to the large degree of degeneracy of matter above saturation
density.  On the other hand, thermal effects can provide a more
significant impact for matter at lower densities, increasing the NS
radius.  In Fig.~\ref{fig:NSequilibria} we report equilibrium
sequences in the mass-radius and mass-central density plane obtained
for the EOS used in this work, considering both a cold (continuous
lines) and an isentropic (dashed lines) EOS with $s=2~{k_{\rm B}~{\rm
baryon^{-1}}}$.  Due to thermal effects, $R_{1.4}$ increases by the
$15.6 \%$ for the LS220 EOS and $36.4 \%$ for the SLy4 EOS.  while for
the BLh and the SFHo EOS the variation is $\sim 21-22 \%$.  The
different relative impacts on the NS radius clearly correlate with the
different values of the nucleon effective mass.

Rotational support also increases the maximum NS mass. For example, in
the limiting case of rigid rotation at the Keplerian limit, the
maximum NS mass is increased by $\sim 20\%$ for all EOS models, as
visible in Fig.~\ref{fig:NSequilibria}, dotted lines.  Since this
affects the whole star, the NS radius is typically increased by $\sim
40\%$, but at the same time the central density is decreased by a
similar amount, if one compares non-rotating and Keplerian NSs of
identical masses.  These properties emphasizes the importance of using
the full EOS (i.e. including thermal effects) in merger
simulations. Thermal (and composition, see \citep{Kaplan:2013wra})
effects are indeed key to quantify the prompt collapse dynamics,
mass-shedding in the remnant and disc properties.

%%%%%%%%%%%%%%%%%%%%%%%%%%%%%%%%%%%%%%		
\section{Simulations}
\label{sec:sim}

\subsection{Methods}
\label{sec:sim:method}

We construct initial data for irrotational binaries in quasi-circular orbit solving
the constraint equations of 3+1 general relativity in presence of a
helical Killing vector and under the assumption of a conformally flat
metric \cite{Gourgoulhon:2000nn}. The equations are solved with the
pseudo-spectral multidomain approach implemented in the
\texttt{Lorene} library~\footnote{\url{http://www.lorene.obspm.fr/}}. The EOS used for the
initial data are constructed from the minimum temperature slice of the
EOS table employed for the evolution assuming neutrino-less
beta-equilibrium. Initial data have a residual eccentricity of
${\sim}0.01$ which is radiated away before merger, e.g. \cite{Thierfelder:2011yi}.

The initial data are then evolved with the 3+1 Z4c free-evolution
scheme for Einstein's equations \cite{Bernuzzi:2009ex,
  Hilditch:2012fp} coupled to general relativistic hydrodynamics.
For the latter, we use the \texttt{WhiskyTHC} code
\cite{Radice:2012cu,Radice:2013hxh,Radice:2013xpa} that implements the approximate neutrino
transport scheme developed in \citet{Radice:2016dwd,Radice:2018pdn}
 and the general-relativistic large eddy
simulations method (GRLES) for turbulent
viscosity \cite{Radice:2017zta}. 
The interactions between the fluid and neutrinos are treated with a 
leakage scheme in the optically thick regions
\cite{Ruffert:1995fs,Rosswog:2003rv,Neilsen:2014hha} while free-streaming neutrinos
are evolved according to the M0 scheme \cite{Radice:2018pdn}. 
The latter is a computationally efficient scheme that 
incorporates an approximate treatment of gravitational and Doppler
effects, is well adapted to the geometry of BNS mergers and free of the 
radiation shock artifact that plagues the M1 scheme~\cite{Foucart:2018gis}.
The turbulent viscosity in the GRLES is parametrized as
$\sigma_T = \ell_{\rm mix} c_s$, where $c_s$
is the sound speed and $\ell_{\rm mix}$  is a free parameter 
sets the intensity of the turbulence. For the simulations of this work
$\sigma_T$ is prescribed as a function of the 
rest-mass (baryon) density using the
results of the high-resolution general relativistic
magnetohydrodynamics simulations results of a NS merger of
\citet{Kiuchi:2017zzg}. A detailed description of the model can be
found in \cite{2020arXiv200509002R}; simulations with this model are also 
presented in \citet{Perego:2019adq,Nedora:2019jhl}.

We remark that the GRLES method introduces parabolic terms for which
there is no maximum characteristic velocity. However, parabolic
equations still have an effective, wavelength dependent, propagation
speed \cite{Weymann:1967,Kostadt:2000ty}. Accordingly, only
disturbances that have spatial scale that are small compared to
the mixing length parameter are propagated with an effective velocity
larger than the speed of light.  These modes are absent in our
simulations because the mixing length is always set to be smaller than
the minimum grid scale in the simulation.  Indeed, the GRLES method
becomes invalid precisely when the mixing length becomes comparable with
the grid scale, which would correspond to turbulent motion on a scale
that is resolved in the simulations and should be included directly, not
through a subgrid model. This is also the reason why it is possible to
integrate the GRLES equations using an explicit time integration scheme.
Moreover, it is possible to show that in the long wavelength and low
frequency limit that is relevant for us, the solution of the parabolic
model are always arbitrarily close to those of an associated hyperbolic
model obtained with the introduction of relaxation terms
\cite{Nagy:1994}. In other words, the parabolic and the (significantly
more complex) hyperbolic models of turbulent viscosity should give the
same results in our context.

\texttt{WhiskyTHC} is implemented
within the \texttt{Cactus} \cite{Goodale:2002a,2007arXiv0707.1607S} framework and coupled to an
adaptive mesh refinement driver and a metric solver.
The spacetime solver is implemented in the
\texttt{CTGamma} code \cite{Pollney:2009yz, Reisswig:2013sqa}, which is
part of the \texttt{Einstein Toolkit} \cite{Loffler:2011ay}.
We use fourth-order finite-differencing for the metric's spatial
derivatives method of lines for the time evolution of both metric and
fluid. We adopt the optimal 
strongly-stability preserving third-order Runge-Kutta scheme
\cite{Gottlieb:2009a} as time integrator. The timestep is set according
to the speed-of-light Courant-Friedrich-Lewy (CFL) condition with CFL factor $0.15$.
While numerical stability requires the CFL to
be less than $0.25$, the smaller value of $0.15$ is necessary
to guarantee the positivity of the density when using the
positivity-preserving limiter implemented in \texttt{WhiskyTHC}.

The computational domain is a cube of $3,024$~km in diameter whose
center is at the center of mass of the binary. Our code uses
Berger-Oliger conservative AMR \citealt{Berger:1984zza} with sub-cycling in time and
refluxing \cite{Berger:1989a, Reisswig:2012nc} as provided by the
\texttt{Carpet} module of the \texttt{Einstein Toolkit}
\cite{Schnetter:2003rb}. We setup an AMR grid structure with 7
refinement levels. The finest refinement level covers both NSs
during the inspiral and the remnant after the merger and has a typical
resolution of $h\simeq 246$~m (grid setup named LR), $h \simeq 185$~m
(SR) or $123$~m (HR).

Black-hole formation is indicated by the appearance of an apparent
horizon (AH) that is computed with the module \texttt{AHFinderDirect}
\cite{Thornburg:2003sf}. With the
gauge conditions employed in the simulations, the BH is formed and simulated as a
puncture \cite{Thierfelder:2010dv,Dietrich:2014wja}. 
In a first series of simulations, the AH finder could not find an AH in
the simulations using the GRLES scheme. We have thus repeated them and
found that in those cases was necessary to (i) increase the number of more
guess spheres for the finder, and (ii) switch off the GRLES scheme in
regions with $\alpha<0.1$ in order to compute the AH robustly.
The latter is analogous to the "hydro-excision"
implemented in many codes to facilitate AH location.
This way we obtained horizon data for all the LR and most of the
SR simulations. We could not rerun the HR simulations for which the AH
was not found initially for lack of computational resources. 
Finally, the employed grid structure is not optimal to follow the
dynamics of the BH+disc remnant; thus simulations are stopped
${\sim}5-10$~ms after BH formation.

\subsection{BNS Models}
\label{sec:sim:models}

We consider 10 binaries with fixed chirp mass ${\cal M}_c\simeq 
1.188\Msun$ and simulate them at different resolutions. 
The chirp mass is ${\cal M}_c=M\nu^{3/5}$ where $\nu=M_AM_B/M^2=q/(1+q)^2$.
The main properties of the BNS initial data are summarized in Tab.~\ref{tab:sim}.
We simulated the equal mass case and mass ratio $q=1.67$ for all the EOS with the GRLES scheme.
The highest mass ratio simulated are $q=1.8$ for the BLh and SLy EOS.
A subset of models were simulated also without turbulent viscosity to directly assess
its impact on the merger dynamics and on the ejecta properties. 
The initial separation between the NS is set to $45\, {\rm km}$,
corresponding to ${\sim}4-6$ orbits to merger.
Note that similar equal mass LS220 and SFHo BNS were already presented in
\citealt{Perego:2019adq}, but the mass here is slightly larger.
An equal mass SLy4 without turbulent viscosity was instead presented in \citet{Endrizzi:2020lwl}.

The table also reports the reduced tidal parameter \cite{Favata:2013rwa}
\be
\label{eq:LambdaT}
\tilde\Lambda = \frac{16}{13}
\frac{(M_\text{A} + 12 M_\text{B}) M_\text{A}^4}{M^5}\Lambda_\text{A} + (A\leftrightarrow B)\ .
\ee
where $\Lambda_i\equiv 2/3 k^i_2 (GM_i/R_ic^2)^{5}$, with $i=(A,B)$, 
are the dimensionless quadrupolar tidal polarizability parameters of the individual 
stars~\cite{Flanagan:2007ix,Damour:2009wj}, $k^i_2$ 
the dimensioless quadrupolar Love
numbers~\cite{Damour:1983a,Hinderer:2007mb,Damour:2009vw,Binnington:2009bb},
and $(M_i,R_i)$ the NS mass and radius. 
The tidal parameter enters at leading-order the post-Newtonian
dynamics and it is directly measurable from the GW \cite{Damour:2009wj,Damour:2012yf}. Its range for
fiducial BNS systems is $\tilde\Lambda\approx(10,2000)$, where
softer EOS, larger masses and higher mass-ratios result in smaller
values of $\tilde\Lambda$. It can be used as a measure of the binary
compactness and correlates with prompt collapsed remnant and disc
masses \cite{Radice:2017lry,Zappa:2017xba}.

\begin{table}
\centering    
  \caption{BNS models considered in this work. $\Mmax$ is the maximum
    gravitational mass for a TOV solution with the specified EOS,
    $C^\text{TOV}_\text{max}=G\Mmax/Rc^2$ is the compactness relative to the
    maximum mass configuration. $M_b$ is the total baryonic mass of
    the BNS, $M_A$ and $M_B$ are the gravitational masses of the individual 
    NSs at infinite separation, $q$ is the mass ratio $M_A/M_B\geq1$ and
    $\tilde{\Lambda}$ is the tidal parameter of Eq.~\ref{eq:LambdaT}.
    $f_\text{GW}(0)$ is the initial GW frequency.
    Masses are expressed in $\Msun$, frequencies in Hertz.} 
\centering    
\scalebox{0.75}{%
  \begin{tabular}{ccc|cccccc|c}        
    \hline\hline
    EOS & $\Mmax$ & $C^\text{TOV}_\text{max}$ & $M_b$ & $M_A$ & $M_B$
    & $q$ & $M$ & $\tilde{\Lambda}$& $f_\text{GW}(0)$\\
         & $[\Msun]$ &  & $[\Msun]$ & $[\Msun]$ & $[\Msun]$
    & & $[\Msun]$ & & $[\text{Hz}]$\\
    \hline\hline
    BLh & $2.103$ & $0.298$ & $2.98$ & $1.364$ & $1.364$ & $1.0$& $2.728$ & $511$ &$565$\\
    BLh & $2.103$ & $0.298$ & $3.14$ & $1.772$ & $1.065$ & $1.67$& $2.837$ & $506$&$574$\\
    BLh & $2.103$ & $0.298$ & $3.21$ & $1.856$ & $1.020$ & $1.8$& $2.876$ & $504$& $576$\\
    \hline
    LS220 & $2.044$ & $0.284$ & $2.98$ & $1.364$ & $1.364$ & $1.0$& $2.728$ & $639$ &$565$\\
    LS220 & $2.044$ & $0.284$ & $3.14$ & $1.772$ & $1.065$ & $1.67$& $2.837$ & $638$&$574$\\
    \hline
    SFHo & $2.059$ & $0.294$ & $3.00$ & $1.364$ & $1.364$ & $1.0$& $2.728$ & $395$ &$565$\\
    SFHo & $2.059$ & $0.294$ & $3.16$ & $1.772$ & $1.065$ & $1.67$& $2.837$ & $386$&$573$\\
    \hline
    SLy4 & $2.055$ & $0.303$ & $3.00$ & $1.364$ & $1.364$ & $1.0$& $2.728$ & $361$&$565$\\
    SLy4 & $2.055$ & $0.303$ & $3.17$ & $1.772$ & $1.065$ & $1.67$& $2.837$ & $358$&$574$\\
    SLy4 & $2.055$ & $0.303$ & $3.24$ & $1.856$ & $1.020$ & $1.8$& $2.876$ & $357$&$577$\\ 
    \hline\hline   
  \end{tabular}
  }
 \label{tab:sim}
\end{table}

%%%%%%%%%%%%%%%%%%%%%%%%%%%%%%%%%%%%%%		
\section{Merger Dynamics \& Remnant}
\label{sec:remnant}

\begin{figure*}
  \centering
  % data/plot_scalars.py
  \includegraphics[width=\textwidth]{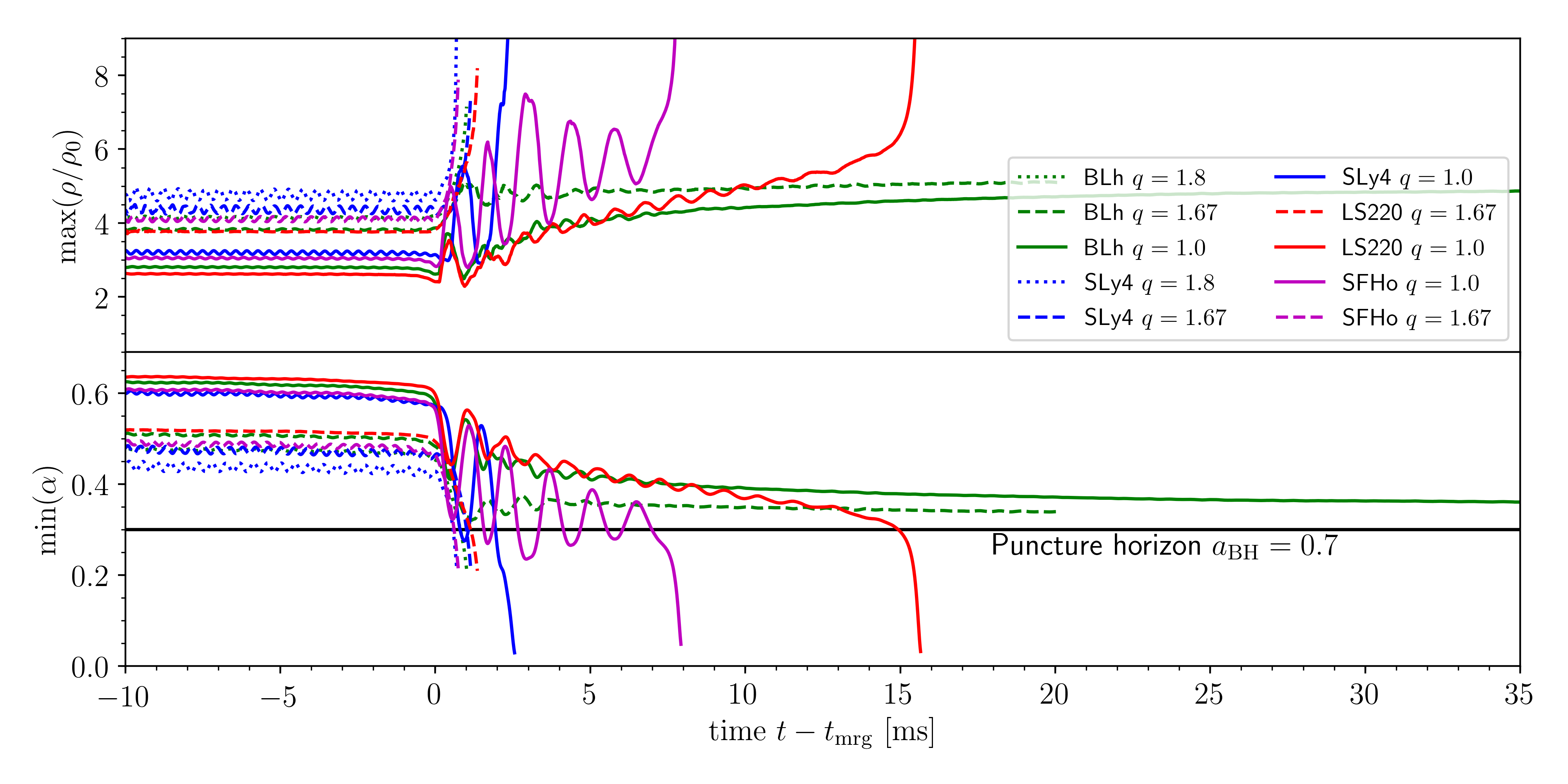}
  \caption{Evolution of the maximum density (normalized to nuclear
  saturation density) and of the minimum lapse in all simulations. The
  horizontal line indicates the value of the lapse at the horizon for
  a puncture with the indicated dimensionless spins
  (Cf. Tab.~\ref{tab:rem_prop}). Data refer to simulations with
  turbulent viscosity and resolution SR.} 
  \label{fig:overview}
\end{figure*}

\begin{figure}
\centering
\includegraphics[width=.5\textwidth]{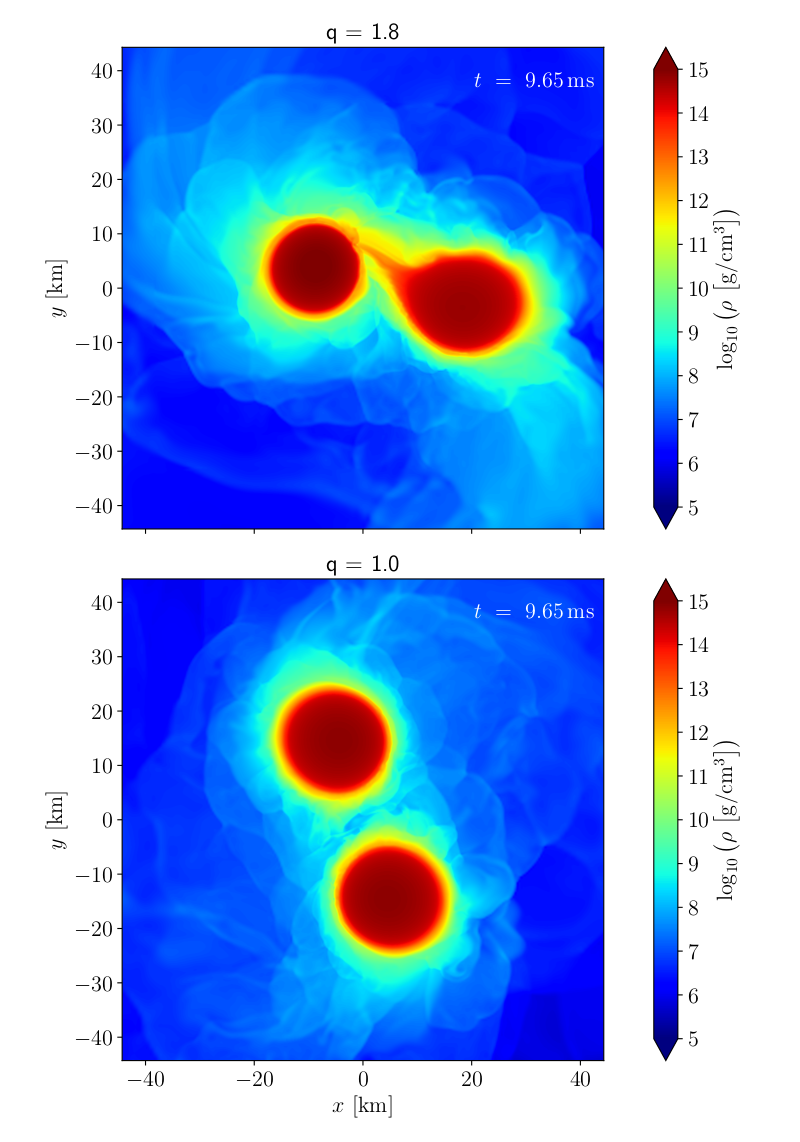}
\caption{Snapshots of premerger dynamics for BLh $q=1.8$ (top) and
  $q=1.0$ (bottom) simulations. Shown is the rest-mass density in the
  orbital plane at ${\sim}9$~ms corresponding to the third orbit from
  the beginning of the simulations and $2$ orbits to the moment
  of merger. The companion in the $q=1.8$ BNS is tidally disrupted and
  a significant accretion onto the primary is taking place. Accretion
  starts approximately after one orbits from the beginning of the simulations.}
\label{fig:rho_2d_blh_acc}
\end{figure}

Starting at a GW frequency of ${\sim}570$~Hz, the binaries revolve for
${\sim}$4-6 orbits before reaching the moment of merger. The latter is 
defined as the peak amplitude of the $(2,2)$ GW mode and
marks the end of the chirp signal.
A summary of the merger dynamics for all the runs is given in
  Fig.~\ref{fig:overview}, which shows the maximum mass density (fluid frame) and the
minimum of the lapse function, $\alpha$. 
Black-hole formation is indicated by the lapse dropping below
$\alpha\lesssim 0.3$. Note that the 1+log slicing of the 
  spherical puncture has lapse function at the horizon
  $\alpha_\text{AH}\simeq0.376$ \cite{Hannam:2006vv,Hannam:2008sg},
  but punctures formed in our simulations have dimensionless spins
  ${\sim}0.7$ for which  $\alpha_\text{AH}\simeq0.3$ (See
  Appendix~\ref{sec:punc}). 
  At the same time the lapse decreases below $\alpha_\text{AH}$, the
  maximum density increases beyond $6\rho_0$ and it is then
  unresolved on the grid due to the gauge conditions
  \cite{Thierfelder:2010dv}. 

The remnants of BLh $q=1.8$, LS220 $q=1.67$, SFho $q=1.67$ and SLy4
$q=1, 1.67,1.8$ collapse to BH within ${\sim}$2-3~ms from merger.  We
call prompt BH collapse mergers those in which the NS cores collision
has no bounce but instead the remnant immediately collapse at
formation, see Tab.~\ref{tab:rem_prop}. This usually happens within
1-2~ms from the moment of merger and can be identified by the maximum
density monotonically increasing to the collapse,
Fig.~\ref{fig:overview}.  Note that this definition of prompt collapse
implies negligible shocked dynamical ejecta because the bulk of this
mass emission comes precisely from the (first) core
bounce \cite{Radice:2018pdn}.
The BH masses of the prompt collapsed remnants are $M_\text{BH}\simeq
2.49,2.44,2.45,2.47,2.52 \Msun$ for BLh $q=1.8$, LS220 $q=1.67$, SFho
$q=1.67$ and SLy4 $q=1.67,1.8$ respectively. The BH spins are
$a_\text{BH}\simeq 0.66, 0.7, 0.68, 0.69,0.66$. 
The remnants of LS220, SFHo and SLy $q=1$ also form BH within the simulated 
time and they have $M_\text{BH}=2.41,2.41, 2.38 $ and spins $a_\text{BH}=0.5,0.75, 0.76$
respectively.
The SLy4 $q=1$ merger was simulated in \citet{Endrizzi:2020lwl} without
viscosity, and in that case the remnant survives for
${\sim}10$~ms. The earlier collapse here is a consequence of the
angular momentum redistribution by the subgrid model for turbulent
viscosity \cite{Radice:2017zta}. Overall these results for the BH
spins consistently indicate an upper limit on the BH rotation of
$a_\text{BH}\lesssim0.8$, also when including $q\sim2$
BNS \cite{Kiuchi:2010ze,Bernuzzi:2013rza,Bernuzzi:2015opx,Dietrich:2016hky}.
In the following, we first discuss the details of the BH formation
highlighting the effect of high mass ratio and the main differences
with respect to the (well-studied) equal mass cases. Then, we
discuss the properties of the remnant discs.

For comparable masses the NS cores
enter in contact before reaching the moment of
merger~\cite{Thierfelder:2011yi} and the last two 2-3 GW cycles before
the amplitude's peak are emitted by the cores collision and remnant
formation. At high mass ratios, a new effect is the tidal disruption of
the companion and its accretion onto the primary NS.
This has been reported also in
previous simulations with a stiff polytropic EOS \cite{Dietrich:2016hky},
and we confirm it here for softer and microphysical EOS. As a
representative example we show in Fig.~\ref{fig:rho_2d_blh_acc} the
cases of BLh $q=1$ vs. $q=1.8$. The accreting material has initially low
temperatures but as soon as the accretion becomes more massive and
faster the temperature raises. At approximately the time of the
snapshot the accreting material shocks against the primary NS core and
there the temperature raises up to ${\sim}100$~MeV. As a consequence
of this shock, some material becomes unbound, although the exact amount
of ejecta cannot be confidently measured in the simulations 
(see Sec.~\ref{sec:ejecta}).

\begin{figure*}
\begin{minipage}{\linewidth}
\begin{minipage}{0.3\linewidth}
\includegraphics[width=\textwidth]{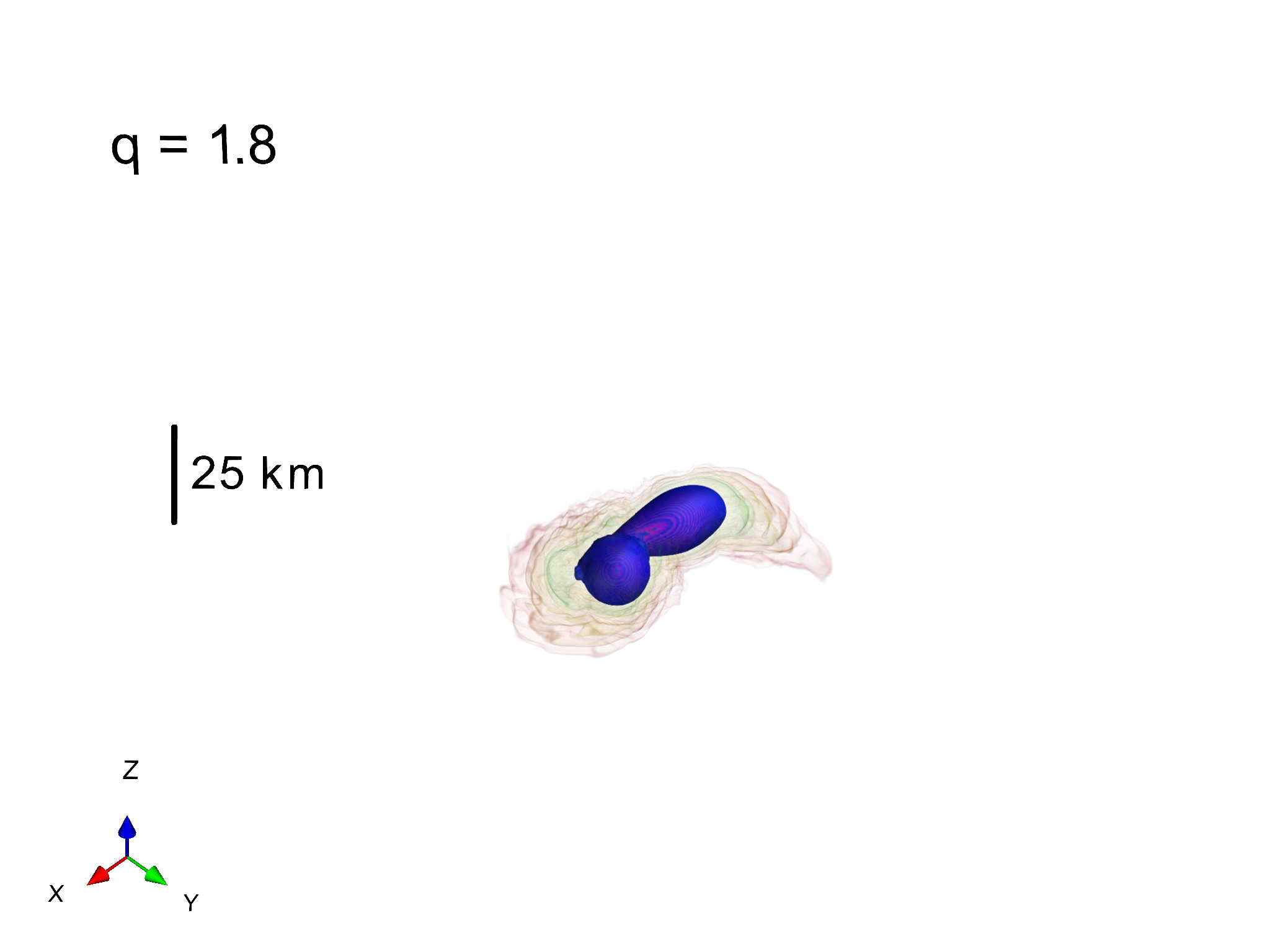}
\end{minipage}
\begin{minipage}{0.3\linewidth}
\includegraphics[width=\textwidth]{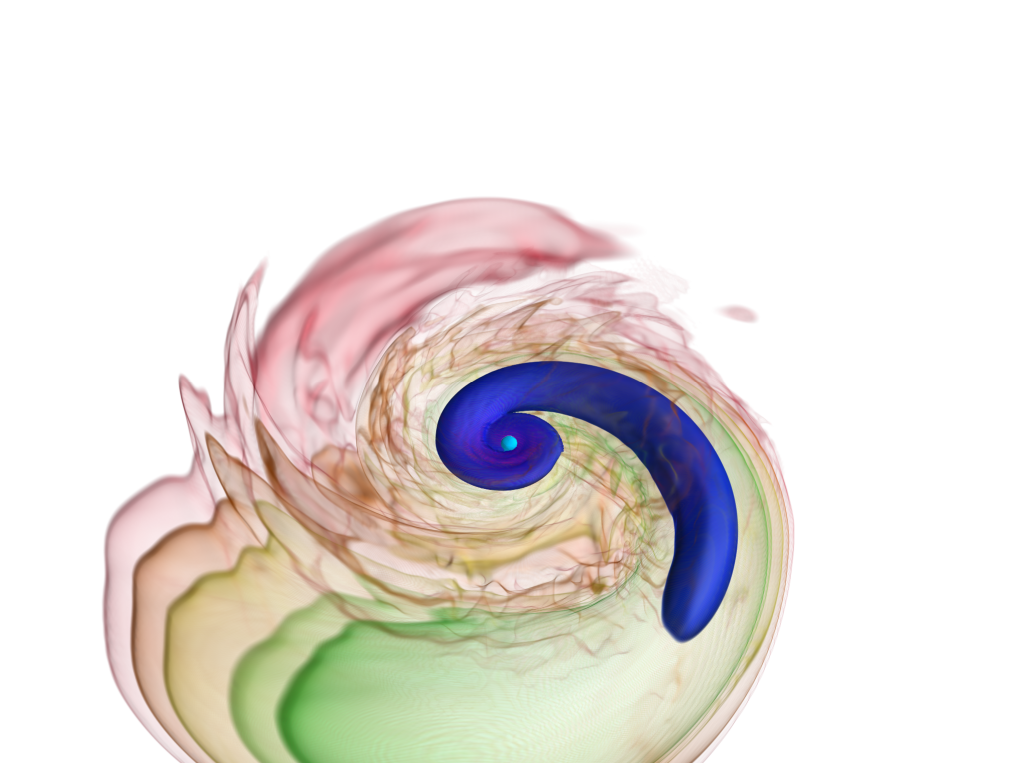}
\end{minipage}
\begin{minipage}{0.3\linewidth}
\includegraphics[width=\textwidth]{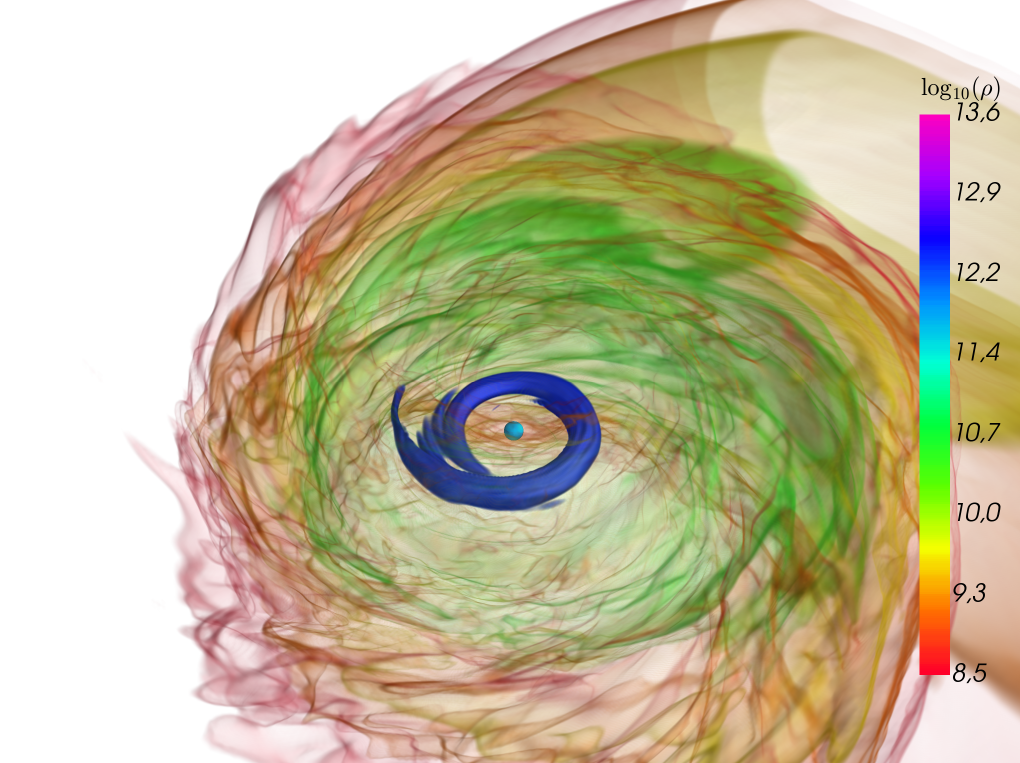}
\end{minipage}
\end{minipage}
\begin{minipage}{\linewidth}
\begin{minipage}{0.3\linewidth}
\includegraphics[width=\textwidth]{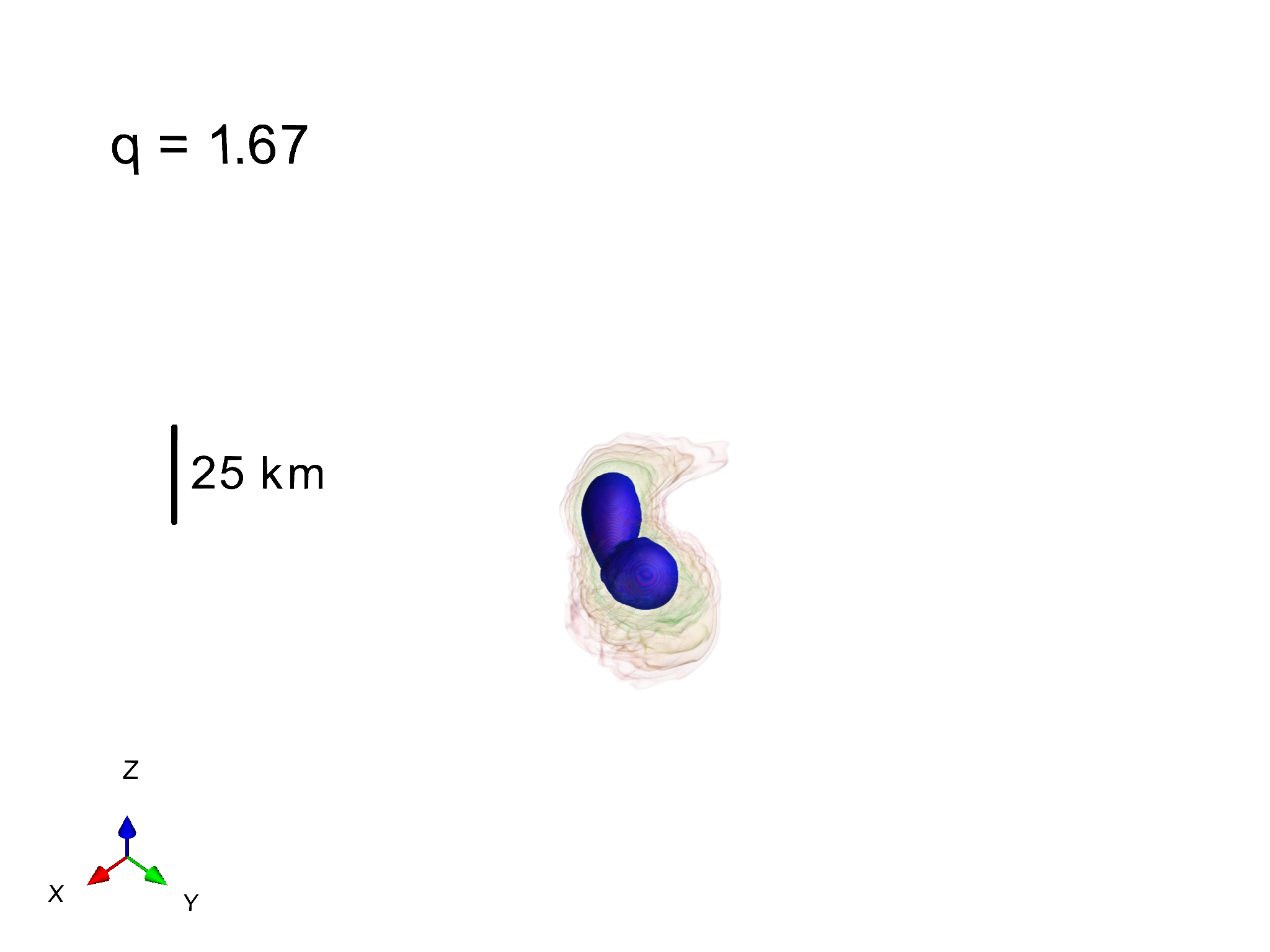}
\end{minipage}
\begin{minipage}{0.3\linewidth}
\includegraphics[width=\textwidth]{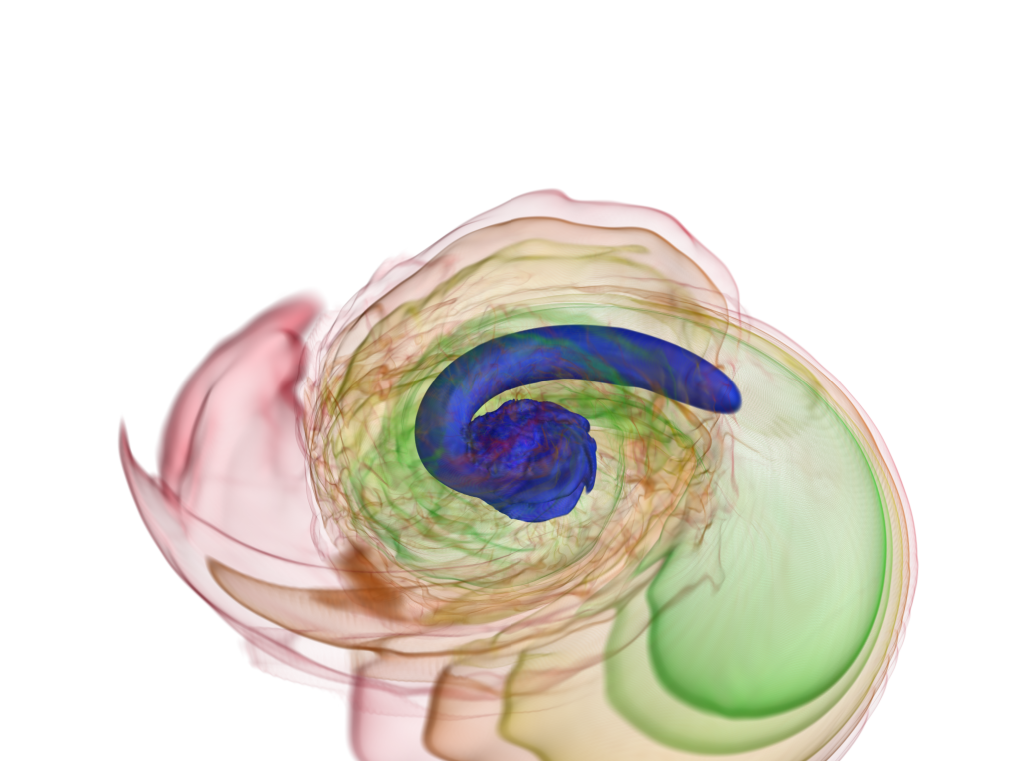}
\end{minipage}
\begin{minipage}{0.3\linewidth}
\includegraphics[width=\textwidth]{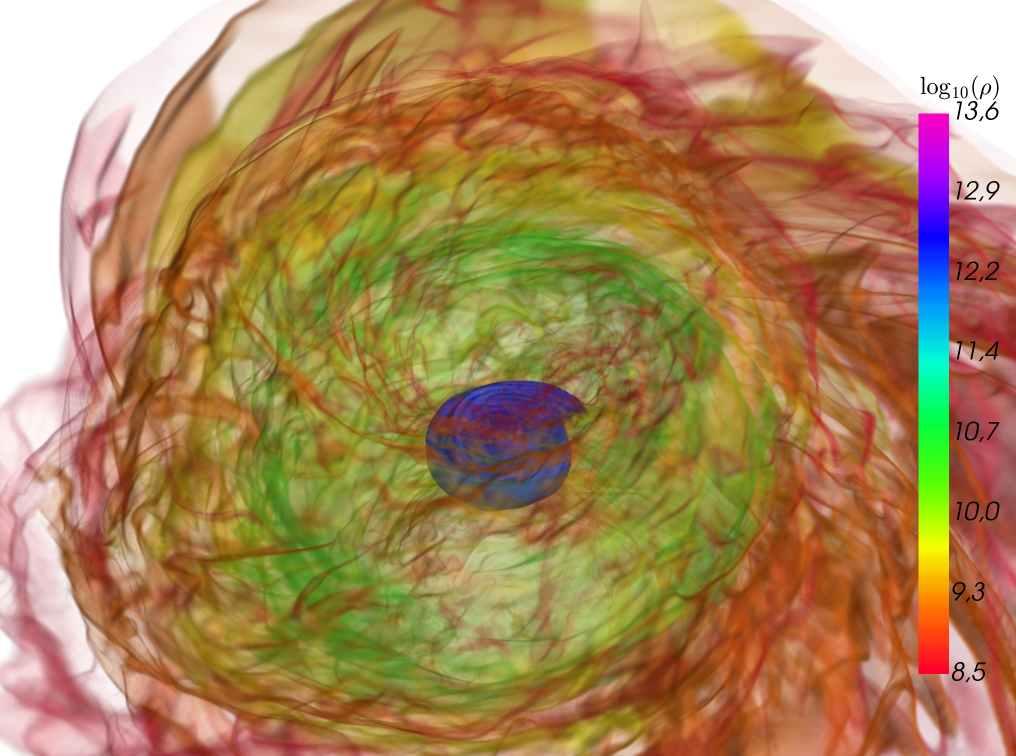}
\end{minipage}
\end{minipage}
\begin{minipage}{\linewidth}
\begin{minipage}{0.3\linewidth}
\includegraphics[width=\textwidth]{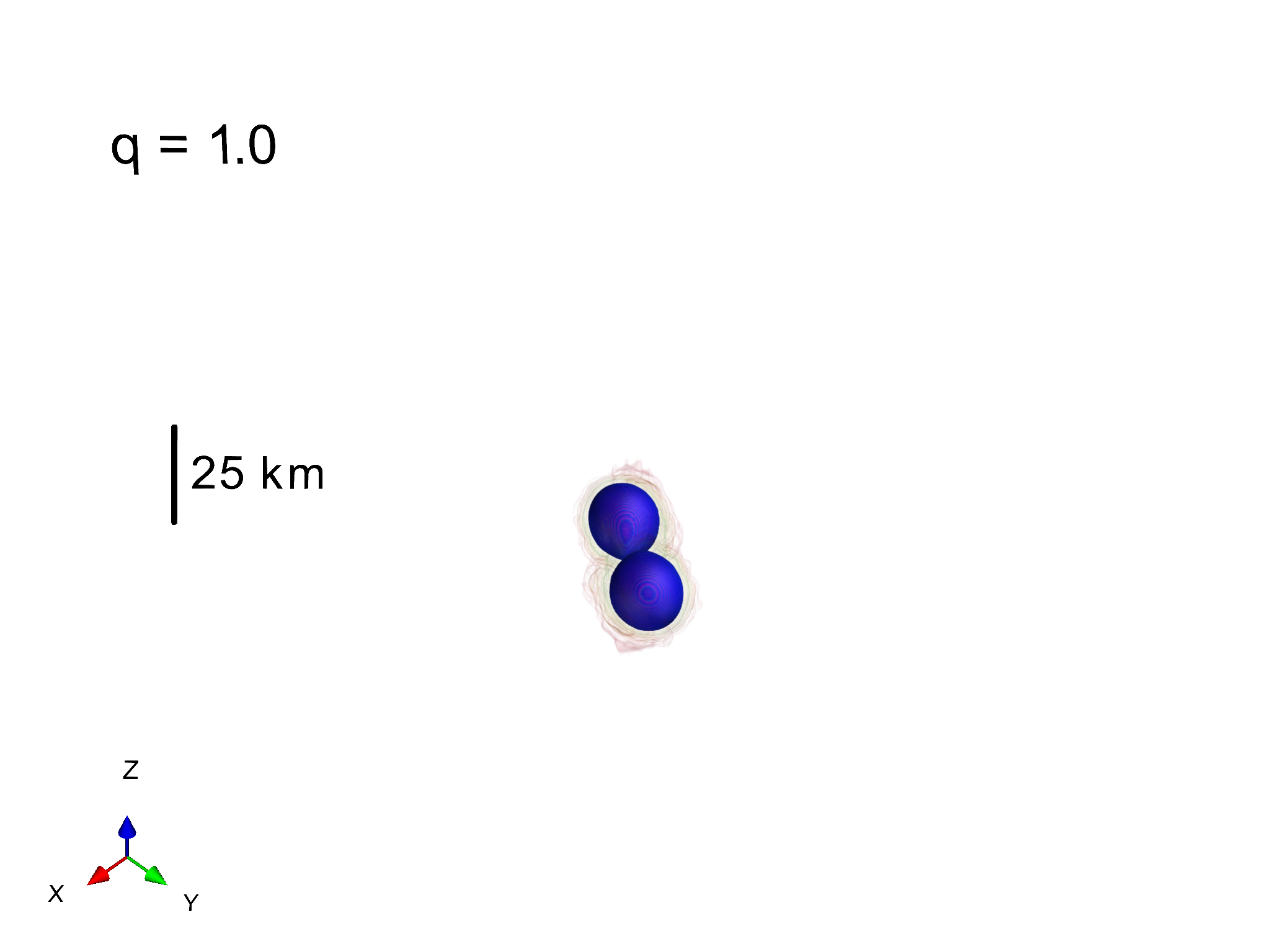}
\end{minipage}
\begin{minipage}{0.3\linewidth}
\includegraphics[width=\textwidth]{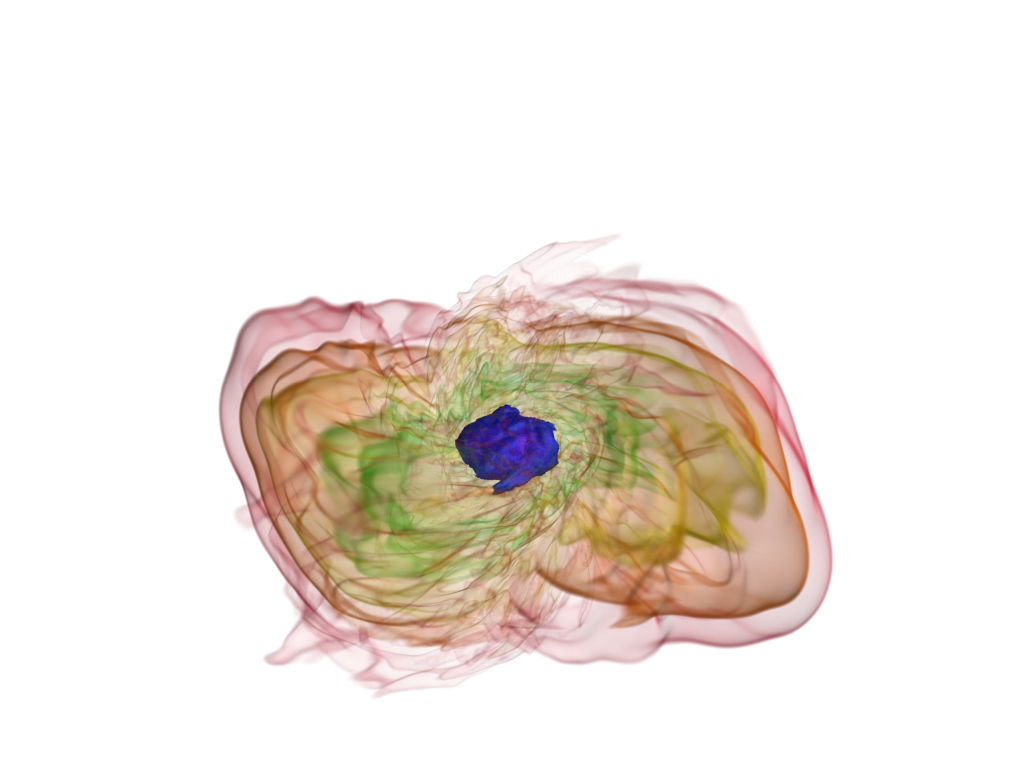}
\end{minipage}
\begin{minipage}{0.3\linewidth}
\includegraphics[width=\textwidth]{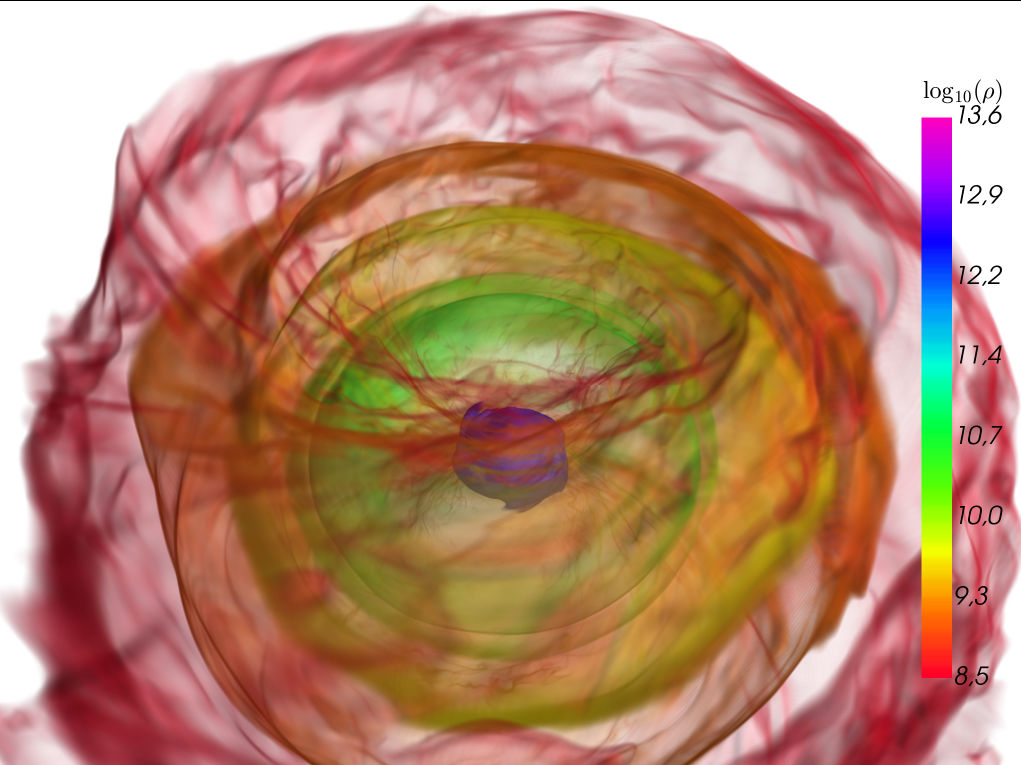}
\end{minipage}
\end{minipage}
\centering
\caption{3D volume rendering of the rest mass density $\rho$ in
  g cm$^{-3}$ expressed in logarithmic scale for the BLh models. Each
  column represents a different time inside the simulation: merger
  time (left), early postmerger ($\sim 2 {\rm ms}$, middle) and
  later stages ($\sim 10 {\rm ms}$, right). In each row we show a
  different mass ratio $q=M_A/M_B$: $q=1.8$ (top), $q=1.67$ (middle) and
  $q=1.0$ (bottom). The BH apparent horizon is shown as a bright
  green isosurface of the lapse function $\alpha=\alpha_\text{AH}$.}
\label{fig:rho_3d_blh}
\end{figure*}

The new aspect highlighted by our simulations is the dynamics of
prompt collapse for high mass-ratio BNS. In a $q \sim1.5-2$
binary the tidal disruption and accretion of the companion NS onto the massive primary NS 
can drive the remnant unstable and causes a prompt collapse to BH. 
The process is shown in Fig.~\ref{fig:rho_3d_blh} (top panels) in a 3D
volume rendering of the rest mass density for the representative case
of the BLh EOS. The BLh $q=1.8$ has a rather massive primary NS with
$M_A=1.856\Msun$ as compared to the maximum TOV mass for the BLh EOS
($M_\text{max}^\text{TOV}=2.103\Msun$), and a companion NS of small
compactness ($M_B=1.020$ and $C_B\simeq0.12$.) The companion
NS is almost completely destroyed by tidal effects and its accretion
results in the prompt formation of a BH surrounded by a massive accretion disc (see below).
By contrast the lower mass-ratio and equal-mass binaries with the same
chirp mass produce a less compact remnant and none of them collapse to
the end of the simulated time (middle and bottom panels). Note that the equal mass 
BLh was evolved beyond $80$~ms postmerger. We checked with a
sequence of simulations at intermediate mass ratios that the behavior
is continuous in the mass ratio parameter (See Appendix~\ref{sec:q-dep}).

\begin{figure}
  % data/kthresh_plot.py
\centering
  \includegraphics[width=.5\textwidth]{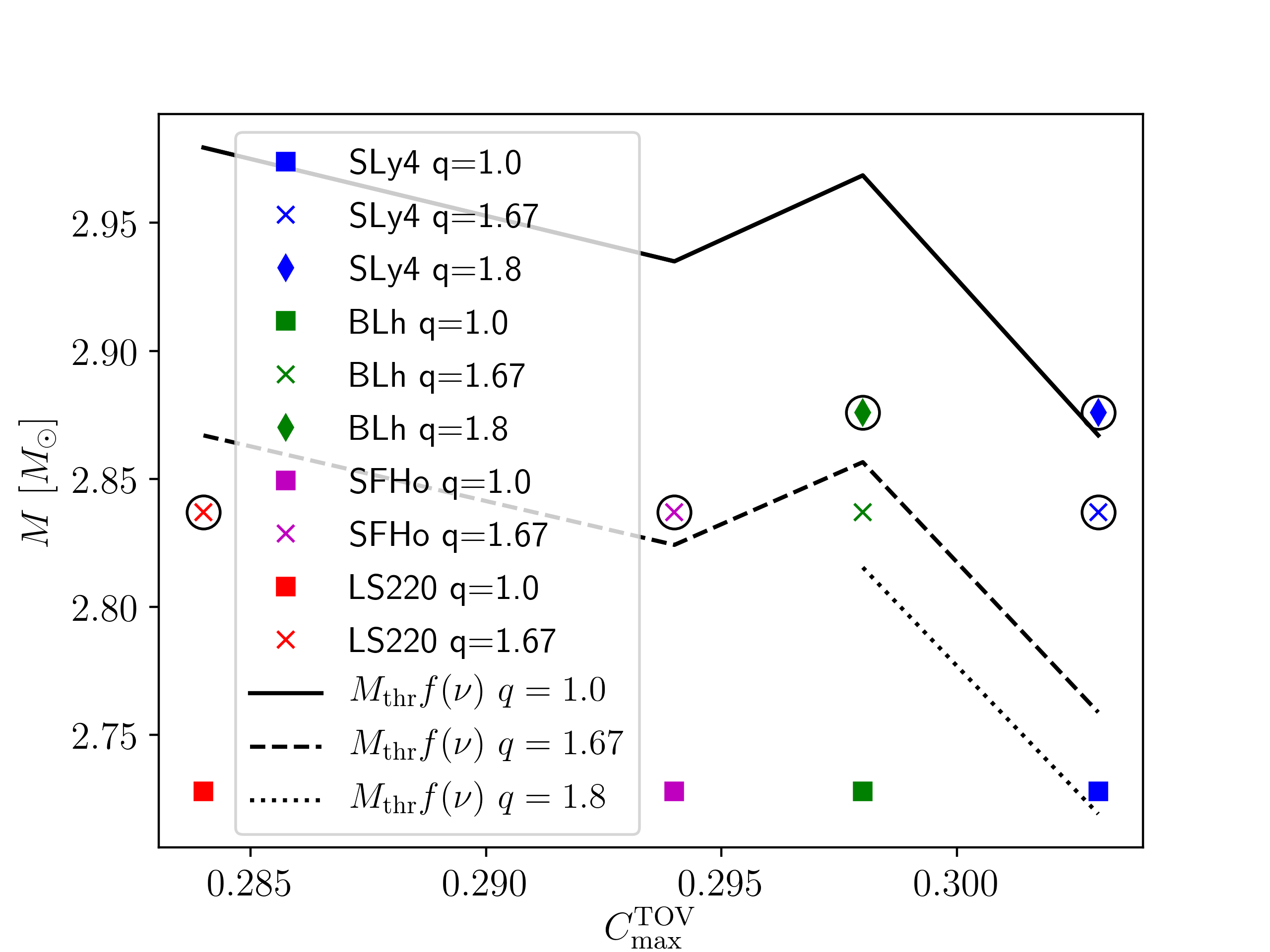}
\caption{Masses of simulated BNS as function of the TOV maximum
  compactness as predicted by the zero-temperature EOS. Prompt
  collapsing remnant are marked with additional black circles. The
  solid line is the mass threshold given by Eq.~\ref{eq:Mthr_fit}
  ($q$-independent): binaries above the line are predicted to
  collapse. The 
  dashed and solid lines are the thredshold model modified by the
  mass-ratio correction in Eq.~\ref{eq:Mthr_fitnu}. The plot refers to
simulations with turbulent viscosity only.}
\label{fig:Mtrh}
\end{figure}

Comparing our results to the numerical-relativity--based models of prompt collapse available in
the literature we find that the current models fail to predict the
behavior at high mass ratio. This is not surprising since all the models 
are calibrated using almost exclusively comparable masses simulations. In
particular, the prompt collapse model proposed in
\citet{Bauswein:2013jpa} predicts prompt collapse for
BNS with masses exceeding a threshold mass 
\be\label{eq:Mthr_fit}
M > M_\text{thr} = k_\text{thr} M_\text{max}^\text{TOV} \ ,
\ee
where the quantity $k_\text{thr}$ can be expressed in an approximately
EOS-independent way in terms of the maximum mass TOV compactness. Using
also data from \citet{Hotokezaka:2011dh,Zappa:2017xba,Koppel:2019pys}, 
a best fit expression was derived in \citet{Agathos:2019sah}:
\be \label{eq:kthr_fit}
  k_\text{thr}(C_\text{max}) = -(3.29\pm0.23) \, C_\text{max} + (2.392\pm0.064)  \ .
\ee
The above model does not include any dependence on the mass ratio and predicts that all
models simulated in our work would produce a NS remnant, except the
BLh $q=1.8$. The prediction is shown as solid line in the $M$ vs. $C_\text{max}$
diagram in Fig.~\ref{fig:Mtrh}; prompt collapse would be expected for
BNS above the solid line. A
possible way to improve the criterion in Eq.~\eqref{eq:Mthr_fit} is to
correct the threshold mass by a function of the mass ratio,
$f(\nu)$. For example, one could look 
for a criterion based on the chirp mass. Letting
\be\label{eq:Mthr_fitnu}
M_\text{thr}\mapsto M_\text{thr}f(\nu) = M_\text{thr}(4\nu)^{3/5} \ ,
\ee
lowers the threshold and approximately reproduces our results (dashed and
dotted lines in the Fig.~\ref{fig:Mtrh}.) The limited data points
available do not allow us more quantitative studies or fitting.

Another criterion for prompt collapse that is 
independent on the EOS is based on the value of the
tidal parameter \cite{Zappa:2017xba,Agathos:2019sah}
\be
\tilde\Lambda > \tilde\Lambda_\text{thr} \sim 338-386 \ .
\ee
Note that the $\tilde\Lambda$ parameter contains the mass
ratio dependence, and for $q\gg1$
\be
\tilde\Lambda = \frac{16}{13}\frac{M_A^5}{M^5}\left(1+12 \dfrac{M_B}{M_A}\right)\Lambda_A +
(A\leftrightarrow B)\approx q^5 \ .
\ee
Comparing to our data, we find that it predicts correctly the prompt
collapse of the highest simulated mass ratio for SFHo and SLy4, but
fails for the LS220 and BLh. This is also expected since $\tilde\Lambda$ 
does not account for tidal disruption but only measures the binary
compactness [Cf. discussion in Appendix~A in \citet{Breschi:2019srl}.]

\begin{table}
  \begin{center}
    \caption{Remnant properties.
      For each simulation the table reports BH properties (if
      applicable) and estimates for the disc properties.
      The BH properties are all reported from simulations at SR except
      the BLh $q=1.8$ and LS220 $q=1$ for which only LR data are available.
      The relative differences between LR and SR data are
    ${\sim}1$\% and ${\sim}3$\% for mass and spin respectively.
      The disc mass is measured at the time it is
      maximum. For remnant collapsing to BH this corresponds  to the
      mass at formation that later accretes 
      onto the BH. For NS remnants the disc can also increase its
      mass over time acquiring matter expelled from the NS. Note the numbers in
      this table are reported from simulations at resolutions LR or SR
      as available, and are affected by uncertainties up to 20-40\%.}
    \begin{tabular}{cc|cccccccc}
      \hline\hline
      EOS & $q$
      & Prompt BH & $M_\text{BH}$ & $a_\text{BH}$ 
      & $M_{\text{disc}}^\text{max}$ $[\Msun]$
      \\ 
      \hline\hline
      BLh & 1.0 &  \xmark & N.A. &N.A. & 0.15\\ 
      BLh & 1.67 & \xmark & N.A. & N.A. & 0.29\\ 
      BLh & 1.8 &  \cmark & 2.49 & 0.66 & 0.17 \\
      \hline
      LS220 & 1.0 & \xmark & 2.41 & 0.55 & 0.12\\
      LS220 & 1.67 & \cmark & 2.44 & 0.70 & 0.16\\ 
      \hline
      SFHo & 1.0 &  \xmark & 2.38 & 0.75 & 0.08\\
      SFHo & 1.67 & \cmark & 2.45 & 0.68 & 0.14\\
      \hline
      SLy4 & 1.0 &  \xmark & 2.41 & 0.76 & 0.05\\
      SLy4 & 1.67 & \cmark & 2.47 & 0.69 & 0.10\\
      SLy4 & 1.8 &  \cmark & 2.52 & 0.66 & 0.15\\
      \hline
      \hline
    \end{tabular}
  \end{center}
  \label{tab:rem_prop}
\end{table}

Let us now discuss disc formation, evolution and properties. Following a
common convention, we define disc the baryon material either outside
the BH's apparent horizon or the one with densitites
$\rho\lesssim10^{13}$~$\gccm$ around a NS remnant. The baryonic mass of
the discs are computed as volume integrals of the conserved rest-mass
density 
$D=\sqrt{\gamma}~W\rho$ from 3D snapshots of the simulations in
postprocessing ($\gamma$ is the 3-metric's determinant and $W$ the
Lorentz factor). 
Estimates for the disc masses are reported in Tab.~\ref{tab:rem_prop}.
The disc mass is reported as measured at the time when it is maximum during the
simulation. For remnant collapsing to BH this can be interpreted as the mass at
BH formation, since the disc mass can only decrease with time due to accretion. 
For NS remnants the disc (remnant at lower densitites) can also increase 
its mass over time as it acquires matter expelled from the higher densities shells.

\begin{figure}
  % discmass/plot_qualitative_discev.py
  \centering
  \includegraphics[width=.49\textwidth]{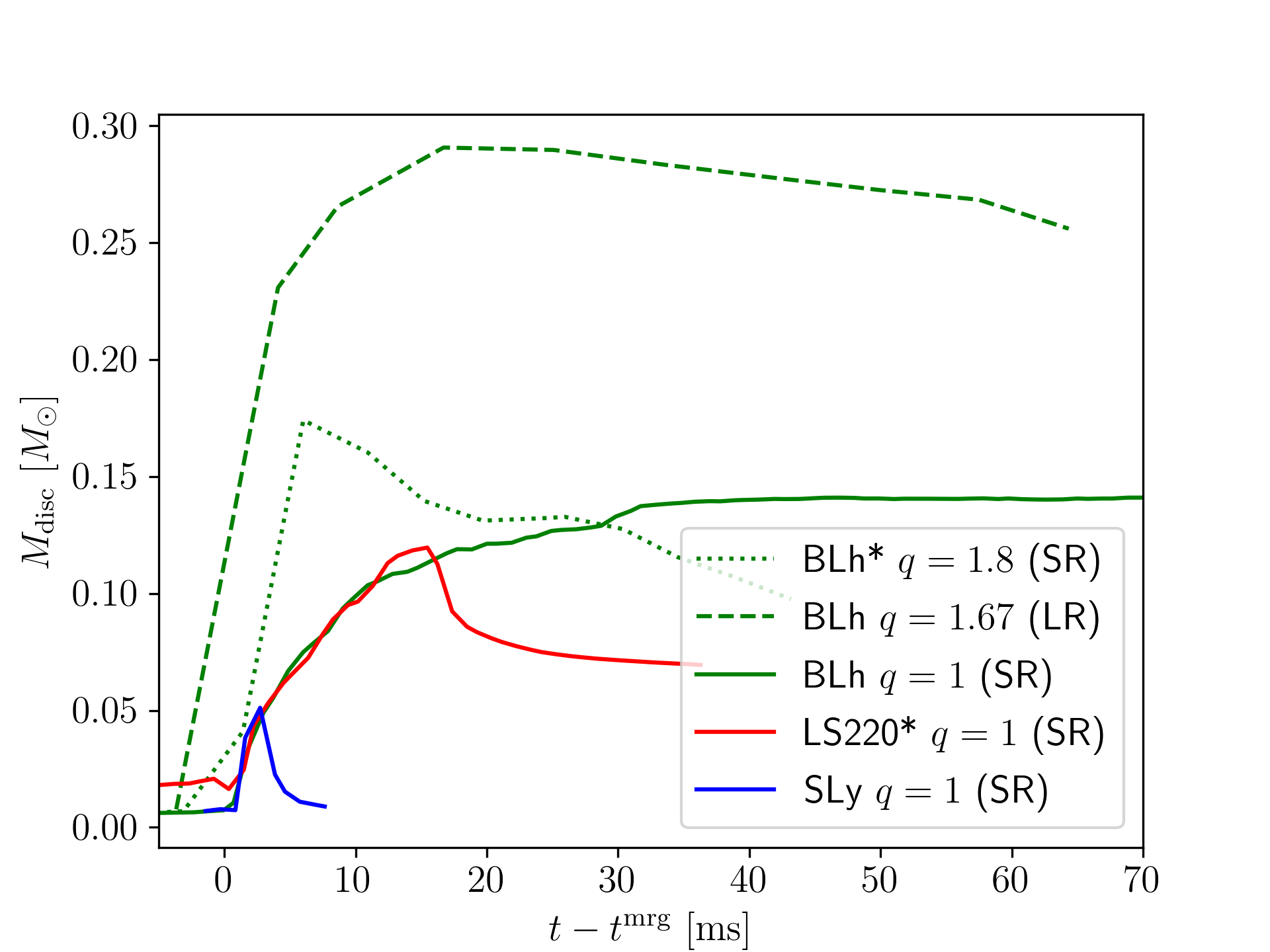}
  \caption{Evolution of the disc rest mass in representative remnants. The disc is defined as
the remnant outside the apparent horizon if the remnant has collapsed to a BH 
or as the portion of the remnant whose rest mass density satisfies $\rho < 10^{13}~\gccm$ otherwise.}
\label{fig:discev}
\end{figure}

Examples of the disc mass evolutions for different remnants are shown in
Fig.~\ref{fig:discev}. Note that we show the BLh $q=1.8$ and LS220
$q=1$ simulations at resolution SR but without turbulent viscosity and
the $q=1.67$ with viscosity but at LR because these are the longer
dataset available to us (see below for a discussion about turbulence.)

In the case of comparable mass BNS the accretion disc is formed
during and after the merger.  
As time evolves, if the remnant does not collapse, 
it continuosly shedes mass and angular momentum increasing the mass of the disc 
and generating outflows \cite{Radice:2018xqa, Nedora:2019jhl}. This is why in 
Fig.~\ref{fig:discev} the accretion disc mass is increasing with time 
for these binaries. These processes terminate with BH formation, which is accompanied 
by the rapid accretion of a substantial fraction of the disc. An important consequence 
is that, in the case of comparable mass ratio binaries, prompt BH formation
results in very small accretion disc masses \cite{Radice:2017lry,Kiuchi:2019lls},
because the mechanism primarily responsible for the formation of the
disc is shut off immediately in these cases.

In high mass ratio BNS mergers the companion star is tidally disrupted
(Fig.~\ref{fig:rho_3d_blh}). In these cases, the bulk of the
accretion disc is constituted by the tidal tail, which is for the most
part still gravitationally bound to the remnant. This tail is launched 
prior to merger. So massive accretion discs are possible even if prompt 
BH formation occurs (see also \citealt{Kiuchi:2019lls}.) 
In general, high $q$ binaries are found to
generate more massive discs than binaries with the same chirp mass but
lower $q$ \cite{Shibata:2003ga, Shibata:2006nm, Kiuchi:2009jt,
Rezzolla:2010fd, Dietrich:2016hky}. The postmerger evolution of these
discs is also very different. While in the massive NS case the central object
pushes material into the disc and drives outflows, in the case of high
mass-ratio binaries forming BHs the fallback of the tidal tail perturbs
the disc and drives rapid accretion onto the BH as evinced by the rapid
decrease of the disc masses with time shown in Fig.~\ref{fig:discev}.

\begin{figure*}
  \centering
  \includegraphics[width=.48\textwidth]{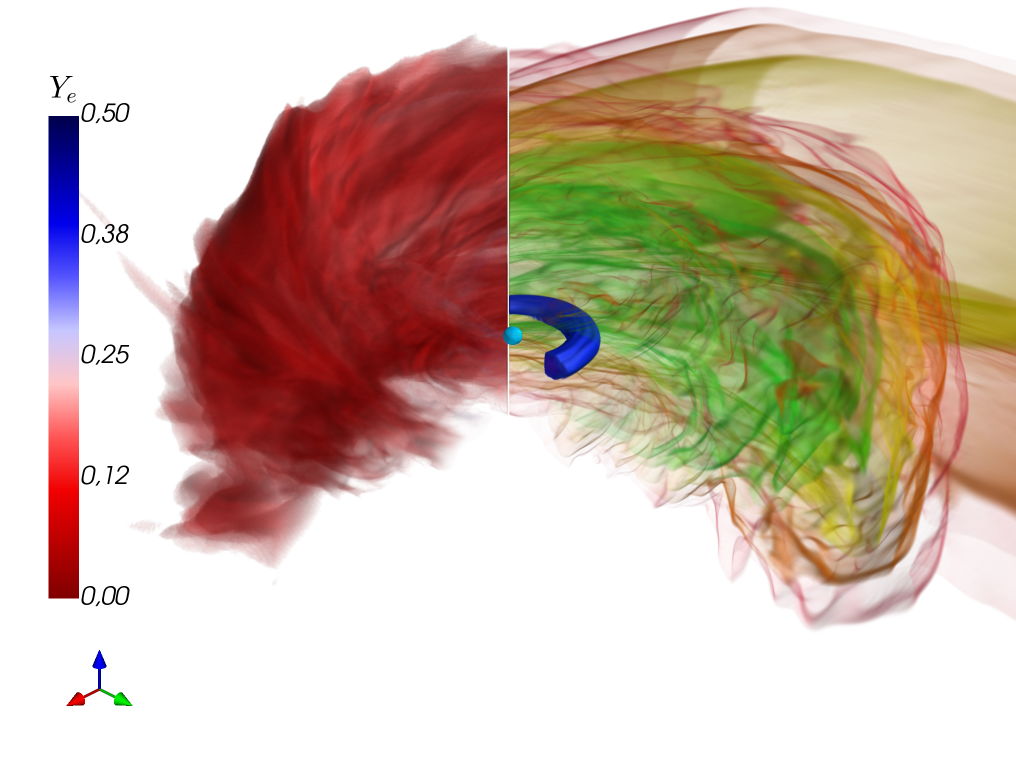}
  \includegraphics[width=.48\textwidth]{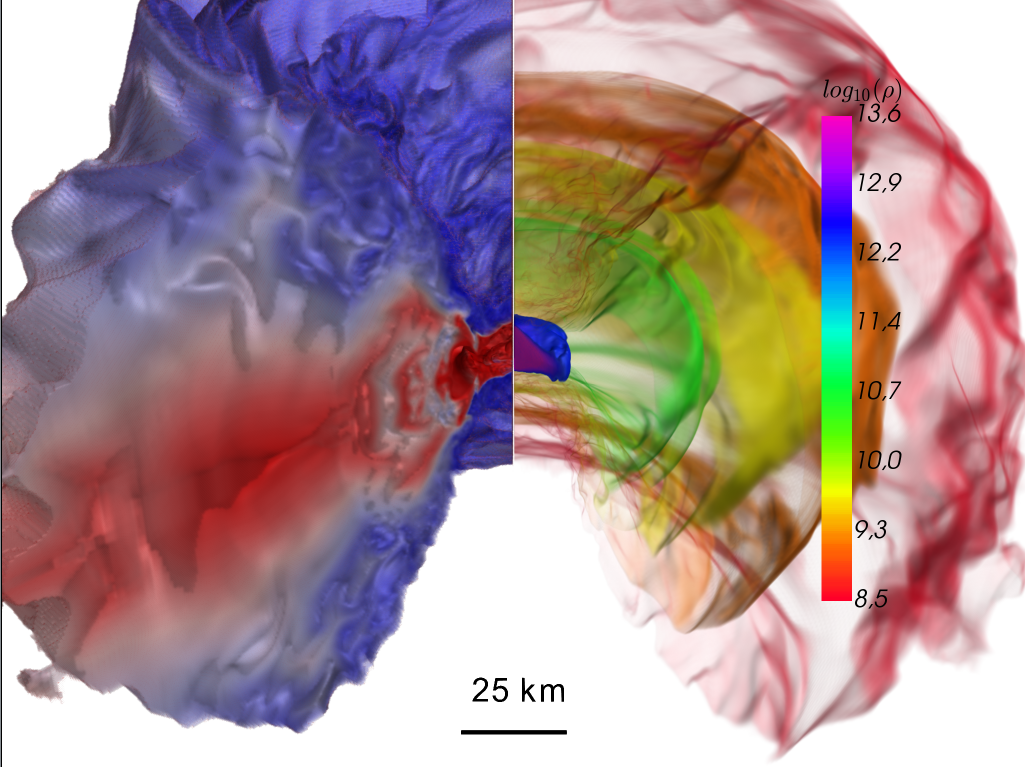}
  \caption{Volume rendering of density at the end of the BLh $q=1.8$
  (left panel) and $q = 1$ (right panel) simulations. In each panel, on
  the left we show the electron fraction distribution, while on the
  right we plot the rest mass density; for both quantities, we show 
  only matter with densities $\rho \geq 3\times 10^{8} \mathrm{g cm}^{-3}$. 
  The spatial scale is the same in the two panels.}
  \label{fig:disc_struct}
  \label{fig:disk_struct}
\end{figure*}

These different formation mechanisms are imprinted in the structure and
composition of the discs formed in comparable and very unequal mass
binaries, as shown in Fig.~\ref{fig:disc_struct}. 
In the case of equal mass binaries, the disk is composed of material squeezed out of
the collisional interface between the NSs \cite{Radice:2018pdn}, see
also Fig.~\ref{fig:rho_3d_blh}. This matter is heated to temperatures of
tens of MeV before being pushed out of the central part of the remnant,
so its electron fraction is reset by pair processes
\cite{Perego:2019adq}, see Fig.~\ref{fig:disc_struct}.
Due to the absence of strong compression and shockes, the discs formed in high mass ratio
binaries are initially colder and more neutron rich (Fig.~\ref{fig:disc_struct}).
Since high-$q$ BNS mergers launch tidal tails to large radii, comparable mass-ratio 
binaries create discs that are initially more compact and have higher $Y_e$'s.
Besides the mass ratio, the structure of the disc is also strongly dependent on the
nature of the remnant. In the case of BH remnants, the discs are
typically more compact and thin than those around massive NS remnants. In the
latter case, because of the additional pressure support, the discs reach
higher densities ${\sim}10^{13}~\gccm$ and become
partly optically thick to neutrinos 
\cite{Perego:2019adq,Endrizzi:2020lwl}.

%%%%%%%%%%%%%%%%%%%%%%%%%%%%%%%%%%%%%%		
\section{Gravitational waves}
\label{sec:gw}

\begin{table}
  \centering    
  \caption{Gravitational-wave data extracted at the moment of merger
    from NR simulations presented in this work, expressed in
    dimensionless units according with the convention
    $G=c=\Msun=1$. The energy $e_b^\text{mrg}$ and the angular
    momentum $j^\text{mrg}$ are defined from the gravitational wave
    data as: $e_b^\text{mrg} = E_b^\text{mrg}/(\nu M)$ and
    $j^\text{mrg} = J^\text{mrg}/(\nu M^2)$. The following values have
    a numerical uncertainty of ${\sim}$5\% due to the
    different grid resolutions.} 
  \begin{tabular}{ccc|cccc}        
    \hline\hline
    EOS & $M$& $q$ & $M f^\text{mrg}_{22}$ &$e_b^\text{mrg}$ & $j^\text{mrg}$ & $A_\text{mrg}/M$    \\
    \hline\hline
    BLh &2.728& $1.0$ & 0.02567 & -0.05900 &  3.443 & 0.2566  \\
    BLh &2.837& $1.67$ & 0.01966 & -0.05538 &3.523 & 0.2110 \\
    BLh &2.876& $1.8$ & 0.01885 &-0.05471 & 3.517 &0.2014   \\
    \hline
    LS220 &2.728& $1.0$ &0.02349 & -0.05714 &  3.469 & 0.2465   \\
    LS220 &2.837& $1.67$ & 0.01804 &-0.05265 &3.582 &  0.1986  \\
    \hline
    SFHo &2.728& $1.0$ &0.02581 & -0.06099 & 3.404 & 0.2670 \\
    SFHo &2.837& $1.67$&0.02063 & -0.05569 & 3.485 &0.2100 \\
    \hline
    SLy4 &2.728& $1.0$&0.02649 & -0.06087 & 3.413 & 0.2766  \\
    SLy4 &2.837& $1.67$&0.02032  &-0.05445& 3.490&0.2097  \\
   SLy4 &2.876& $1.8$& 0.01998 & -0.05693 & 3.478 &0.1796   \\
    \hline\hline   
  \end{tabular}
 \label{tab:gw}
\end{table}

\begin{figure}
  \includegraphics[width=.5\textwidth]{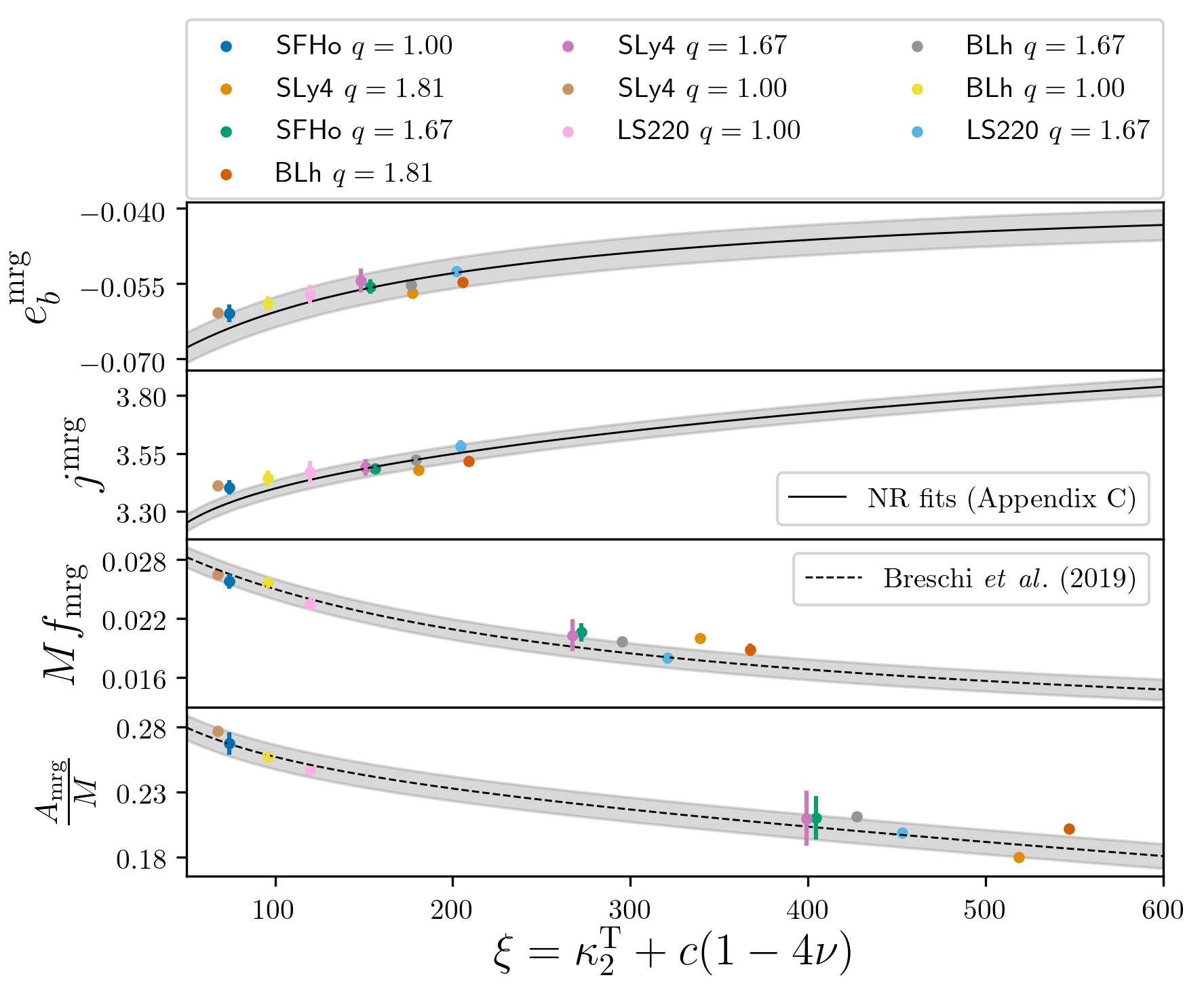}
\centering
\caption{Comparison between GW data at moment of merger extracted from
  the simulations introduced in this work (and listed in
  Tab.~\ref{tab:gw}) and fits calibrated using NR simulations with
  $q \le 1.5$.
  The fits for energy and angular momentum are reported in
  Appendix~\protect\ref{sec:EJ}, while frequency and amplitude fits are
  from~\protect\citet{Breschi:2019srl}.
  The quantities $e_b^{\rm mrg}$ and
  $j^{\rm mrg}$ are defined from the GW data as $e_b^{\rm mrg} =
  E_b^{\rm mrg}/(\nu M)$ and $j^{\rm mrg} = J^{\rm mrg}/(\nu 
  M^2)$. The results extracted from high mass ratio simulations are
  consistent with the current fits in the limit of the uncertainties.}  
\label{fig:fits_highq}
\end{figure}

\begin{figure}
\centering
\includegraphics[width=.5\textwidth]{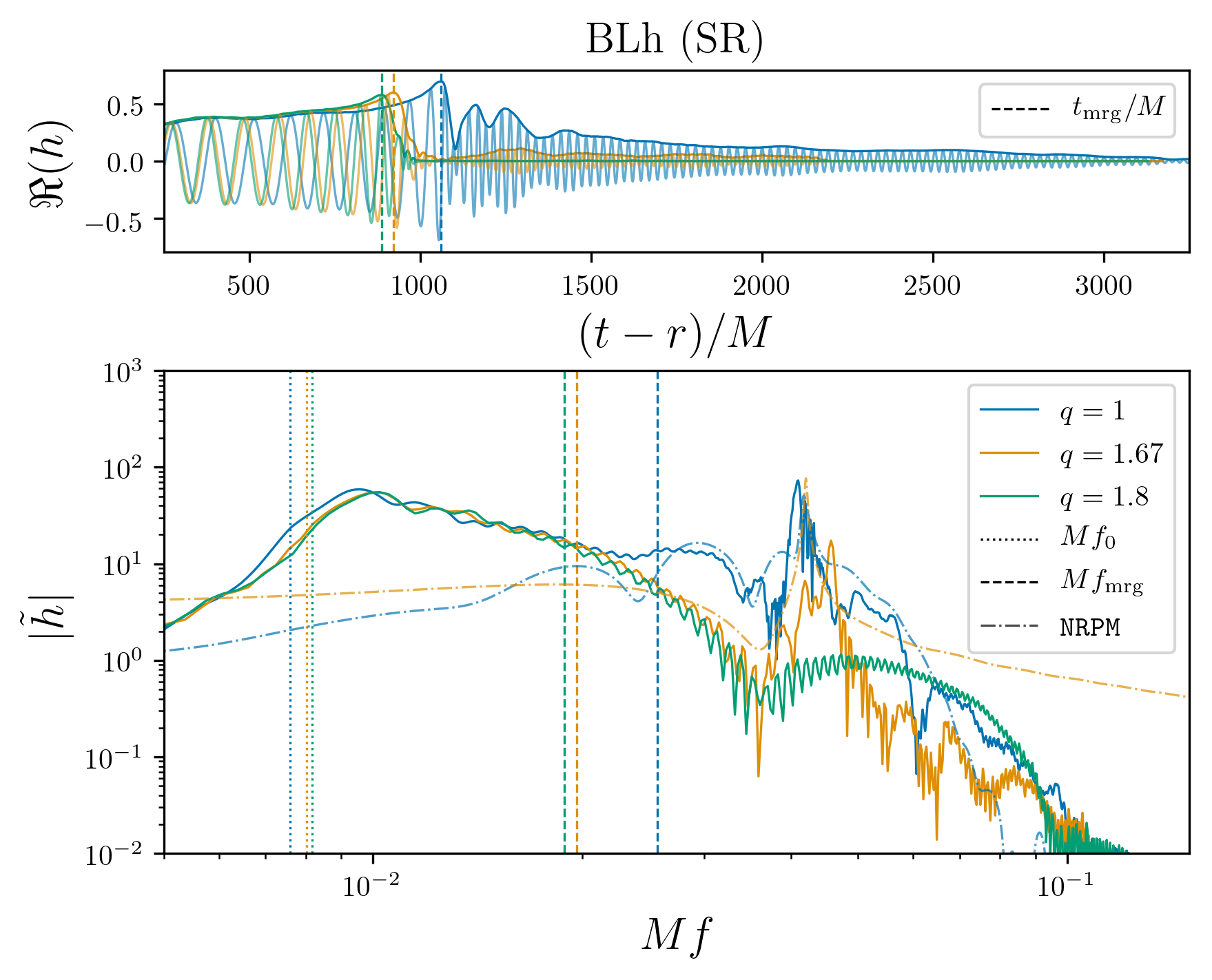}
\caption{Example of GW signals. Amplitude and real part of the $(2,2)$ mode
  strains (top panel) and respective frequency-domain spectra (bottom panel) for
  the three simulations with BLh EOS at standard resolution. The
  vertical dashed lines mark the merger times, both in time- and
  frequency-domain, while the dotted lines in the bottom panel show
  the initial point of the simulations. The frequency-domain spectra
  are compared with {\tt NRPM} postmerger model evaluated with the
  same physical values, except for the prompt-collapse case (BLh,
  $q=1.8$).}
\label{fig:gw_strain}
\end{figure}

In this section we analyze the GW signals computed from the
simulations. The latter are too short for a quantitative comparison
with inspiral-merger models,
e.g.~\cite{Akcay:2018yyh}. Hence, we focus on the
merger and postmerger signal. The new simulations allow us to verify
(and extend) the quasiuniversal relations characterizing the merger
and study the postmerger waveform \cite{Bernuzzi:2015rla,Breschi:2019srl}.

Following \citet{Damour:2011fu} (see also \citealt{Bernuzzi:2012ci})
we compute the reduced binding energy $e_b^\text{mrg}=
E_b^\text{mrg}/(\nu M)$ and the angular momentum $j^\text{mrg}=
J^\text{mrg}/(\nu M^2)$ at the moment of merger from the multipolar
GW. Those and other quantities at merger are approximately EOS-independent functions
of the tidal parameter \cite{Bernuzzi:2014owa}. To find these
relations it is best to use, instead of $\tilde\Lambda$, the parameter
\be\label{eq:kappa2T}
 \kappa_2^\text{T}
 = \dfrac{3}{2} \left[ \Lambda_2^\text{A} \left(\frac{M_\text{A}}{M}\right)^4
 M_\text{B} + (A\leftrightarrow B) \right] \ , \ee determining both
 tidal dynamics and tidal waveform at leading post-Newtonian
 order \cite{Damour:2009wj,Damour:2012yf}.  High mass-ratio effects
 are included by further considering the parametrization %\cite{}
\be\label{eq:xi}
\xi = \kappa_2^T + c (1-4 \nu) \ ,
\ee
where $c$ is a fitting parameter \cite{Zappa:MSc,Breschi:2019srl}.
Binding energy, angular momentum and the waveform key quantities at
merger are reported in Tab.~\ref{tab:gw} for all simulations. From the
table one notices that the binding energy and the angular momentum
increase (binding energy is less negative) as $\tilde\Lambda$
decreases and $q$ increases (Tab.~\ref{tab:sim}); consequently the
merger GW frequency and amplitude decreases.  The dimensionless BH
spin of the remnants is $a_\text{BH}\sim0.7$
(Tab.~\ref{tab:rem_prop}), and it can be compared to the angular
momentum available at merger considering its reduced value
$j_\text{BH}(\nu)=a_\text{BH}\nu$. The angular momentum at merger is
partly radiated in GW and partly gives the disc angular momentum and
BH spin. For the $q=1.8$ prompt collapse remnants
($\nu\simeq0.22959$, BLh and SLy4) we obtain
$j_\text{BH}(0.22959)=0.66/0.22959\simeq2.87$ to be compared to
$j^\text{mrg}\simeq3.5$. For the $q=1$ ($\nu=0.25$) SLy4 and SFHo with
BH formation we obtain $j_\text{BH}(0.25)=0.76/0.25\simeq3.04$ to be
compared to $j^\text{mrg}\simeq3.4$. These estimates, obtained using
gauge invariant quantities, indicate that discs around BHs generated
by prompt collapse $q=1.8$ binaries have a reduced angular momentum
that is larger by about $60\%$ than that of discs around equal BHs
resulting from the prompt collapse of equal mass NS binaries.  This
observation is strengthened by the fact that the postmerger GW is
weaker if the BH is promptly formed (see below).

Figure~\ref{fig:fits_highq} compares the new NR data of this paper
(Tab.~\ref{tab:gw}) with the fits of simulations of the {\tt CoRe}
collaboration~ for $q \le1.5$ proposed in \citet{Breschi:2019srl}. The
fits are consistent with the new data with $q>1.5$ within the 
the uncertainties, indicating the robustness of the model (and
especially the ansatz Eq.~\ref{eq:xi}). The fits for the binding
energy and angular momentum at merger were not presented in
\cite{Breschi:2019srl} and are thus are given here in Appendix~\ref{sec:EJ}.
  
Regarding the postmerger waveform, Fig.~\ref{fig:gw_strain} (top
panel) shows a comparison between the waveforms from the BLh BNS for
the three mass ratio considered here. The figure clearly indicates that
for similar (though not identical) initial frequencies, the moment of
merger occurs earlier for unequal mass simulations due to tidal
disruption of the high-$q$ binaries, where the companion has larger
radius than the primary NS (See also Fig.~\ref{fig:rho_2d_blh_acc} and
Fig.~\ref{fig:rho_3d_blh}). As expected, the dependency of the
waveform on $q$ is smooth as shown explicitely in
Appendix~\ref{sec:q-dep}. Note that, in general, the postmerger
amplitude is smaller for high-$q$ than for equal mass due to a less
violent shock between the two NS cores and either a less compact
remnant or the formation of a BH that quickly rings down to a
stationary state. 

The only new unequal mass simulation with a long postmerger GW signal is the BLh with $q=1.67$. 
For this case, the value of the characteristic postmerger frequency
$f_2$ is properly captured from NR fits presented
in~\citet{Breschi:2019srl}: from the simulation we get 
$f_2\approx 3.31$~kHz, while for the same binary the NR fit
predicts $f_2^\text{NRPM}\approx 3.01$~kHz, which is within 
the uncertainty of the fits (${\sim}12\%$).
This result is inline with the interpretation of
\cite{Bernuzzi:2015rla,Radice:2016rys}: the postmerger $f_2$ frequency
is mostly determined by $\kappa_2^\text{T}$ and the merger physics.
Figure~\ref{fig:gw_strain}
(bottom panel) shows the comparison between the spectrum of this NR
simulation and the respective spectrum generate with the {\tt NRPM}
model of \cite{Breschi:2019srl}. While the {\tt NRPM} model captures well the
characteristic frequencies, it does not reproduce the morphology of
these high-mass-ratio waveforms due to imperfect modeling of the
characteristic amplitudes and damping times. This fact further stresses the need
of new simulations to improve postmerger models and/or of more agnostic
approaches to kiloHertz GW modeling (Cf. \citet{Breschi:2019srl}.)

\begin{figure}
\centering
\includegraphics[width=.5\textwidth]{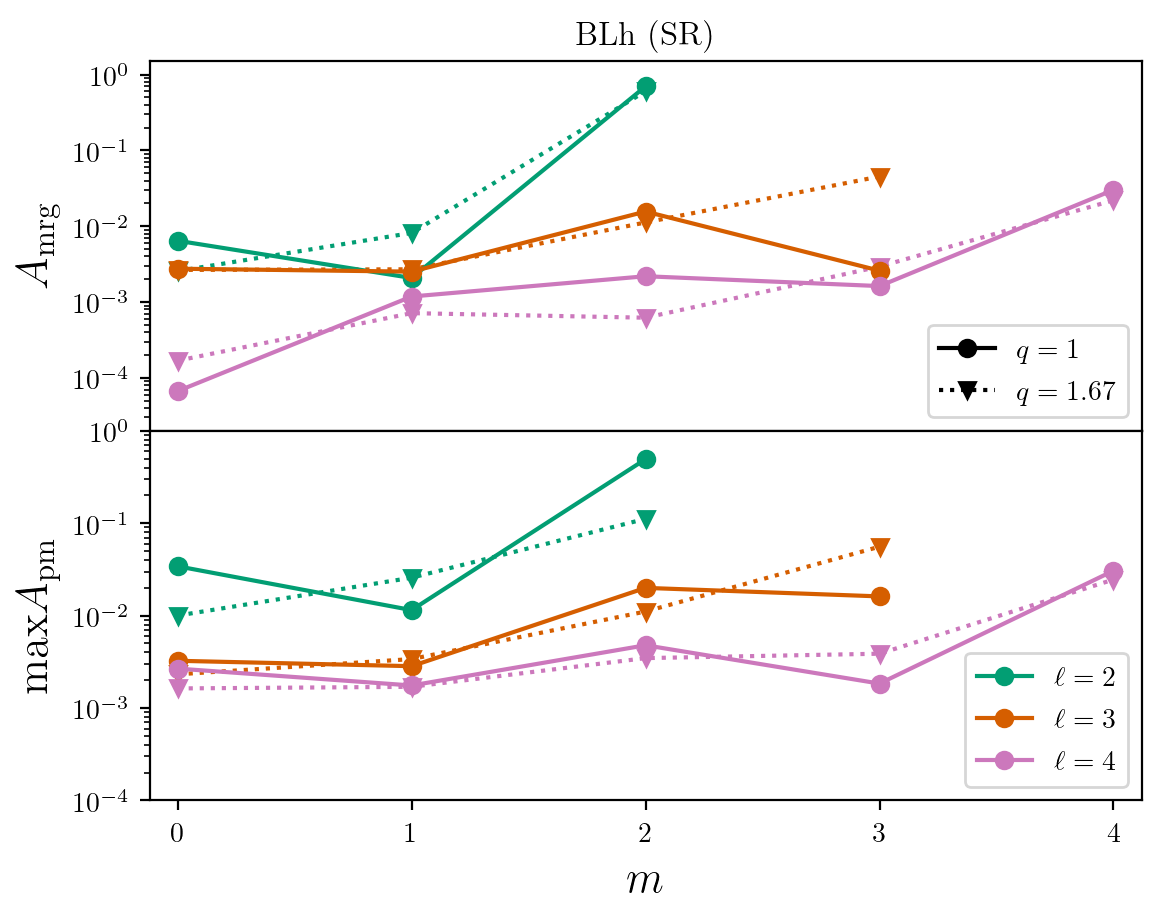}
\caption{Amplitude at merger and maximum postmerger amplitude 
of different modes of the GW strain computed with BLh EOS.
The upper panel shows the equal-mass binary, while the lower one is a $q=1.67$ case.
For unequal-mass coalescences, odd-parity higher-order modes
are boosted, e.g. $(2,1)$ and $(3,3)$ modes.}
\label{fig:hm_amp}
\end{figure}

In the context of high mass ratio binary coalescences, higher-order modes
could play an important role. The GW strain $h(t,\mathbf{x})$ is the
sum of the contribution of the several modes $h_{\ell m}(t,r)$ times
the spin-weighted spherical harmonics $^{(s)}Y_{\ell m}(\theta,\phi)$
with $s=-2$ that contain the dependence on the source's sky position, 
 \be
 h(t,\mathbf{x}) =h_+ -\i h_\times  = \sum_{\ell,m}
 %\sum_{m=-\ell}^\ell
 h_{\ell m}(t,r)\,{}^{(-2)}Y_{\ell m}(\theta,\phi)\ .
 \ee
The maximum amplitudes $A_{\ell m}=|h_{\ell m}|$ for the different
modes at merger and postmerger are shown in Fig.~\ref{fig:hm_amp}.
In equal-mass postmerger waveform $m=1$ are suppressed at merger
(top panel, solid lines), and the dominant modes are, in order,
$(\ell,m)=(2,2), (3,2)$ and $(4,4)$. The contribution of the odd modes
and $(2,0)$ increases in the postmerger. The $(2,0)$ mode, in
particular, is relevant
in the early postmerger times and its amplitude could reach the 15\% of the $(2,2)$ amplitude. 
This is sometimes interpreted as due to radial oscillations of the
remnant which contribute to the emitted signal and could generate a
coupling with the dominant mode~\cite{Stergioulas:2011gd} in analogy to what happens
with nonlinear perturbations of equilibrium NS
\cite{Dimmelmeier:2005zk,Passamonti:2007tm,Baiotti:2008nf,Stergioulas:2011gd}. 
The waveform mode hierarchy for high-$q$ binaries is similar to that of
the $q=1$ binaries. However, the odd modes have a larger relative
contribution to the signal during the late inspiral at merger (Cf.~\citealt{Dietrich:2016hky}).
The amplitudes of these modes can be up to the 20\% of the $(2,2)$
amplitude before merger and in the late postmerger. 
However, inspection of the waveforms show that during the very
dynamical early postmerger phase the amplitude of the $(2,1)$ and
$(3,3)$ modes can instantanously reach the same order of the $(2,2)$.  
The contributions of the subdominant modes in the GW correlates
to density modes in the NS remnant triggered in asymmetric mergers,
see e.g. \cite{Stergioulas:2011gd,Bernuzzi:2013rza}. 

%%%%%%%%%%%%%%%%%%%%%%%%%%%%%%%%%%%%%%		
\section{Dynamical ejecta}
\label{sec:ejecta}

\begin{table}
\centering
    \caption{Dynamical ejecta average properties for each simulation and
      for different resolutions.
      $M_\text{ej}$ is the total mass of the ejecta;
      $\langle \theta_{\text{ej}} \rangle$ and
      $\langle \phi_{\text{ej}} \rangle$ are the mass weighted 
      rms of the polar and azimuthal angle, respectively;
      $\langle \upsilon_{\text{ej}} \rangle$ and $\langle Y_e \rangle$,  
      are the mass-averaged electron fraction and speed. 
      The last column is the ratio $X_{s}=M_{\rm ej}^{\rm shocked}/M_{\rm ej}$, 
      where the shocked and tidal ejecta are
      defined by those with entrpy respectively above and below the
      threshold of $10k_B$ per baryon.
      Simulations without turbulent viscosity are
      indicated with *.}
\scalebox{0.8}{
    \begin{tabular}{ccc|cccccc}
      \hline\hline
      EOS & $q$ & Grid &
      $M_{\text{ej}}$ & $\langle \theta_{\text{ej}} \rangle$ & $\langle \phi_{\text{ej}} \rangle$ &
      $\langle \upsilon_{\text{ej}} \rangle$ & $\langle Y_e \rangle$ &
      $X_{s}$\\
      &   &   &  $[10^{-2}M_{\odot}]$ & $[{}^\circ]$ & $[{}^\circ]$ & $[c]$ & & \\ 
      \hline\hline

BLh & 1.0 & SR & 0.136 &  39 & 101 & 0.18 & 0.263 & 0.82 \\ 
BLh & 1.0 & LR & 0.131 &  40 & 103 & 0.16 & 0.268 & 0.87 \\ 
\hline
BLh & 1.67 & HR & 0.507 &  19 &  58  & 0.13 & 0.083 & 0.21 \\ 
BLh & 1.67 & SR & 0.451 &  24 &  53 & 0.13 & 0.114 & 0.30 \\ 
BLh & 1.67 & LR & 0.386 &  24 &  63 & 0.11 & 0.106 & 0.28 \\ 
\hline
BLh & 1.8 & HR & 0.830 &   6 &  26 & 0.11 & 0.025 & 0.01 \\ 
BLh & 1.8 & SR & 0.762 &   7 &  29 & 0.11 & 0.030 & 0.02 \\ 
BLh & 1.8 & LR & 0.841 &   7 &  30 & 0.12 & 0.043 & 0.02 \\ 
\hline
BLh* & 1.8 & HR & 1.014 &   6 &  26 & 0.12 & 0.020 & 0.00 \\ 
BLh* & 1.8 & SR & 1.056 &   6 &  28 & 0.12 & 0.029 & 0.01 \\ 
BLh* & 1.8 & LR & 1.142 &   7 &  30 & 0.12 & 0.035 & 0.02 \\ 
\hline
LS220 & 1.0 & SR & 0.137 &  38 & 104 & 0.16 & 0.260 & 0.71 \\ 
LS220 & 1.0 & LR & 0.170 &  35 & 106 & 0.17 & 0.233 & 0.64 \\ 
\hline
LS220* & 1.0 & HR & 0.105 &  37 & 103 & 0.16 & 0.223 & 0.71 \\ 
LS220* & 1.0 & SR & 0.171 &  33 &  98 & 0.16 & 0.217 & 0.64 \\ 
LS220* & 1.0 & LR & 0.215 &  35 & 105 & 0.18 & 0.218 & 0.59 \\ 
\hline
%LS220 & 1.67 & SR & 0.874 &   9 & 58 & 0.12 & 0.046 & 0.06 \\ 
LS220 & 1.67 & SR & 0.842 &  12 & 60 & 0.14 & 0.060 & 0.10 \\ 
LS220 & 1.67 & LR & 1.383 &  14 & 56 & 0.15 & 0.070 & 0.15 \\ 
\hline
LS220* & 1.67 & SR & 0.859 & 8 & 58 & 0.13 & 0.033 & 0.03 \\ 
LS220* & 1.67 & LR & 1.047 & 7 &  65 & 0.13 & 0.050 & 0.04 \\ 
\hline
SFHo & 1.0 & HR & 0.354 &  31 & 106 & 0.20 & 0.211 & 0.69 \\ 
%SFHo & 1.0 & SR & 0.320 &  33 & 138 & 0.22 & 0.233 & 0.73 \\ 
%SFHo & 1.0 & SR & 0.663 &  34 & 123 & 0.18 & 0.226 & 0.78 \\ 
SFHo & 1.0 & SR & 0.451 &  34 & 106 & 0.19 & 0.208 & 0.72 \\ 
SFHo & 1.0 & LR & 0.698 &  36 &  73 & 0.11 & 0.332 & 0.90 \\ 
\hline
SFHo & 1.67 & SR & 0.140 &  12 &  52 & 0.12  & 0.069 & 0.13 \\ 
SFHo & 1.67 & LR & 0.146 &  10 &  56 & 0.13  & 0.071 & 0.08   \\ 
%\hline
%SFHo & 1.67 & LR &  &  &  & & & \\ 
%SFHo & 1.67 & LR & 0.146 &  10 &  59 & 0.13 & 0.071 & 0.08 \\ 
\hline
%SLy4 & 1.0 & SR & 1.141 &  28 & 181 & 0.16 & 0.372 & 0.96 \\ 
SLy4 & 1.0 & SR & 0.072 &  33 & 94 & 0.25 & 0.240 & 0.84 \\ 
SLy4 & 1.0 & LR & 0.102 &  29 & 103 & 0.28 & 0.210 & 0.66 \\ 
\hline
%SLy4 & 1.67 & SR & 0.281 &   7 &  43 & 0.12 & 0.049 & 0.03 \\ 
SLy4 & 1.67 & SR & 0.310 &   7 &  41 & 0.12 & 0.047 & 0.03 \\ 
SLy4 & 1.67 & LR & 0.305 &  10 &  52 & 0.13 & 0.067 & 0.08 \\ 
\hline
%SLy4 & 1.8 & SR & 0.004 & 16.1 & 67.42 & 0.32 & 0.214 & 0.62 \\ 
SLy4 & 1.8 & SR & 0.729 & 6 & 40 & 0.13 & 0.047 & 0.01 \\ 
SLy4 & 1.8 & LR & 0.595 &   7 &  35 & 0.13 & 0.053 & 0.03 \\ 
      \hline\hline
    \end{tabular}
    }
  \label{tab:ejecta}
\end{table}

\begin{figure*}
  % data/ejecta_histo.py
\centering
  \includegraphics[width=\textwidth]{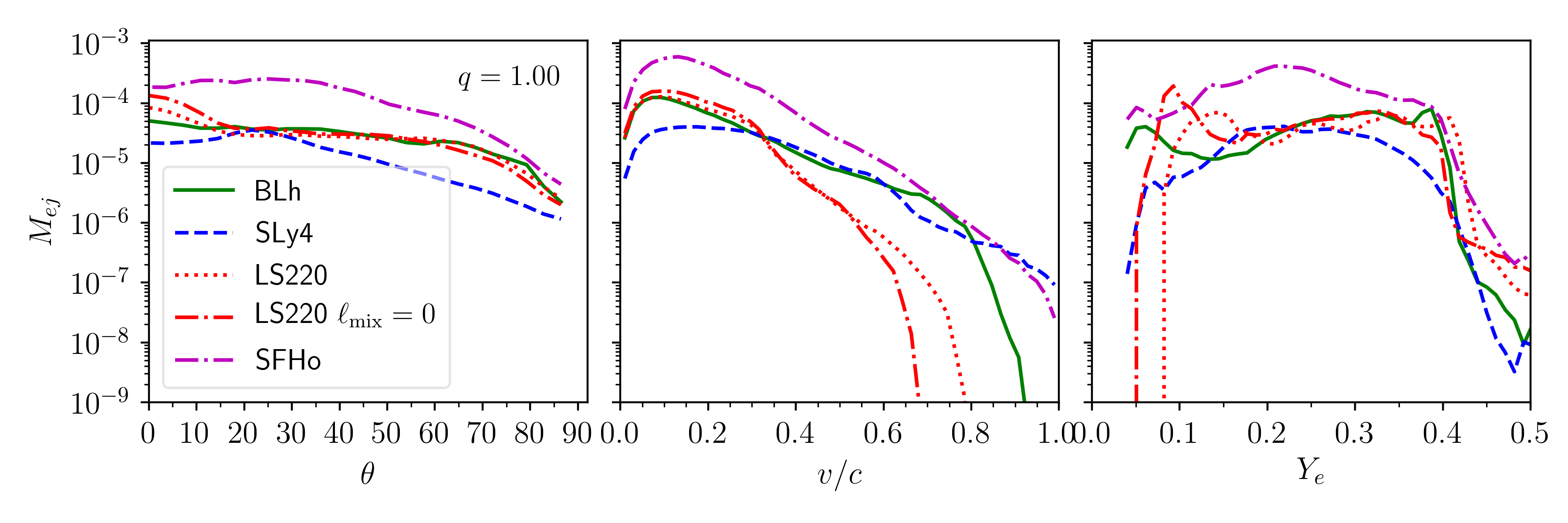}\vspace{-0.3 cm}
  \includegraphics[width=\textwidth]{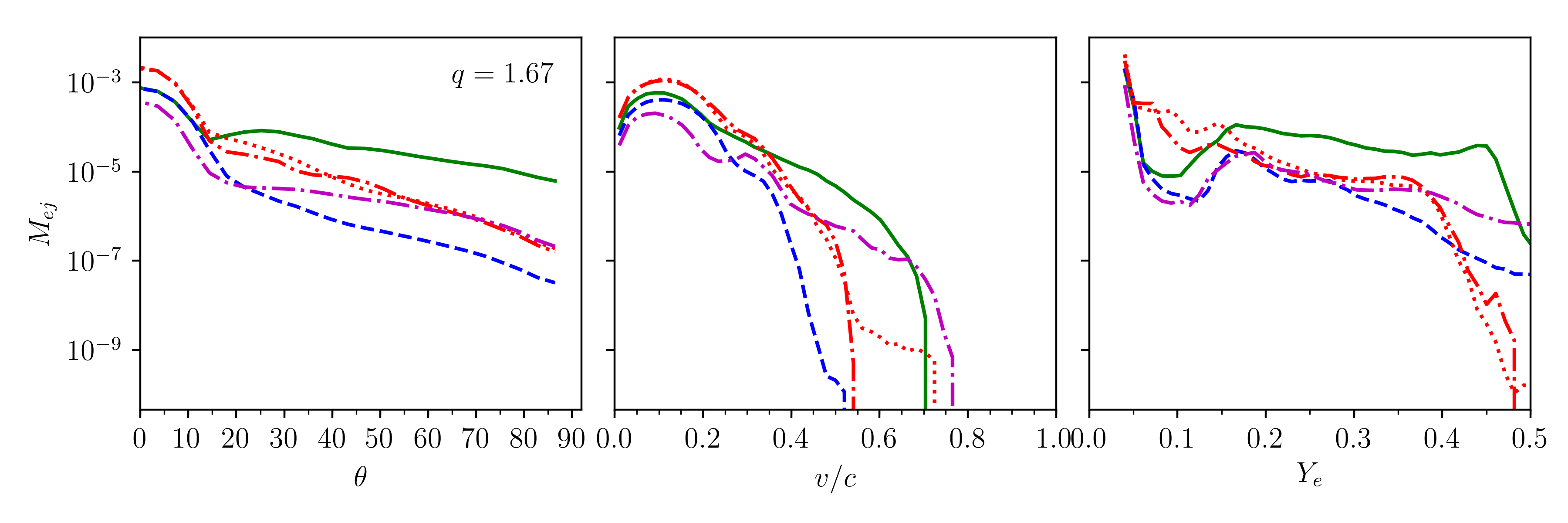}\vspace{-0.3 cm}
  \includegraphics[width=\textwidth]{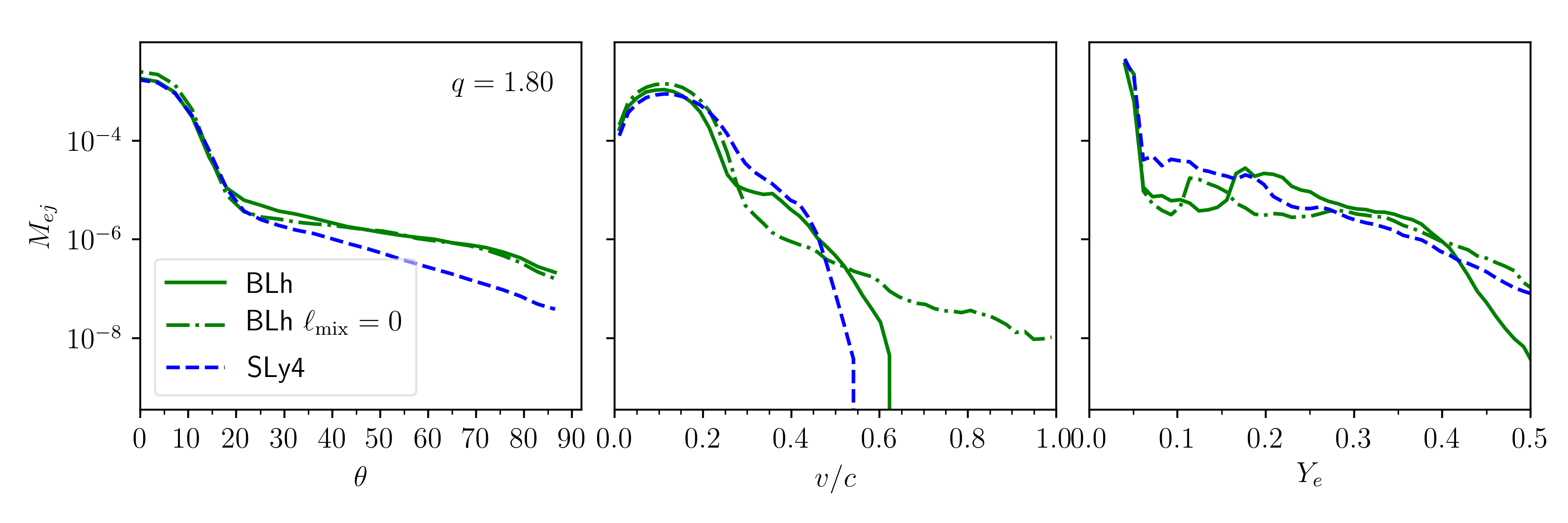}
\caption{Distributions of the ejecta mass in the polar angle (left), velocity
  (middle) and electron fraction (right) of the dynamical ejecta. Each
  row refers to a different mass ratio, from top to bottom
  $q=1,1.67,1.8$. Note that the angle $\theta=0^\circ$ identifies the orbital
  plane, while $\theta=90^\circ$ is the pole above the remnant. Data
  refer to resolution SR; data from simulations without turbulent viscosity are also shown.}
\label{fig:ej_histosummary}
\end{figure*}

\begin{figure}
  \centering
  % data/ejecta_phi.py
  \includegraphics[width=.48\textwidth]{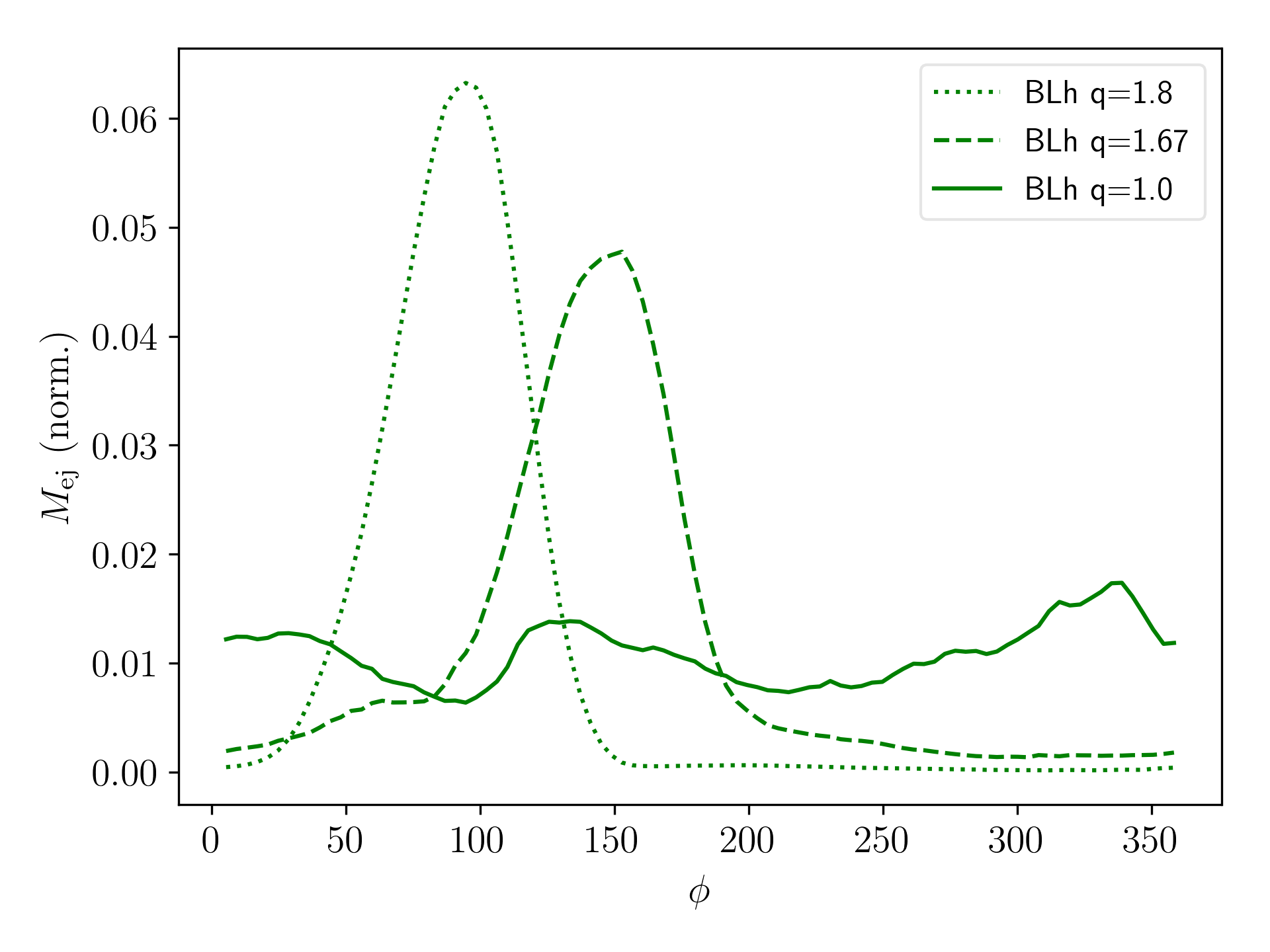}
\caption{Distribution of the cumulative dynamical mass ejecta in the
  azimuthal angle for the BLh BNS. The ejecta of the high-$q$ binaries
  expands with a crescent-like geometry similar to what found in
  simulations of black-hole--neutron-star binaries. The mass is
  normalized to the total ejecta mass.} 
\label{fig:ej_phi}
\end{figure}

\begin{figure*}
  \centering
  % data/ejecta_profile_summary.py
  \includegraphics[width=.48\textwidth]{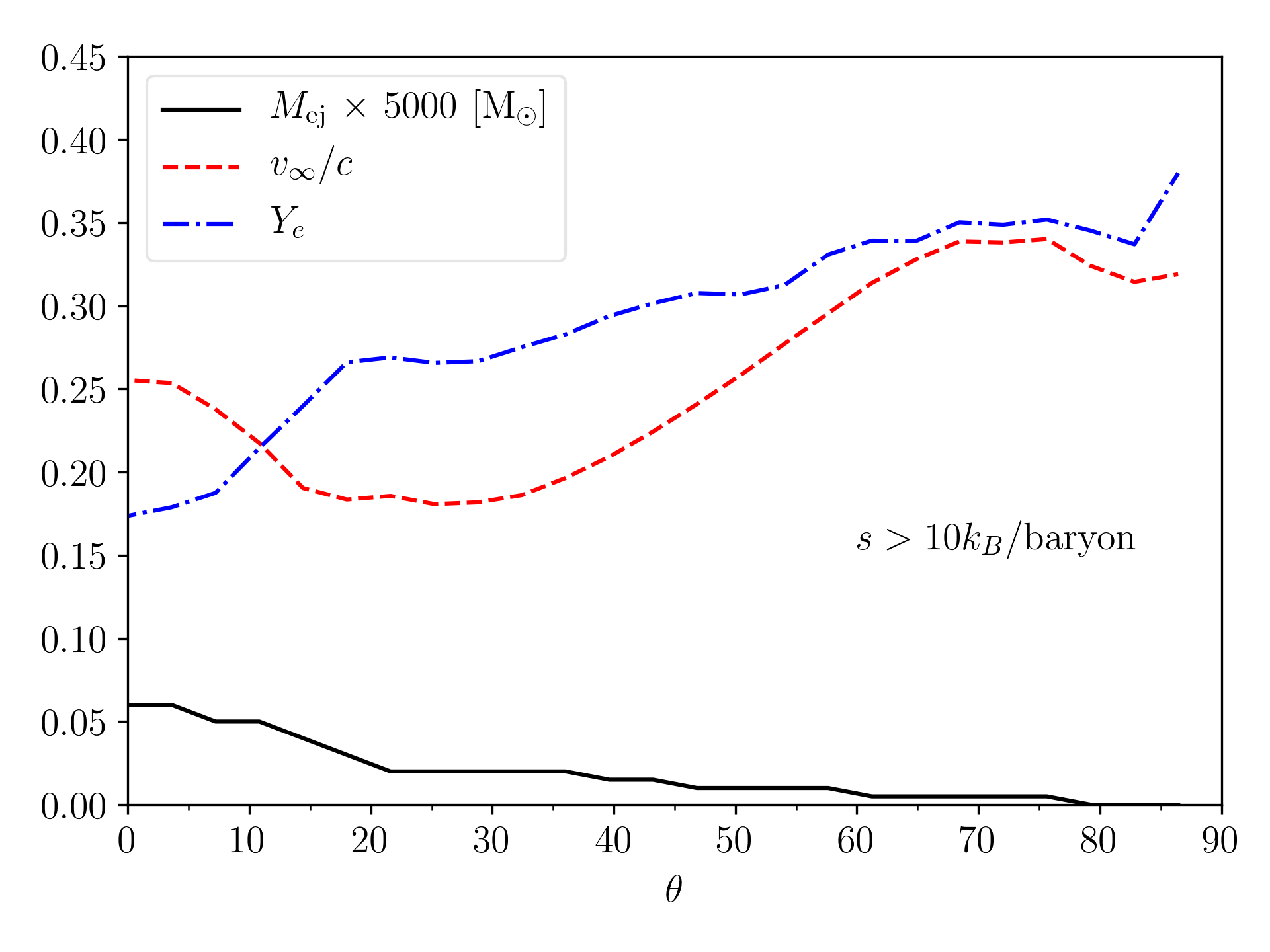}
\includegraphics[width=.48\textwidth]{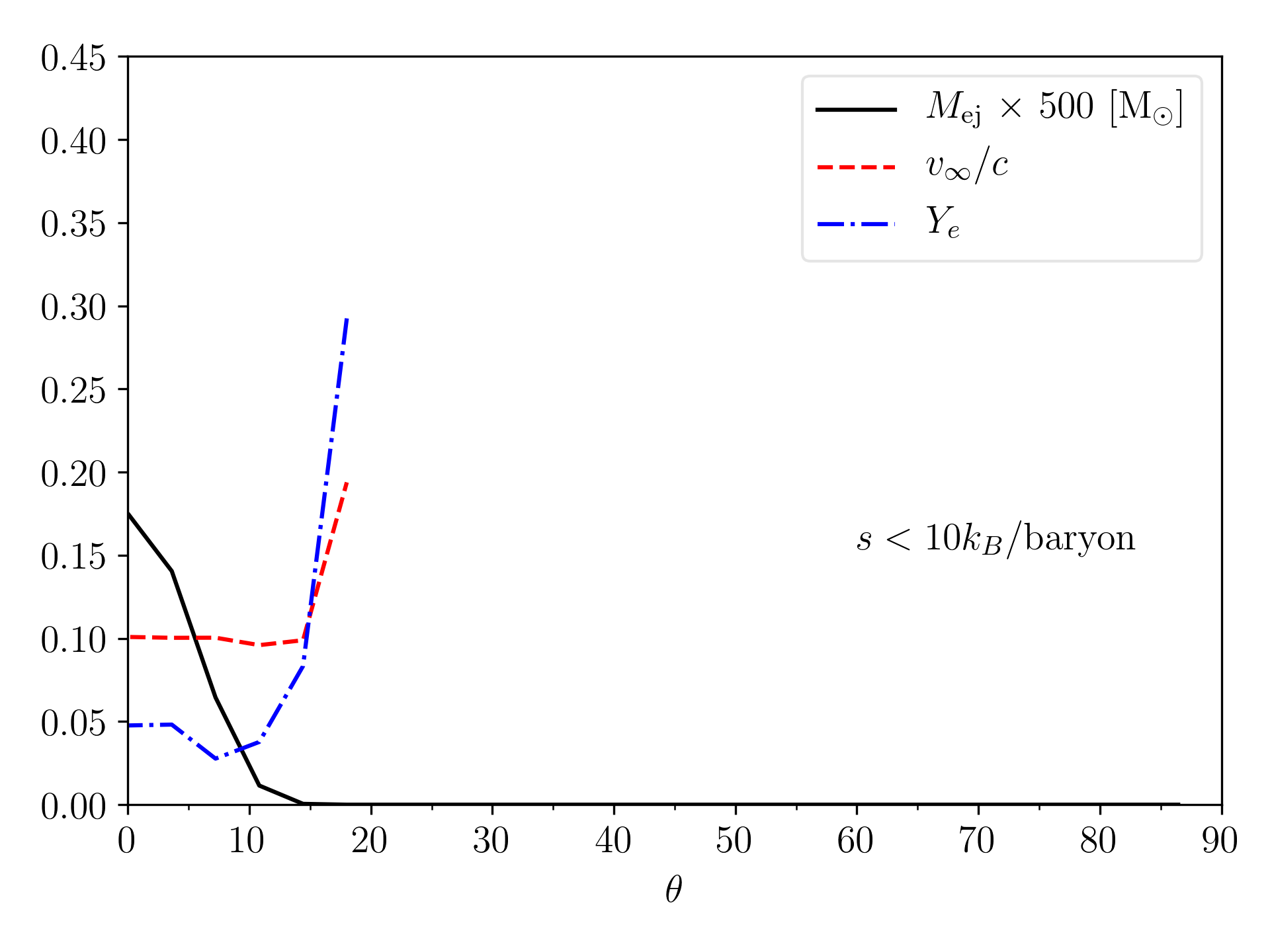}
\caption{Dynamical ejecta cumulative profiles of shocked (left) and tidal (right)
  components for BLh $q=1.67$ (SR) as a function of the polar
  angle. The different lines are the mass, the velocity and the
  electron fraction. Tidal and shocked components are conventionally
  separated by flagging fluid elements with specific entropy
  smaller or larger than $10k_B$ per baryon, respectively. 
  Note that the angle $\theta=0^\circ$ identifies the orbital
  plane, while $\theta=90^\circ$ is the pole.}
\label{fig:ej_tidalshock}
\end{figure*}

Mass ejecta are calculated on coordinate spheres at $r\simeq300$~km
assuming stationary spacetime and flow, 
and flagging the unbound mass according to the geodesic criterion. A particle on
geodesics is unbound if the 4-velocity component $u_t\leq -1$
and thus it reaches infinity with velocity 
$v_\infty\simeq(1-u^2_t)^{1/2}$. This geodesic criterion neglects the
fluid's pressure, thus potentially underestimating the mass, but it is considered
appropriate for the dynamical ejecta that are moving on ballistic
trajectories, e.g. \cite{Kastaun:2014fna}.

We compute the mass-histograms of the ejecta main properties and show them in
Fig.~\ref{fig:ej_histosummary}. In the case of equal mass BNS (top panels), dynamical ejecta
are distributed all over the solid angle and composed of both the 
tidal and the shocked component, (e.g.,~\citealt{Hotokezaka:2012ze,Bauswein:2013yna,Sekiguchi:2015dma,Radice:2018pdn}).
The velocity of the material peaks at $v\sim0.1-0.2$~c and has high speed tails extending up to
$v\sim0.6-0.8$~c. The largest tail velocities are reached by the 
softest EOS and in the polar regions, where baryon pollution is minimal, as
a consequence of the NS cores' bounce (Fig.~3 of \citealt{Radice:2018pdn}).
Note however, masses ejecta ${\ll}10^{-5}$ and velocities
${\gtrsim}0.9$~c cannot be trusted and can
suffer of large numerical errors due to the atmosphere treatment and
imperfect mass conservation (Appendix~\ref{sec:q-dep}).
The ejecta's composition is characterized by a wide range of $Y_e$;
for the LS220 and the BLh EOS two peaks at ${\sim}0.1-0.15$ and at
${\sim}0.4$ are clearly visible; they roughly correspond to the 
shocked and tidal components, although the former has also a 
significant amount of material with low $Y_e$ material. Note the SFHo model
peaks instead at ${\sim}0.25$.
Comparing the two equal mass LS220 BNS, we
find a small effects of turbulent
viscosity: the viscous ejecta have a more prominent peak at lower
$Y_e$ and a slightly reduced tidal component, possibly due to the
difference in the early-postmerger dynamics around the moment of core bounce.

The dynamical ejecta of asymmetric BNS with $q\gtrsim1.67$ 
(middle panels of Fig.~\ref{fig:ej_histosummary}) are
quantitatively different from symmetric BNS. The ejected material
is distributed more narrowly about the
orbital plane and decreases almost monotonically to the polar latitudes.
The dependence on the azimuthal angle is also very different from the
equal-mass cases. Because the matter is almost entirely expelled by 
tidal torques, the ejecta is distributed over a fraction of the azimuthal angle 
around its ejection angle and has a crescent shape, Fig.~\ref{fig:ej_phi}.  This is
similar to what observed in black-hole--neutron-star binaries~\cite{Kyutoku:2015gda,Kawaguchi:2016ana}.
Hence, the ejecta for high-$q$ BNS are not formed isotropically.
Most of the unbound mass has low $Y_e\lesssim
0.1$, although several $q=1.67$ BNS have a second peak at
${\gtrsim}0.4$.
Thus, while the tidal component is the dominant for asymmetric BNS, 
a small shocked component persists. %, specially for stiff EOS.
The velocity distributions have comparable peak values indicating that 
the tidal component has velocities comparable to those of shocked
component (Cf.~Fig.~6 of \citealt{Dietrich:2016hky}).
Note that the fast tails are suppressed for increasing mass ratio.
This is because of the less violent merger and bounce experienced by
these binaries.
These features are even more extreme for the
$q=1.8$ case (bottom panels of Fig.~\ref{fig:ej_histosummary}).
The above results appear consistent with those reported in
\cite{Sekiguchi:2015dma,Lehner:2016lxy}, although different EOS and
more moderate mass ratio were used there.

The mass-averaged properties of the dynamical ejecta computed from the
histograms are reported in Tab.~\ref{tab:ejecta}. We show results for
all the resolutions available in order to convey an idea of the
uncertainties. The latter are difficult to precisely quantify since
strict convergence is not observed in the data. However, the results
are robust for a large variation of the grid resolution with mass 
variations at the ${\sim}20\%$ level between SR and HR and less than a
factor two between LR and SR. Note there is a factor 2 (1.5) between
the spacing of LR and SR (SR and HR) grids. 
The following discussion mostly refers to highest resolutions available, as
the LR is not always sufficient to properly resolve the composition (see below). 

The large mass asymmetry can boost the mass ejecta by up to a factor four with respect to
the equal mass cases. The average electron fraction of the
dynamical ejecta from asymmetric BNS is ${\sim}0.11$, a factor
two smaller than for the respective equal mass BNS.
The mass-distribution is concentrated around the
equatorial plane. The rms of the polar angle is
${\sim}5-15^\circ$ for asymmetric BNS with
$q=1.8-1.67$, while it is ${\sim}35^\circ$ for symmetric BNS.
Overall these results show that while the tidal component of the
dynamical ejecta is dominant with respect to the shocked ejecta in
high mass ratio binaries, a delayed collapse can produce unbound
mass with electron fractions that can extend to $Y_e\sim0.4$. 
The rms of the azimuthal angle is reduced from ${\lesssim}106^\circ$ of symmetric BNS to less
than half, ${\gtrsim}50^\circ$, for asymmetric BNS.
We recall that the rms of a uniform distribution with support on
the segment $2\alpha\in(0,2\pi]$ is
$\langle\phi\rangle=\sqrt{3}/3(\pi-\alpha)$, thus giving 
$\langle\phi\rangle\simeq 104^\circ$ if the
support is the full interval ($360^\circ$) and $\langle\phi\rangle\simeq54^\circ$ if the
support is half of the interval ($180^\circ$). A similar argument holds also for
the polar angle support around the equator, $\pi/2-\alpha \leq \theta \leq \pi/2+\alpha$, 
for which $\langle\theta\rangle=\left( \sqrt{3}/3 \right) \alpha $. 
This is correct as far as the ejecta is emitted uniformly over a small portion
around the equator (a good approximation in the case of high mass ratio BNS).

The tidal and shocked contributions to the dynamical ejecta
are calculated by conventionally distinguishing the unbound matter with
specific entropy smaller or larger than  $10k_B$ per baryon, respectively ~\cite{Radice:2018pdn}. 
The last column of Tab.~\ref{tab:ejecta} reports the ratio
\be
X_s = \frac{M_{\rm ej}^{\rm shocked}}{M_{\rm ej}} = \frac{M_{\rm
    ej}^{\rm shocked}}{M_{\rm ej}^\text{tidal}+M_{\rm ej}^{\rm
    shocked}} \ ,
\ee
indicating the mass fraction of the shocked ejecta to the total value. 
For the BLh EOS $X_s$ increases from $0.01$ to $0.3$ and $0.9$ for
$q=1.8$ to $1.67$ and $q=1$, respectively. For the SLy EOS $X_s\simeq0.01$
for $q=1.8$ and $q=1.67$ that have a similar dynamics charcaterized by
the accretion-induced BH formation and prominent tidal ejecta, and
$X_s\simeq0.8$ for $q=1$. The other two $q=1$ mergers with short-lived
NS remnants have $X_s\simeq0.7$ that reduces to $0.1$ for $q=1.67$. 

As an example, we discuss mass-histograms for the shocked and
tidal components separately for the BLh $q=1.67$, Fig.~\ref{fig:ej_tidalshock}.
The tidal component is confined within an  
angle of $\theta\lesssim10^\circ$ from the orbital plane; most of the
mass has $Y_e\sim0.05$ with the largest
electron fractions $Y_e\sim0.15$ reached at those latitudes. The 
velocities are uniformly distributed $v\sim0.1$~c. The shocked
component, instead, has mass mostly distributed at angles
$\theta\sim25^\circ$ but it extends to polar latitudes. The ejecta 
has electron fraction $Y_e\sim0.17-0.25$ for $\theta\lesssim 25^\circ$
and $Y_e\sim0.30-0.35$ for $\theta> 60^\circ$.
The velocity of the bulk ejecta at the orbital latitudes is $v\lesssim0.25$~c, minimal
at around $\theta\sim 27^\circ$, and has a peak $v\lesssim0.3$~c at
polar latitudes.
In general, the shocked component is slightly delayed with respect the tidal
component because it is generated when the NS cores' bounce
\cite{Radice:2018pdn}.

Table~\ref{tab:ejecta} also highlights a dependency on resolution, especially for
high mass ratios BNS. This is expected since resolving NS with
different sizes is more challenging than with equal sizes for the
box-in-box AMR. In particular, the LR resolutions does not seem
sufficient to deliver quantitatively robust results for all the cases,
especially at high $q$ and with viscosity. Note, for example, that ejecta mass
decreases with resolutions indicating numerical dissipation plays a
role enhancing the ejecta. Moreover, $Y_e$ raises very rapidly from
the NS surface; in the case the latter is not well resolved the tidal
ejecta might be spuriously composed of material from the interior, as
observed in the BLh $q=1.8$ LR simulation. 

\begin{figure}
% data/ejecta_visc.py
  \includegraphics[width=.5\textwidth]{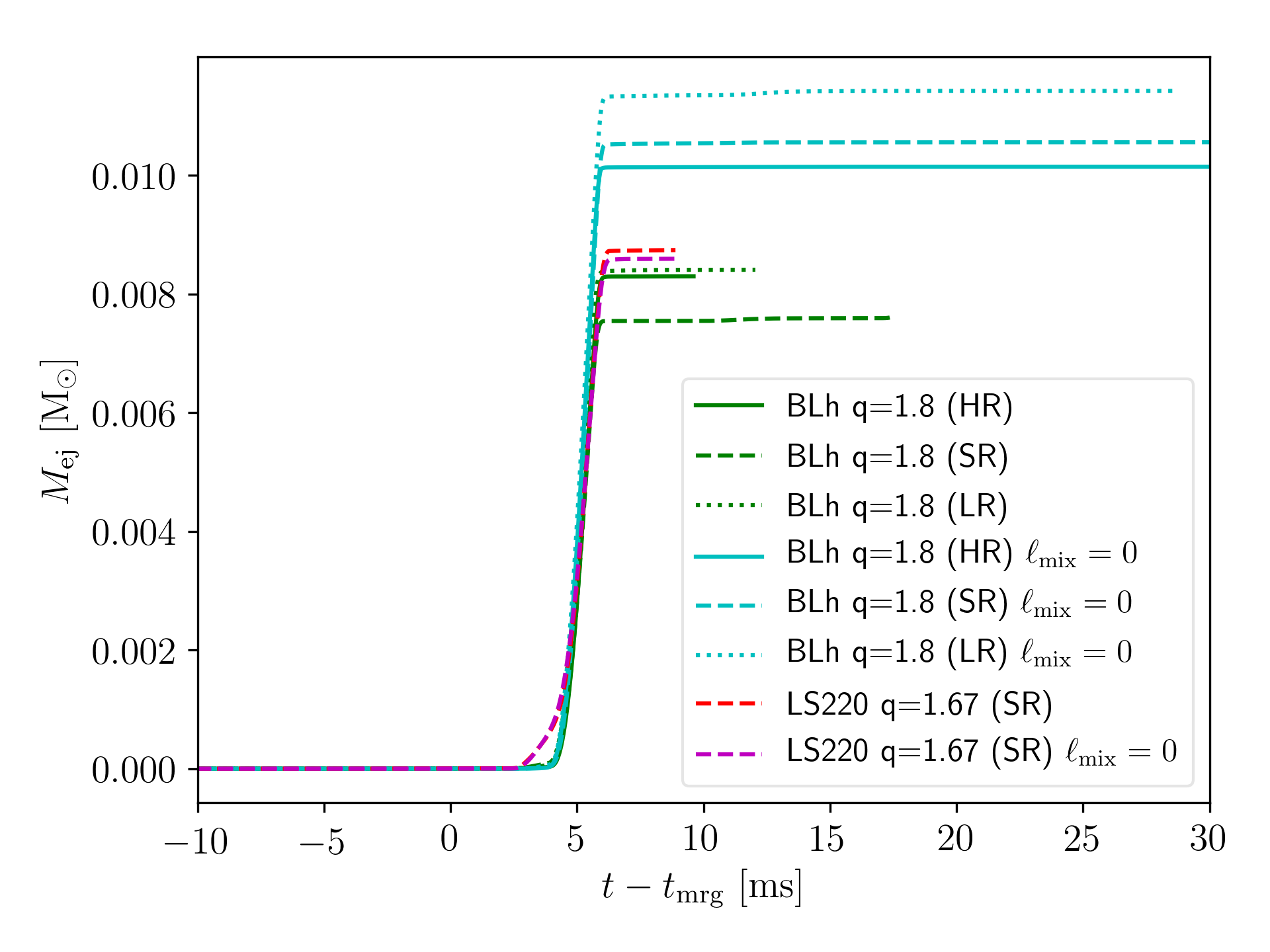}
\centering
\caption{Dynamical mass ejecta in viscous and nonviscous simulations.
The viscous dynamical ejecta reported in \citet{Radice:2018ghv} is
here not present because the shocked component ejecta is
negligible. Furthermore, the angular momentum distribution introduced
by the subgrid model $\ell_{\rm mix}(\rho)$ appears to be dependent on
the EOS model. Note the turbulent viscosity subgrid model employed
here is different from the $\ell_{\rm mix}$ employed in previous simulations.} 
\label{fig:ej_visc_nonvisc}
\end{figure}

We finally comment on the effect of viscosity on the dynamical
ejecta. \citet{Radice:2018ghv} pointed out that the dynamical ejecta in
asymmetric BNS can be enhanced by the thermalization of mass accretion 
streams between the secondary and the primary neutron star. This
viscous component of the dynamical ejecta are characterized by large
asymptotic velocities and have masses that depend on the efficiency of
the viscous mechanism. 
Figure~\ref{fig:ej_visc_nonvisc} shows the ejecta mass for the BLh
$q=1.8$ and LS220 $q=1.67$ BNS. The viscous dynamical ejecta is not present
because the shocked component ejecta is negligible. Actually, the 
turbulent viscosity here can reduce the tidal dynamical ejecta as a consequence
of the different angular momentum distribution due to
turbulence.
Note the effect is significant and robust with respect to the
variation of the grid resolution.
The effect of viscosity is much reduced in the LS220 $q=1.67$ BNS and
practically negligible considering the numerical uncertainties (only
the SR is shown for clarity). This might be related to the differences
in the EOS at low density (Sec.~\ref{sec:eos}). 
The simulations of \cite{Radice:2018ghv}
employed the GRLES scheme as those presented here, but using
$\ell_\text{mix}=const$ and varying systematically the constant for the
turbulent parameter.
We cannot currently exclude that the
specific subgrid model $\ell_\text{mix}(\rho)$ built
from \cite{Kiuchi:2017zzg} determines a different
effect with respect to the $\ell_\text{mix}=const$ model.
A detail investigation of the viscous dynamical ejecta with the
subgrid model $\ell_\text{mix}(\rho)$ for intermediate values of $q$
will be presented elsewhere.
%

%%%%%%%%%%%%%%%%%%%%%%%%%%%%%%%%%%%%%%		
\section{Synthetic kilonova light curves}
\label{sec:kn}

We compute synthetic kilonova light curves for each of the BNS mergers
presented in this work.  We use a semi-analytical multicomponent, anisotropic kilonova
model that takes into account the angular distribution of the ejecta
properties as well as the presence of different kinds of
ejecta \cite{Perego:2017wtu,Radice:2018xqa,Radice:2018pdn,Barbieri:2019kli}.
The latter differ by the mechanisms that cause the ejection and the
timescales over which they operate.  Within this framework, the
homologously expanding ejecta is discretized in velocity space and the
photon diffusion time is estimated by timescale arguments.  Radiation
is assumed to be in local thermodynamical equilibrium up to the
relevant photosphere and photon emission is modelled as a
superposition of blackbody spectra.  The different ejecta components
comprise the dynamical ejecta discussed in Sec.~\ref{sec:ejecta} and
possibly winds expelled by the remnant disc on longer timescales
(0.1-1s) by means of neutrino irradiations and turbulent viscosity of
magnetic origin.  The kilonova emission produced by each
component depends mainly on three quantities that characterize the
ejecta, namely the amount of mass, $M_{\rm ej}$, its average expansion
velocity, $\langle v_{\rm ej} \rangle$, and an (effective) grey photon
opacity, $\kappa_{\rm ej}$. In all our kilonova models, we locate the
merging BNS at a distance of 40 Mpc and we consider a reference
viewing angle of $\pi/6$ with respect to the rotational axis of the
binary. If not otherwise specified, the model parameters and input
physics are assumed to be as in the best fit model named ${\rm BF}$ to
AT2017gfo of \cite{Perego:2017wtu}.

\begin{figure*}
\centering
\begin{minipage}{\linewidth}
\begin{minipage}{0.35\linewidth}
\includegraphics[width=\textwidth]{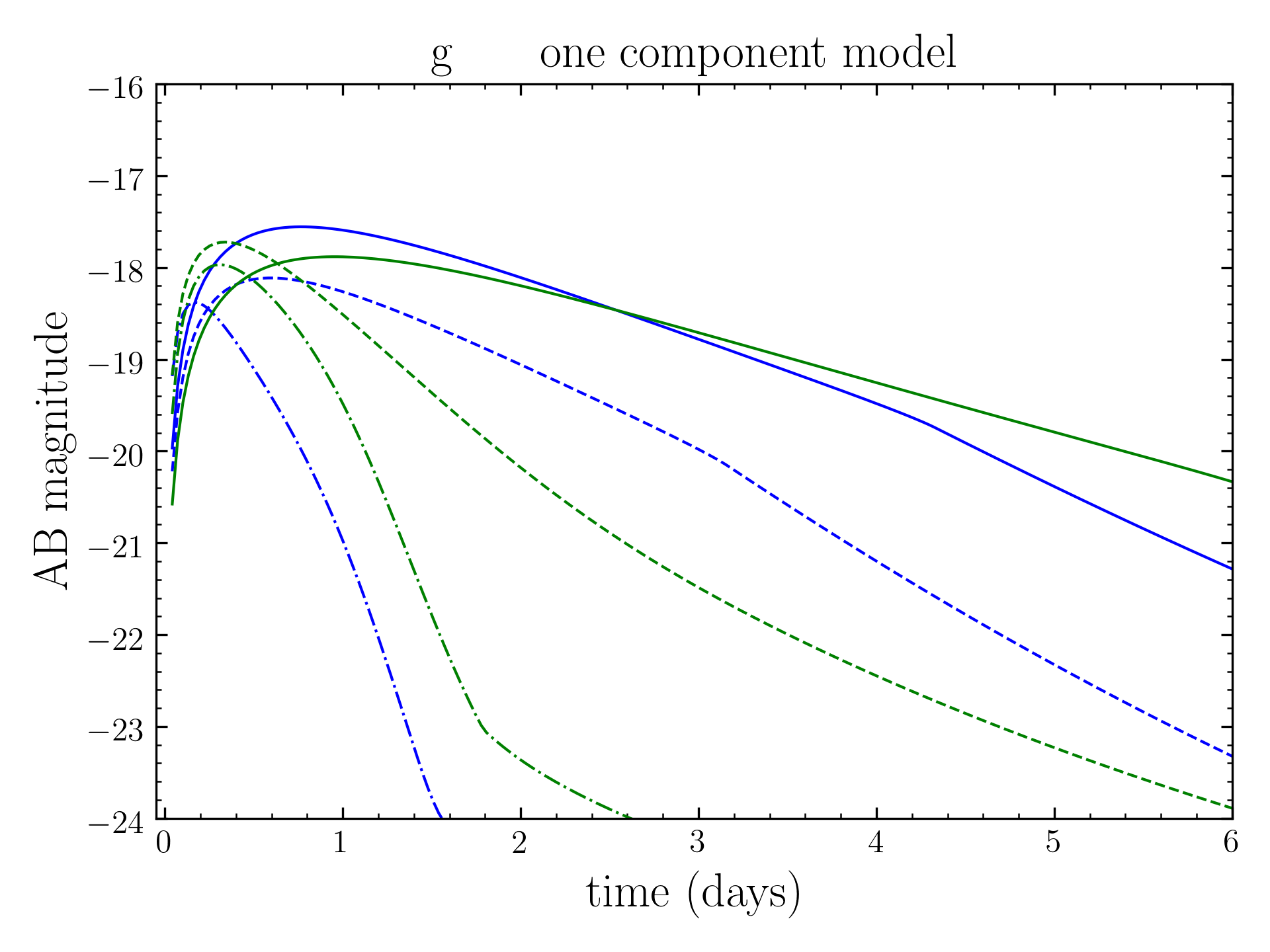}
\end{minipage}
\begin{minipage}{0.35\linewidth}
\includegraphics[width=\textwidth]{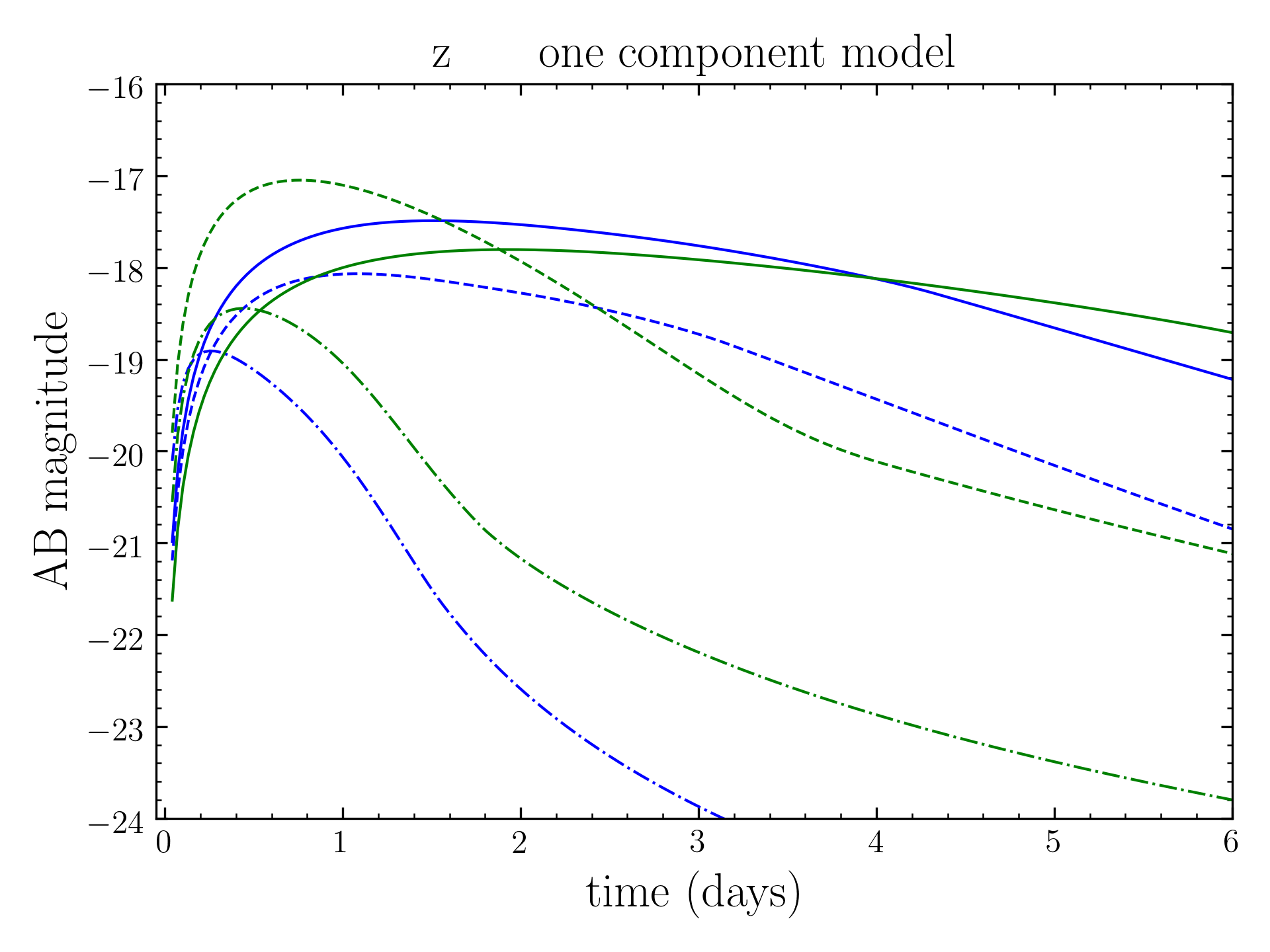}
\end{minipage}
\begin{minipage}{0.35\linewidth}
\includegraphics[width=\textwidth]{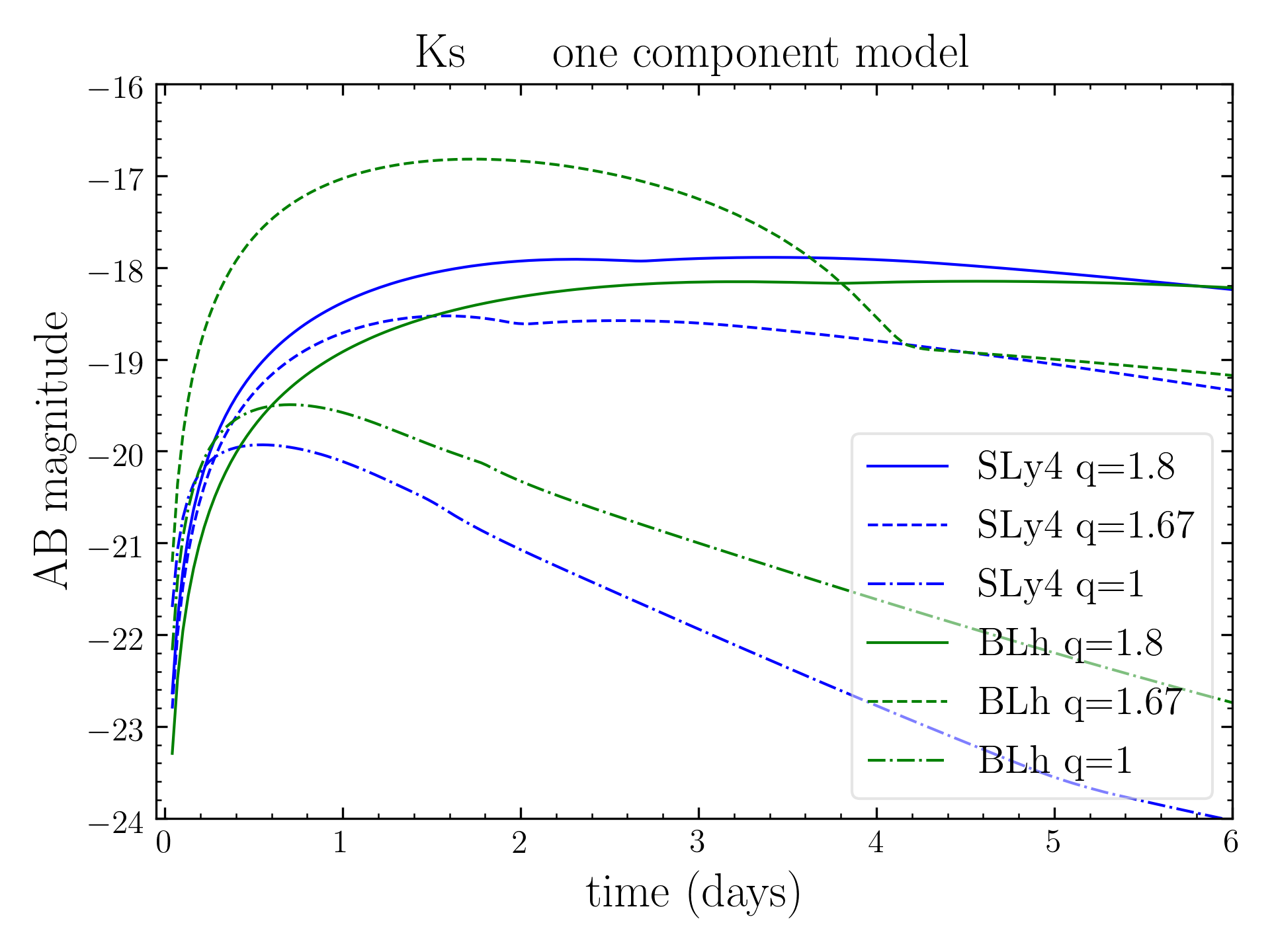}
\end{minipage}
\end{minipage}
\caption{Kilonova light curves from the simulated dynamical ejecta (one component) for the BLh and SLy BNS. The ejecta properties are taken from the simulation at the highest grid resolution. The light curves are computed with the axisymmetric models of \citet{Perego:2017wtu} for $q=1$ and of \citet{Barbieri:2019kli} for $q=1.67,1.8$. Binaries are always assumed to be located at a distance of 40~Mpc and to be observed under a viewing angle of $30^\circ$ with respect to the BNS rotational axis. The bump observed in the $K_s$ band for the BLh $q=1.67$ model results from the radial emission from the crescent pointing towards the observer.}
\label{fig:1comp_lightcurves}
\end{figure*}

We first examine the kilonova emission obtained by considering only
the dynamical ejecta discussed in Sec.~\ref{sec:ejecta}. In the case
of $q=1$ mergers, matter is expelled over the entire solid angle and
we follow the model presented
in \citet{Perego:2017wtu,Radice:2018xqa,Radice:2018pdn}.  We assume
the ejecta to be axisymmetric and the photon diffusion to proceed
mostly radially. In these cases, we discretize the polar angle in 30
slices over the whole solid angle.  We use azimuthal averages of the
angular distribution of the ejected mass, electron fraction and of
mean expansion velocity directly extracted from the latest stages of
our NR simulations.  While the ejecta mass and mean velocity are
directly input into the kilonova model, the electron fraction is used
to assign the ejecta opacity according to $\kappa_{\rm dyn} = 1~{\rm
cm^2~g^{-1}}$ for $ \langle Y_e \rangle > 0.25$, $\kappa_{\rm dyn} =
20~{\rm cm^{2}~g^{-1}}$ otherwise
(Cf.~\citet{Kasen:2013xka,2019arXiv190608914T,Fontes:2019tlk}.)
Alternatively, for the $q=1.67$
and $q=1.8$ cases the dynamical ejecta is confined (in very good
approximation) within a crescent across the equatorial plane (see
Sec.~\ref{sec:ejecta}) and we emply the model described
in \citet{Barbieri:2019kli} (see also \citealt{Kawaguchi:2016ana}) in
which the photon emission is the combination of radial and lateral
emissions from an optically thick disc. In this case, we use the total
ejecta mass, $M_{\rm ej}$, and mean velocity, $\langle v_{\rm
ej} \rangle$, obtained by our NR simulations to initialize a
vertically homogeneous, radially expanding disc.  For the grey
opacity, we assume always $\kappa_{\rm dyn} = 20 {\rm cm^2~g^{-1}}$
since in these cases $ \langle Y_e \rangle < 0.25$ (often $ \langle
Y_e \rangle < 0.10$).  For the disc half-opening angle in the polar
direction we use $\theta_{\rm disc} = \sqrt{3} \langle \theta_{\rm
ej} \rangle$, while for the azimuthal disc opening we set $\phi_{\rm
disc} = 2 \sqrt{3} \langle \phi_{\rm ej} \rangle $ (See Sec.~\ref{sec:ejecta}).
The crescent shape breaks the axisymmetry of the emission. In our calculations,
we always assume the dynamical ejecta to be emitted toward the observer.
For small polar opening angles, this assumption is not very relevant, since
the radial emission is subdominant. In the case of larger discs (as in the
BLh $q=1.67$ case) the radial emission can be relevant and our model assumptions
can be more questionable. 

In Fig.~\ref{fig:1comp_lightcurves}, we present light curves in three
different photometric bands ($g$, $z$, and $K_s$) to span the relevant
wavelength interval from visible to near-infrared radiation, for the
three different models obtained with the BLh and SLy4 EOS.  We first
notice that, even in the case of prompt collapse, BNS mergers can
power bright kilonovae \cite{2020ApJ...889..171K,Kyutoku:2020xka}. In
particular, in the high-$q$ models the 
light curves from the dynamical ejecta are possibly brighter, with
wider light curves peaking at later times compared with the $q=1$
mergers. This is due to the crescent-like configuration of the
expanding ejecta (Cf.~\cite{Tanaka:2013ixa,2020arXiv200400102K}.)
On the one hand, when matter is emitted over a large
portion of the solid angle (as it usually happens for $q\sim 1$) the
hotter ejecta is buried inside the optically thick region and high
energy photons have to diffuse and thermalize before being emitted in
the kilonova. On the other hand, thanks to the disc-like geometry of
the crescent, the innermost, hotter portion of the disc provides a
significant contribution to the kilonova emission at any time,
explaining the brighter and more substained emission.  These effects
are visible in all bands, but the increase in magnitude moving from
$q=1$ to higher $q$'s is more pronunced in the infrared band as a
consequence of the lower electron fraction (and thus of the higher
opacity) of the dynamical ejecta in the crescent. This effect is even
amplified by the larger amount of dynamical ejecta observed in the
high-$q$ models (with the only exception of the SFHo models).

\begin{figure}
\centering
\includegraphics[width=.48\textwidth]{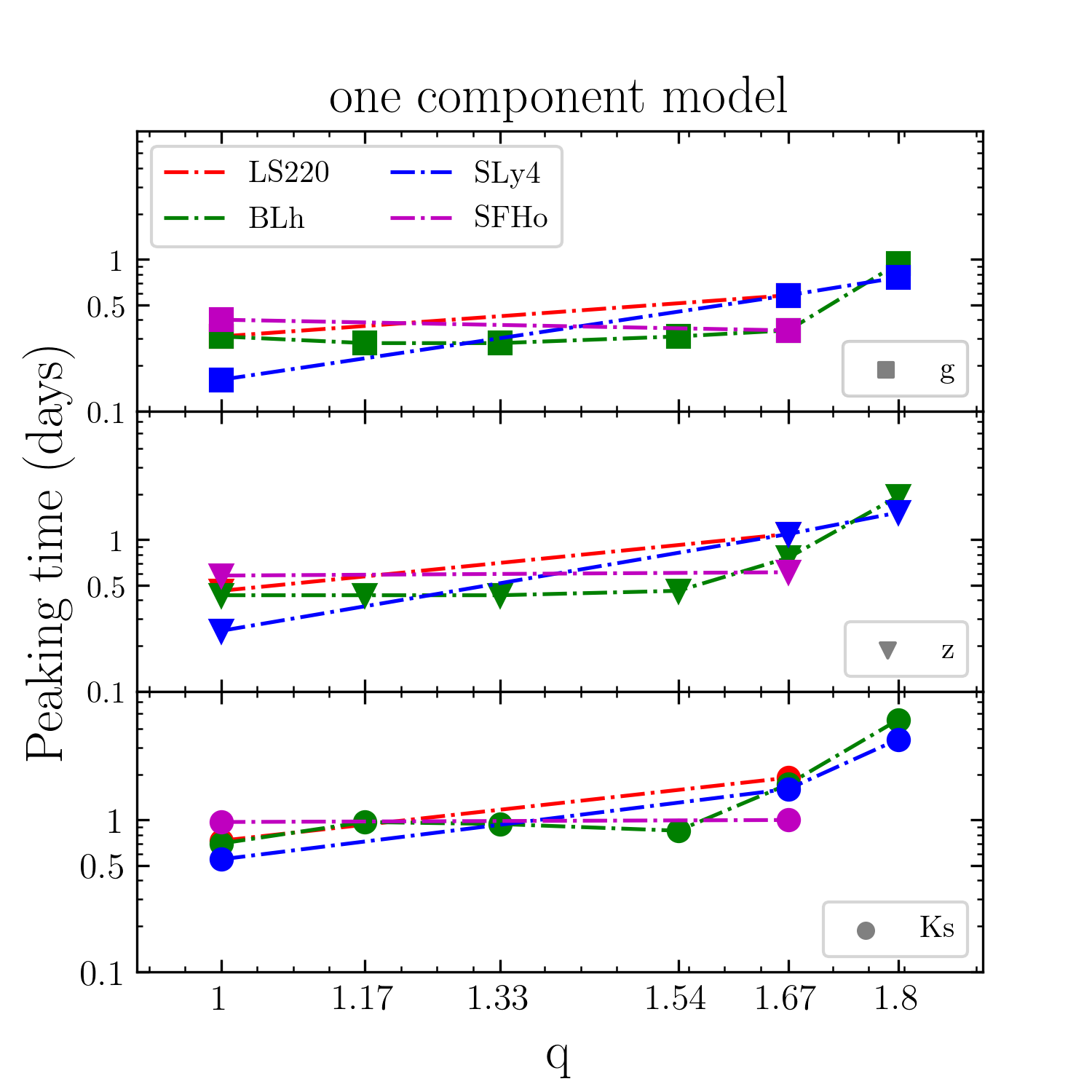}
\caption{Peak time of the one component kilonova models as a function
of the mass ratio for all the simulated BNS. Data from BLh simulations at
intermediate mass ratios are also included (See Appendix~\ref{sec:q-dep}). 
Note the log scale and that straight lines
connecting the points are due to the limited number of simulations
available and do not represent accurately the functional behaviour in
the mass ratio.}
\label{fig:tpeaks_vs_q}
\end{figure}

The peak times of all kilonova models are shown as a function of the mass
ratio in Fig.~\ref{fig:tpeaks_vs_q}. In addition to the models
presented in Tab.~\ref{tab:sim}, we include here also a few more LR models
computed with the BLh EOS (See Appendix~\ref{sec:q-dep}) to better explore 
the dependence on $q$. The kilonova peak times of mergers
undergoing accretion-induced prompt collapse are significantly delayed
with respect to the $q=1$ cases. For the BLh merger, the emission in
$g$, $z$ and $K_s$ bands peaks between few hours and within a day
respectively if $q=1$, and and between a day and a week if
$q=1.8$. The near-infrared frequencies are those that vary most as a
function of the mass ratio. The SLy4 light curves shows a similarly behaviour,
although less data points are available. Less variation in the peak
times is observed in the LS220 and SFHo mergers between $q=1$ and
$q=1.67$, but note that in those cases the dynamical ejecta mass also
vary less with the mass ratio.

We tested that the features described above do not depend on the specific velocity
profile for the homologously expanding ejecta, in which most of the
mass resides in the innermost part of the disc. 
Indeed, a flat distribution 
in the expanding velocity as employed in \citet{Kawaguchi:2016ana}, provides 
very similar results.
This is due to a compensation effect between the larger amount of decaying
material and the denser (thus, optically thicker) vertical profile of
the disc in our models. These features are robust also with respect to
the uncertainties on the ejecta properties of numerical origin.
Considering the ejecta properties extracted from simulations at
different resolutions gives some quantitative changes that mostly
affect the light curves' luminosity. Here is worth to remark that a
factor two of uncertainty in the ejecta mass can translate in up to an
order of magnitudes in luminosity. Moreover, current light curve
models suffer of larger systematic uncertainties in
nuclear (e.g. mass models, fission fragments and $\beta$-decay
rates) and atomic (e.g. detailed wavelength dependent opacities for
$r$-process element)
physics~\citep{Eichler:2014kma,Rosswog:2016dhy,Gaigalas:2019ptx}.

\begin{figure*}
\begin{minipage}{\linewidth}
\begin{minipage}{0.35\linewidth}
\includegraphics[width=\textwidth]{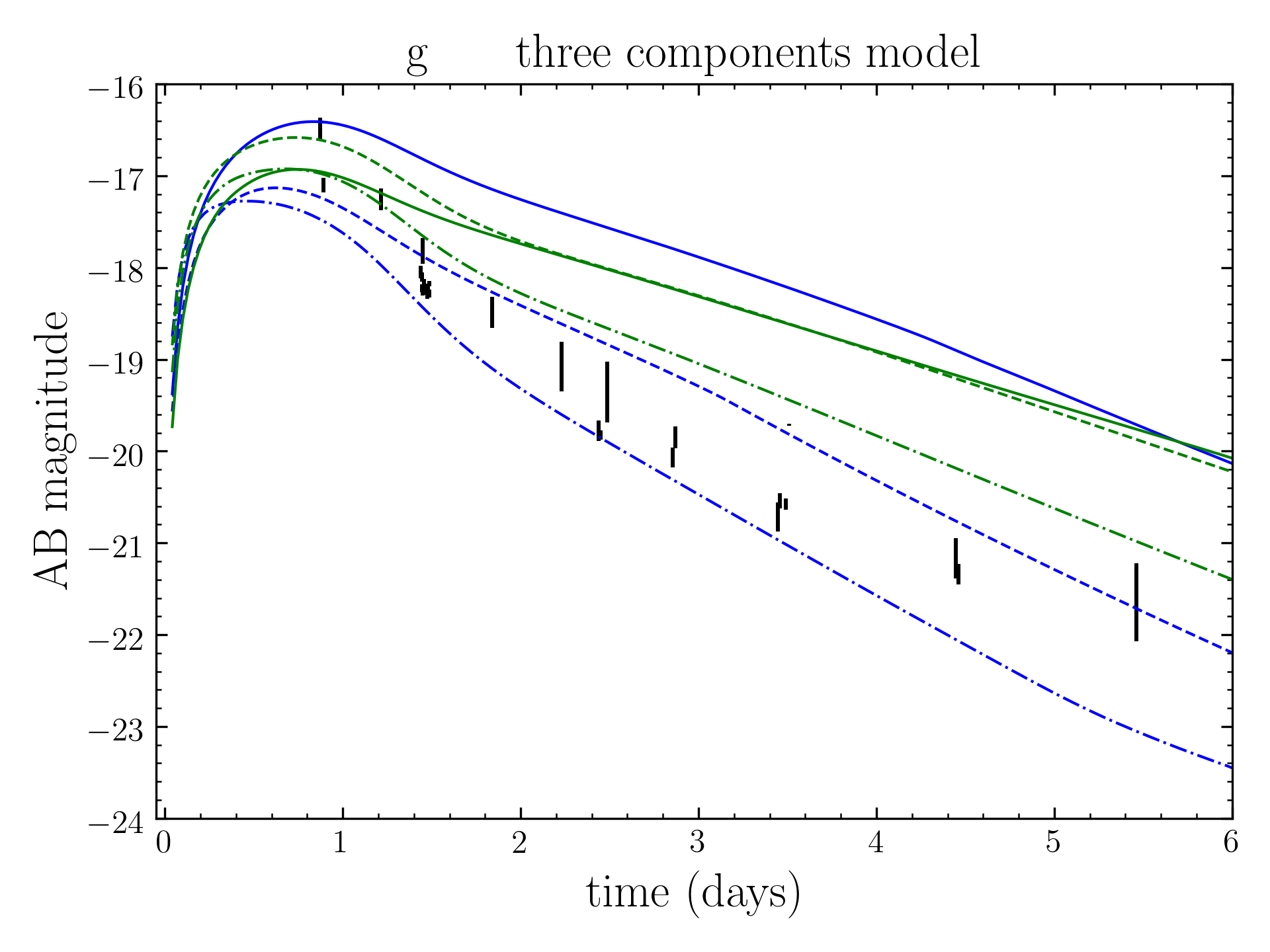}
\end{minipage}
\begin{minipage}{0.35\linewidth}
\includegraphics[width=\textwidth]{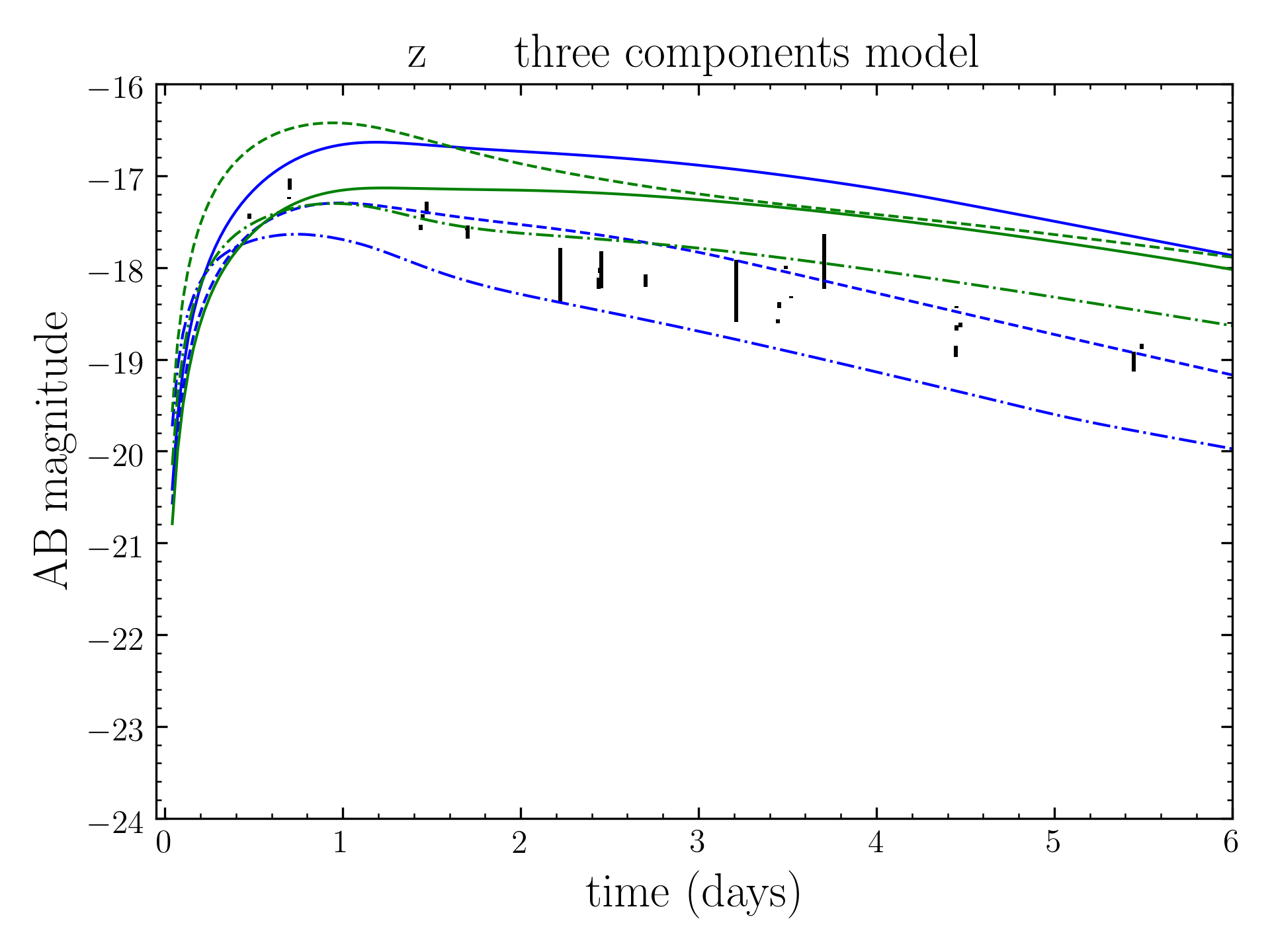}
\end{minipage}
\begin{minipage}{0.35\linewidth}
\includegraphics[width=\textwidth]{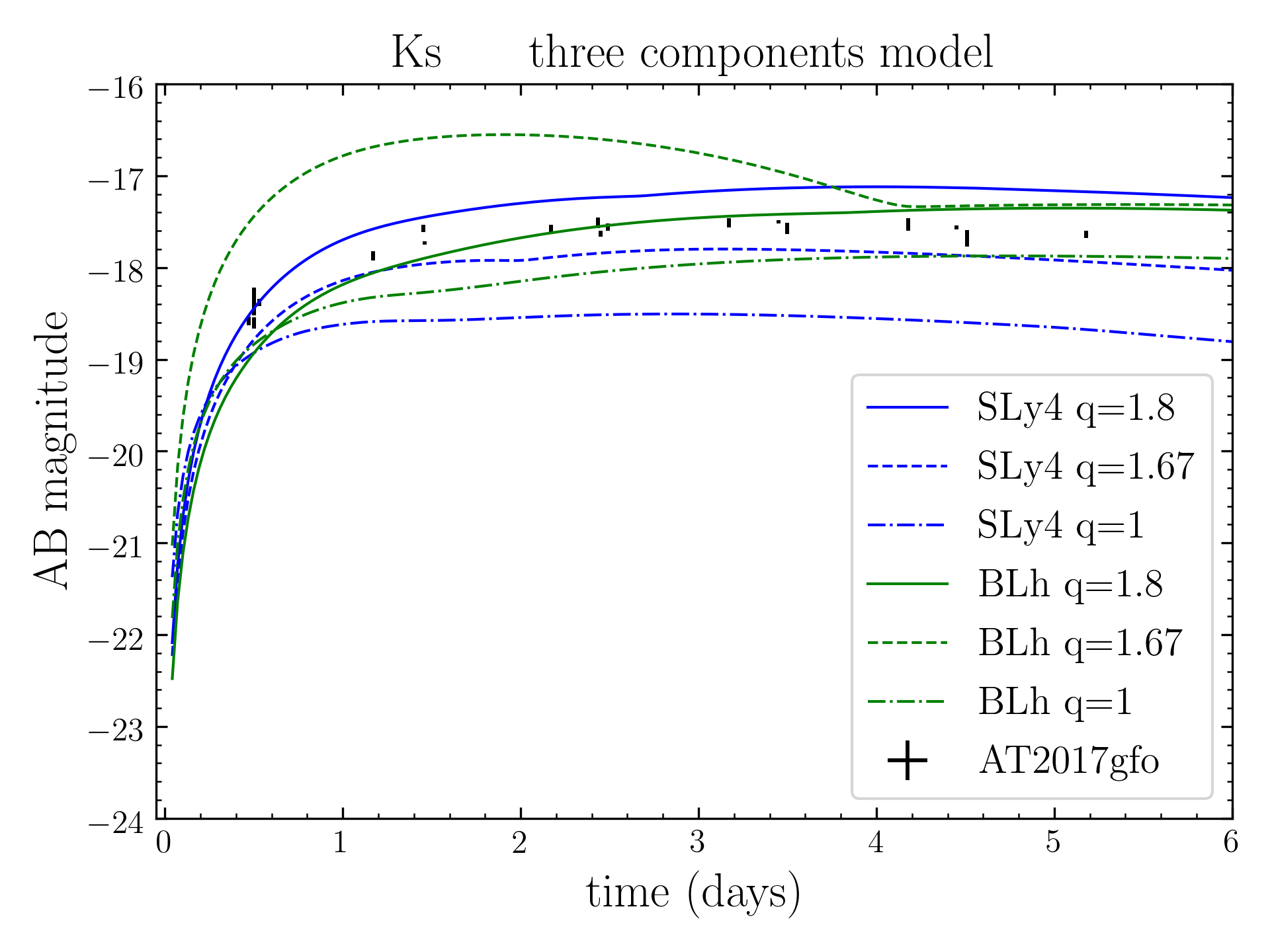}
\end{minipage}
\end{minipage}
\centering
\caption{Kilonova light curves as in Fig.~\ref{fig:1comp_lightcurves}, but employing a three component 
model for the BLh and SLy BNS. The dynamical ejecta component is taken as in
Fig.~\ref{fig:1comp_lightcurves}. The other two components are assumed
from a neutrino-driven and a viscosity-driven wind. The neutrino-wind
mass is assumed 5\% (1\%) of the disc mass if the remnant is a
long-lived (short-lived or promptly collapsing) massive NS; the
effective grey photon opacity is set $\kappa_{\rm w} = 1~{\rm
cm^2~g^{-1}}$, while the wind expands within a $\pi/4$ angle around
the polar axis with an average speed of $\langle v_{\rm w}\rangle =
0.08c$. The viscos-wind mass is assumed 20\% of the disc mass, expanding with
an average speed of $\langle v_{\rm v} \rangle = 0.06c$, and with a
grey opacity is set to $\kappa_{v}=5~{\rm cm^2~g^{-1}}$. The
observational data of AT2017gfo are shown as black markers for
comparison (see discussion in text).}
\label{fig:3comp_lightcurves}
\end{figure*}

The models presented in Fig.~\ref{fig:1comp_lightcurves} do not
contain potentially relevant contributions to the total ejecta coming
from disc winds. Thus, the resulting light curves could be considered
as lower limits for the kilonova emission. To estimate the potential
impact of the disc wind emission on our results, in
Fig.~\ref{fig:3comp_lightcurves} we also present light curves obtained
by considering a three component kilonova model for the same three
photometric filters and models of Fig.~\ref{fig:1comp_lightcurves}.
The dynamical ejecta profiles are NR informed as previously discussed.
For the disc winds, we consider both a neutrino-driven and a
viscosity-driven wind. Since wind ejection is expected over a wide
portion of the solid angle, we model the related kilonova emission
using again the framework described
in \citet{Perego:2017wtu,Radice:2018xqa,Radice:2018pdn}.  For the
neutrino-driven wind component, the amount of ejecta is assumed to be
5\% (1\%) of the disc mass if the remnant is a long-lived (short-lived
or promptly collapsing) massive NS. Due to the effects of neutrino
irradiation, the effective grey photon opacity is set $\kappa_{\rm w}
= 1~{\rm cm^2~g^{-1}}$, while the wind expands within a $\pi/4$ angle
around the polar axis with an average speed of $\langle v_{\rm
w}\rangle = 0.08c$.  For the viscous wind component, the amount of
ejecta is always assumed to be 20\% of the disc mass, expanding with
an average speed of $\langle v_{\rm v} \rangle = 0.06c$, while the
grey opacity is set to $\kappa_{v}=5~{\rm cm^2~g^{-1}}$.  To compute
the masses of the wind ejecta we consider the disc masses presented in
Sec.~\ref{sec:remnant}.

Since the disc ejecta is usually more
relevant than the dynamical one (see, e.g.,~\citealt{Radice:2018pdn}),
the large differences between $q=1$ and high-$q$ models in the
kilonova light curves observed in the one component models reduce for
the multi-component cases.  Nevertheless, since BNS mergers with
higher mass ratios tend to produce also more massive discs, also these
possibly more complete models confirm that BNS mergers undergoing
prompt merger can power bright kilonovae and high-$q$'s can possibly
produce kilonovae that are brighter and charaterized by wider peaks in
all relevant bands, compared to more symmetric mergers mergers that
have the same chirp mass.  More specifically, in the case of high-$q$
binary models for which the dynamical ejecta has a relatively large
mass (up to $10^{-2} \Msun$) and is highly anisotropic (e.g. BLh and
SLy4 $q=1.8$), the emission from the crescent is significant at all
time and possibly dominant for mergers forming discs of not too large
masses ($M_{\rm disc} \lesssim 0.1 \Msun$).  The opposite scenario is
realized in symmetric binaries: in all $q=1$ models, irrespectively of
the EOS, the low mass, widely distributed dynamical ejecta has a
visible impact on the light curves only at very early times and in the
blue portion of the kilonova spectrum. At later times, and especially
at red and infra-red frequencies, the emission is dominated by disc
winds.

%\bs{[Added here comparison with AT2017gfo. moved some text from
%conclusion to here.]}

The observations of AT2017gfo (\citealt{Villar:2017wcc} and refs20200807 therein) 
are also included in Fig.~\ref{fig:3comp_lightcurves} and can be qualitatively compared to
the lightcurves from the simulations (note that the simulated BNS have
chirp mass consistent with GW170817). The light curves from high-$q$
mergers are generically flatter and more extended in time than
those of AT2017gfo. 
Assuming these particular light-curve models, the
observation of AT2017gfo would exclude high-$q$ and stiff EOS with
$\tilde\Lambda\gtrsim600$ (long-lived NS remnants) consistently wih the
low-spin prior GW analysis \cite{Abbott:2018wiz,LIGOScientific:2018mvr}. 
The plots also highlight that the light curves in different bands favour 
different mass ratio, thus anticipating systematics (and degeneracies) 
between the multicomponents light curves and the binary parameters. 
We finally remark that the kilonova model employed here 
avoids the solution of the challenging radiative transfer problem in multi-dimension 
\citep[e.g.][]{Wollaeger:2017ahm,2020ApJ...889..171K,Bulla:2019muo} and approximates 
the time- and frequency-dependent $r$-process opacities with constant, gray opacities. 
This procedure likely introduces systematic uncertainties that are 
not easy to quantify. Direct comparisons between simplier analytical models and the 
outcome of radiative transfer calculations indicate that 
the former tend to predict lower luminosities and later peaks, especially for 
$\kappa \gtrsim 100~{\rm cm^2~g^{-1}}$ \citep{Wollaeger:2017ahm}. The usage of input 
parameters gauged on AT2017gfo and of opacities $\lesssim 25 ~{\rm cm^2~g^{-1}}$ 
possibly limits these uncertainties. We conservatively estimate a residual uncertainty 
of $\pm 0.5$ magnitude at peak. Even including these uncertainties, the qualitative 
differences between AT2017gfo and the light curves obtained for high-$q$ 
and stiff EOSs still hold.

%%%%%%%%%%%%%%%%%%%%%%%%%%%%%%%%%%%%%%		
\section{Conclusions}
\label{sec:concl}

In this paper, we explored in a systematic way the dynamics, the ejecta, and the
expected kilonova light curves of highly asymmetric BNS mergers by means of detailed
simulations in NR. The latter employed different finite-temperature, composition dependent 
EOS, and numerical resolutions.
The prompt collapse dynamics discussed here for high-$q$ BNS has a
underlying mechanisms different from the equal-masses prompt collapse: 
in the former case, the collapse is driven by the accretion of the companion onto
the massive primary star. For binaries with increasing mass ratio and
fixed chirp mass, the companion NS undergoes a progressively more
significant tidal disruption. Thus, in these BNS sequences
accretion-induced prompt collapse should be always present after a
critical mass ratio in connection to the maximum NS mass. For example,
for the BLh EOS the critical mass ratio should fall in the interval $1.54<q_\text{thr}<1.67$.

The remnan BH in these high-mass-ratio mergers is surrounded by a
massive accretion disc in contrast to comparable masses prompt
collapse merger that have no significant disc left outside the BH.
The accretion discs of high-mass ratio mergers are primarily
constituted of tidally ejected material, hence they are initially
cold and neutron rich. The simulations show that fallback of the tidal
tail perturbs the disc and affect its accretion. The long-term disc
and fall back dynamics is relevant to understand the complete kilonova
emission and also for GRB afterglow (extended)
emission \cite{Rosswog:2006rh,Metzger:2009xk,Desai:2018rbc}. 
This study is left for the future.

Perhaps the most relevant astrophysical consequence of our work 
is the possibility of having massive dynamical ejecta from these 
accretion-induced prompt collapsing remnant.
The ejecta mass can reach $M_\text{ej}\sim0.007-0.01\Msun$ and are mostly
emitted within $10^\circ-20^\circ$ about the orbital plane and in a portion
of $100^\circ-180^\circ$ in the azimuthal angle. The ejecta are neutron
rich with $Y_e\lesssim0.1$ and with velocities $v\lesssim0.1$~c.
The related kilonova light curves are predicted to be usually significantly 
brighter than the equal masses case (at fixed chirp mas)
in all the bands 
as a consequence of the crescent-like geometry of the
expanding dynamical ejecta. The light curves peak at later times and
are powered by the sustained emission of the innermost, hotter portion
of the crescent especially in the infrared bands.  

We suggest that the confident detection (or confident nondetection)
of an electromagnetic counterpart for a high-mass binary
can directly inform us about the 
binary mass ratio. Because the latter is currently poorly constrained
by GW analysis, the kilonova counterpart can deliver significant
complementary information. Multimessenger analysis of high-mass events
are thus particularly relevant. They will %multimessenger analysis will
require a precise numerical relativity characterization of the ejecta 
in terms of the binary parameters that is not currently available, as
well as improved nuclear and atomic physics input or suitably
parametrized models for the light curves.

Our results can help interpreting GW190425 in the scenario that the GW
was produced by an asymmetric binary with $q\gtrsim1.6$ (Note the
chirp mass for GW190425 is even larger than the one simulated here,
while large mass ratios are excluded for GW170817 if spins are small).
Using the methods developed in \citet{Agathos:2019sah}, \citet{Abbott:2020uma} estimated that 
the probability for the remnant to prompt collapsed to BH is
${\sim}97\%$.
The NR fitting models used in \citet{Agathos:2019sah} refer to equal
masses and thus are to be considered conservative for $q\gtrsim1.6$.
Hence, if GW190425 is interpreted as a such asymmetric BNS merger, the
BH remnant scenario is further strengthened by our results.
Moreover, a bright and temporally extended red kilonova could have
been expected as a counterpart if GW190425 was produced by a high-$q$
merger [Cf. \cite{2020arXiv200200956F}]. The kilonova signal in this case
could be similar to the one produced in BH-NS binaries \cite{Radice:2018xqa,Kyutoku:2020xka}.

All of our GW waveforms and ejecta data will be publicly available as
part of the \texttt{CoRe} database at 

{\small \url{http://www.computational-relativity.org/}}

%%%%%%%%%%%%%%%%%%%%%%%%%%%%%%%%%%%%%%		

\section*{Acknowledgements}
%%\fi
  SB, MB, BD, NO and FZ acknowledge support by the EU H2020 under ERC Starting
  Grant, no.~BinGraSp-714626.
  MB aslo acknowledges support from the Deutsche
  Forschungsgemeinschaft (DFG) under Grant No.
  406116891 within the Research Training Group RTG
  2522/1.
  Numerical relativity simulations were performed 
  on the supercomputer SuperMUC at the LRZ Munich (Gauss project
  pn56zo),
  on supercomputer Marconi at CINECA (ISCRA-B project number HP10BMHFQQ); on the supercomputers Bridges, Comet, and Stampede 
  (NSF XSEDE allocation TG-PHY160025); on NSF/NCSA Blue Waters (NSF
  AWD-1811236); on ARA cluster at Jena FSU.
  This research used resources of the National Energy Research
Scientific Computing Center, a DOE Office of Science User Facility
supported by the Office of Science of the U.S. Department of Energy
under Contract No. DE-AC02-05CH11231.
  Data postprocessing was performed on the Virgo ``Tullio'' server 
  at Torino supported by INFN.
  The authors gratefully acknowledge the Gauss Centre for Supercomputing
e.V. (\url{www.gauss-centre.eu}) for funding this project by providing
computing time on the GCS Supercomputer SuperMUC at Leibniz
Supercomputing Centre (\url{www.lrz.de}).

%%%%%%%%%%%%%%%%%%%%%%%%%%%%%%%%%%%%%%		
\appendix

%%%%%%%%%%%%%%%%%%%%%%%%%%%%%%%%%%%%%%		
\section{Experimental estimate of the remnant BH}
\label{sec:punc}

\begin{figure}
  \centering
  \includegraphics[width=.5\textwidth]{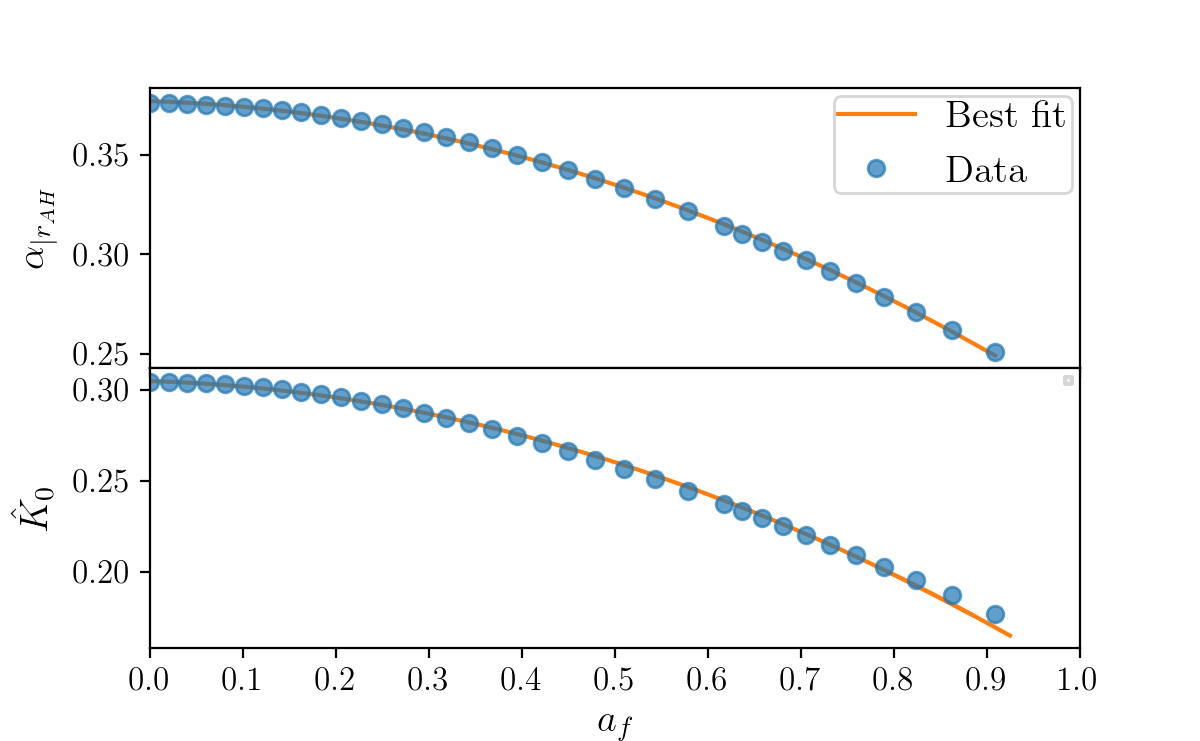}
  \caption{Dependence on the final black hole dimensionless spin of
    the lapse calculated at the apparent horizon (top) and the
    extrinsic curvature multiplied by $M_{\rm ADM}$ at the puncture
    (bottom) and relative best fits.}
  \label{fig:one_spinning}
\end{figure}

We perform single puncture experiments with the gauge conditions used
in the BNS simulations, and study the behaviour of the
lapse $\alpha$ and the extrinsic curvature trace $K$ close to the
puncture. We show that the
evaluation of the extrinsic curvature $K$ at the puncture (origin) 
allows us to estimate the BH spin and the lapse at the AH.
Further assuming an approximate value for the BH mass as given by the
quasiuniversal relation $M_\text{BH}\approx
M|e_b^\text{mrg}(\kappa^T_2)|\nu$ (upper bound), leads to a simple
estimate of both mass and spin of the BH.
Hence, these results could be useful as a simplified criterion for estimating BH
formation without a AH.

The gauge conditions for lapse and shift vector $\beta^i$ employed in
the simulations are \cite{Baker:2005vv,Campanelli:2005dd,vanMeter:2006vi,Brugmann:2008zz} 
\begin{align}
\p_t \alpha - \beta^i\p_i \alpha &=- \alpha^2 \mu_L K \\
\p_t\beta^i - \beta^j\p_j \beta^i &= \mu_S  B^i \\
\p_tB^i - \beta^j\p_j B^i &= \p_t \tilde{\Gamma}^i - \eta B^i \ ,
\end{align}
where $\tilde{\Gamma}^i$ are the conformal variables of
Z4c~\cite{Bernuzzi:2009ex,Hilditch:2012fp}, 
$\eta=1$ is a damping term, $\mu_S=3/4$, $\mu_L=2/\alpha$ are
the characteristic speeds. For simplicity  the initial data for one
puncture with different spins are prepared solving for two
punctures~\cite{Ansorg:2004ds} and
setting one mass much maller than the other $q\sim10^{12}$ and at a
distance smaller than the evolution grid spacing.
These simulations are performed with the \texttt{BAM} code
\cite{Brugmann:2008zz} with $6$ refinement levels and maximum
resolutions of $h\simeq 4.6875\times10^{-2},2.34375\times10^{-2}$.
During the evolution the system quickly
settles to a stationary solution with mass $M_\text{BH}$ and
dimensionless spin $a_\text{BH}$, both measured with the apparent
horizon finder. We then measure the lapse at the horizon
$\alpha_\text{AH}$ and the curvature at the puncture $K_0$.

Figure~\ref{fig:one_spinning} show the puncture's lapse at the horizon 
(top) and $\hat{K}_0 \equiv M_{\rm ADM} K(r=0)$ at the puncture (bottom) calculated for
various spin values. These quantities can be fit to
\be
\alpha_\text{AH} = 0.377 - 0.0146~a_\text{BH} - 0.139~a_\text{BH}^2
\ee
and 
\be
a_\text{BH} = \frac{-0.0161 + \sqrt{2.57\times 10^{-4} +
    0.585\left(0.305 -~\hat{K}_0\right)}}{0.292} \ .
\ee
The second fit was proposed also in \citet{Dietrich:2014cea} 
and the two results agree within the numerical precision of the data.

%%%%%%%%%%%%%%%%%
\section{Continuous dependency of dynamics on mass ratio}
\label{sec:q-dep}

\begin{figure}
    \includegraphics[width=.5\textwidth]{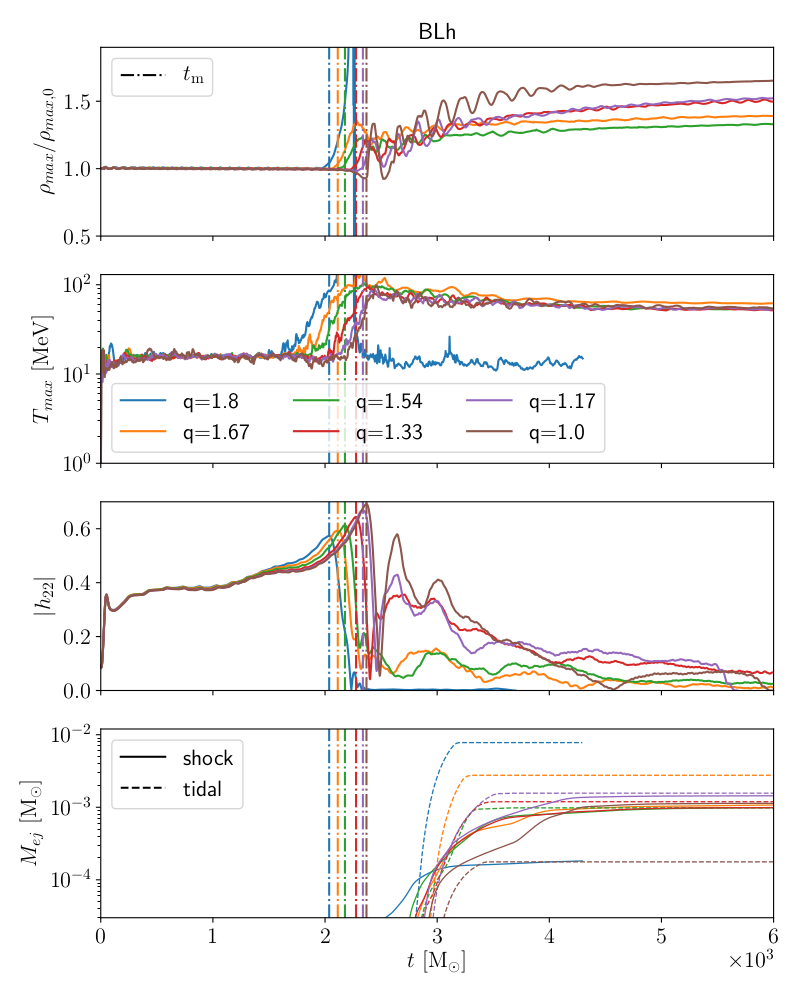}
	\caption{Main scalar quantities for several different
          mass ratio models with BLh. Each simulation presented here
          is run at grid setup LR. In the first
          panel we show the evolution of the maximum density
          ($\rho_\text{max}/\rho_0$), in the second panel the evolution of the
          maximum temperature ($T_{\rm max}$), in the third the 
          gravitational wave amplitude. The last panel shows the
          evolution of the total mass of the dynamical ejecta: with
          solid and dashed lines we highlight the contribution 
          of shock- and tidal-driven components respectively. The
          vertical dashed lines in all  
          panels indicate merger time for each simulation.}
        \label{fig:smoothq}
\end{figure}

We consider here simulations of a sequence of BNS with the BLh
EOS, fixed chirp mass and increasing mass ratio. Note all simulations 
discussed in this Appendix are performed at LR. 
Figure~\ref{fig:smoothq} shows (from top to bottom) the evolution of 
the maximum value of density and temperature, the gravitational wave
amplitude of the dominant $l=m=2$ mode and the dynamical ejecta, split
into shock- (solid) and tidal-driven (dashed) components.
For increasing mass ratio the dynamics smoothly converge towards the 
prompt collapse of the $q=1.8$ binary. This can be observed for both 
density and temperature maxima, as well as for the moment of merger.
On the contrary, the mass ejecta do not show a smooth dependence on the 
mass ratio. The highest mass ratios ($q=1.8,1.67$) exhibit large tidal-to-shocked
ratio, with the $q=1.8$ BNS showing almost no shocked ejecta. This is 
reversed in the equal mass model, where the shocked component is an order 
of magnitude larger than its tidal counterpart. The outlier models are 
the ones with mass ratios $1.17<q<1.54$. For these, the contributions 
from both components are comparable, with the $q=1.17$ model having overall 
the most amount of dynamical ejecta between the three BNS from both channels.
In the extreme $q$ cases, disruption of the lighter NS companion 
leads to tidally-dominated ejecta, while for equal mass NSs that reach 
merger only slightly tidally deformed the shocked components dominates. 

As a complement to the results, we show the violation of hamiltonian
constraint and the total baryonic mass conservation for these
simulations. The hamiltonian constraint violation is under control for
all simulations at all times, and violations are of the same order of magnitude.
The total rest-mass is conserved up to fractional level ${\sim}3\times
10^{-5}$ (approximately floating point precision) before 
merger for all the simulations. We stress that we use the
refluxing scheme \cite{Berger:1989a, Reisswig:2012nc} and that these
simulations are low resolution, thus the results should be considered
conservatives upper limits for the errors in SR and HR, which are
indeed smaller. The rest-mass drops affer merger mainly as a
consequence of the dynamical ejecta, that are typically one-to-two
order of magnitudes larger than the numerical errors.

\begin{figure}
    \includegraphics[width=.5\textwidth]{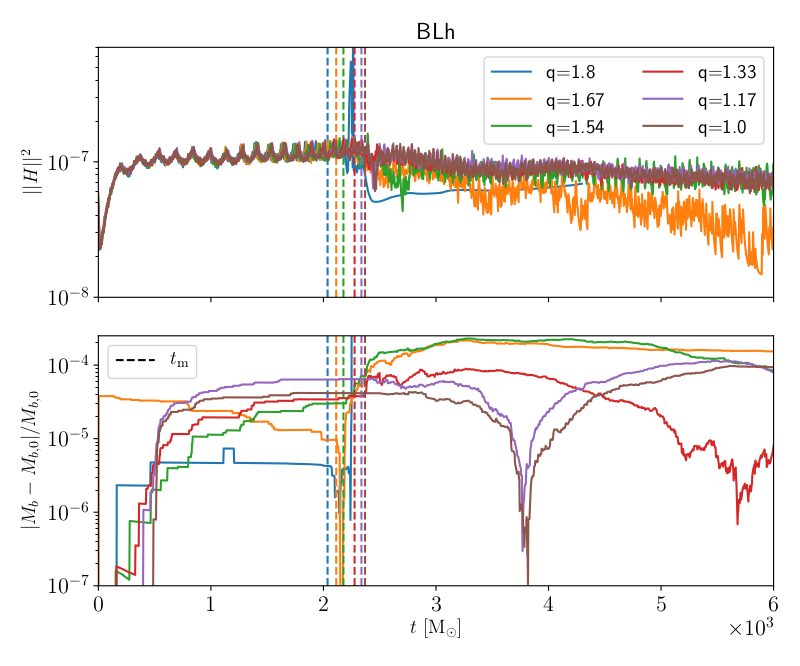}
	\caption{Simulation control quantities for several different
          mass ratio models with BLh. In the first panel we show the 
          2-norm of the Hamiltonian constrain ($||H||^2$), 
          while in the second panel we show the deviation of the total 
          baryonic mass ($M_b$) from the initial data value
          ($M_{b,0}$).}
        \label{fig:simcontrolq}
\end{figure}

%%%%%%%%%%%%%%%%%%%%%%%%%%%%%%%%%%%%%%		
\section{Quasiuniversal relations of binding energy and angular momentum at merger}
\label{sec:EJ}

In this appendix, we introduce NR fit formulae for binding energy
$e^{\rm mrg}_b$ and angular momentum $j^{\rm mrg}$ for BNS at the
moment of merger.  The fits are calibrated on 172 NR simulations with
$q\le 1.5$ extracted from the {\tt CoRe}
database~\cite{Dietrich:2018phi,Radice:2018pdn}. The fitted relations
are a rational functions parametrized with $\xi$, introduced in
Eq.~\eqref{eq:xi},
\be\label{eq:nrfitfunc}
F(\xi) = F_0\,\frac{1 + n_1 \xi+ n_2 \xi^2}{1 + d_1 \xi+ d_2 \xi^2}\,.
\ee
For the binding energy $e^{\rm mrg}_b$, the analyzed data span a range
from -0.065 to -0.043, and the best-fit coefficients are $F_0=
0.20179$, $n_1= -114.42$, $n_2= -0.39976$, $d_1= 286.19$, $d_2=2.2687$
and $c=1285.2$, where $c$ is defined in Eq.~\eqref{eq:xi}.  The
calibration has $\chi^2 = 6.8\times 10^{-3}$ and the intrinsic
uncertainty of the fit corresponds to ${\sim}7\%$ of the estimation,
referring to the 90\% credible regions.  Regarding the angular
momentum $j^{\rm mrg}$, the data have values between 3.3 and 3.8 and
the best-fit coefficients are $F_0= 0.028862$, $n_1=40.884 $, $n_2=
0.072754$, $d_1= 0.352$, $d_2=0.0004703$ and $c=1325.2$.  In this
calibration, we obtain $\chi^2 = 1.9\times 10^{-2}$ and the fit has an
uncertainty of ${\sim}3\%$ within the 90\% credible region.

\begin{figure}
\includegraphics[width=.5\textwidth]{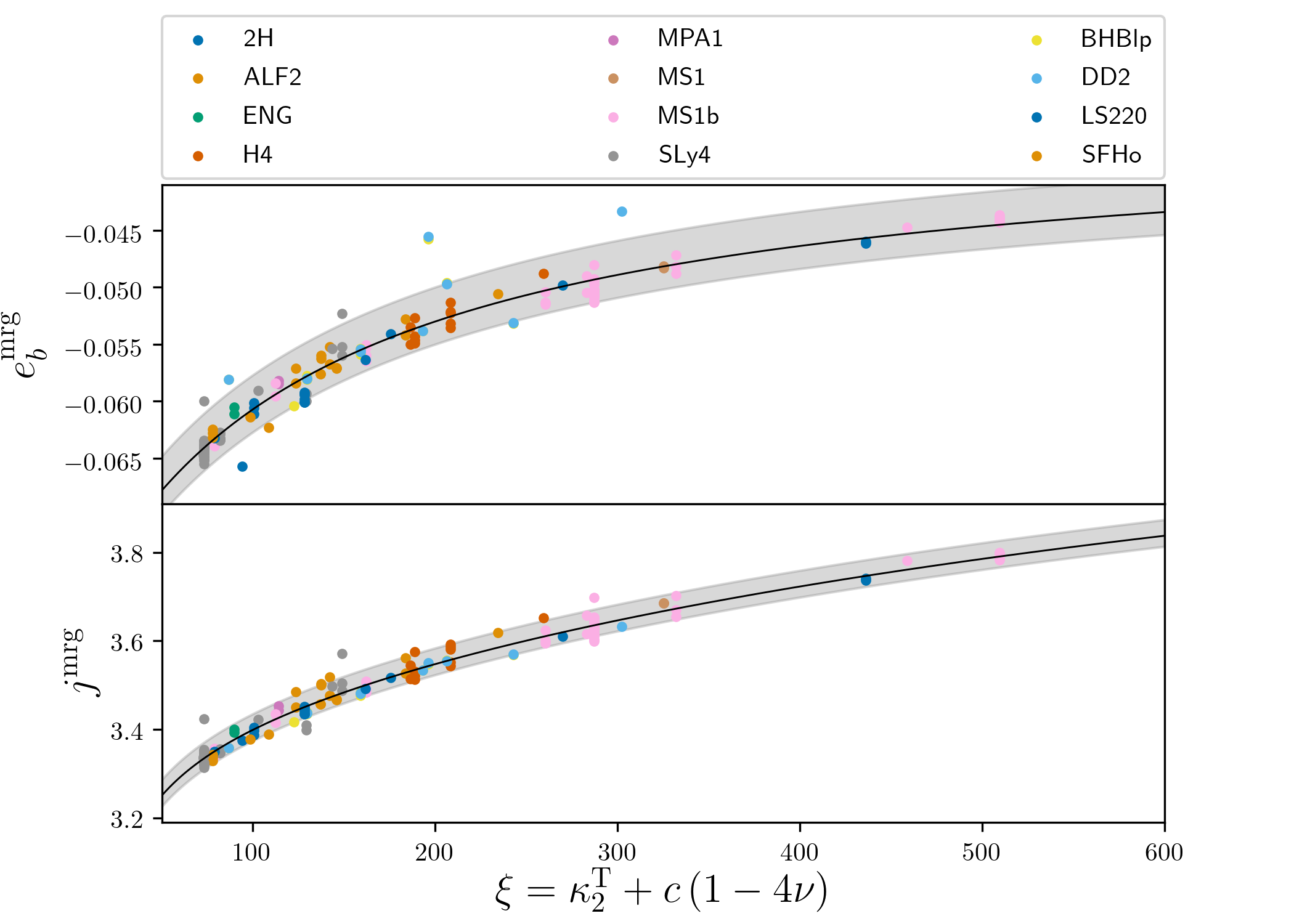}
\centering
\caption{NR fits for binding energy $e^{\rm
    mrg}_b=E^{\rm mrg}_b/(\nu M)$ and angular momentum $j^{\rm
    mrg}=J^{\rm mrg}/(\nu M^2)$ of a BNS at the moment of merger.  The
    employed data are extracted from NR simulations of BNS with
    $q<1.5$ included in the \texttt{CoRe} database.}
\label{fig:bam_thc_fits}
\end{figure}

%________________________________________________________________

%\bibliography{refs20200807,local}

\end{document}